\documentclass[useAMS]{mn2e}

\usepackage{graphicx}
\usepackage{amsmath}

\title[TeV gamma-ray emission from  globular clusters]
{TeV gamma-ray emission initiated by the population or individual millisecond pulsars within globular clusters}
\author[W. Bednarek, J. Sitarek \& T. Sobczak]
{W. Bednarek, J. Sitarek \& T. Sobczak \\ 
Department of Astrophysics, The University of \L \'od\'z,
ul. Pomorska 149/153, 90-236 \L \'od\'z, Poland,\\
bednar@uni.lodz.pl}
\begin{document}

\date{Accepted . Received ; in original form }

\pagerange{\pageref{firstpage}--\pageref{lastpage}} \pubyear{2015}

\maketitle

\label{firstpage}

\begin{abstract} 
Two energetic  millisecond pulsars (MSPs) within globular clusters (GC), J1823-3021A in NGC~6624 and PSR B1821-24 in M28, have been recently 
discovered to emit pulsed GeV gamma-rays. These MSPs are expected to eject energetic leptons. 
Therefore, GCs have been proposed to produce GeV-TeV gamma-rays as a result 
of the comptonization process of the background radiation within a GC. We develop this general scenario by 
taking into account not only the diffusion process of leptons within a GC but also their advection with 
the wind from the GC. Moreover, we consider distribution of MSP 
within a GC and the effects related to the  non-central location of the dominating, energetic MSP. Such more complete scenario
is considered for the modelling  of the GeV-TeV gamma-ray emission from the core collapsed GC M15 and also for GCs which contain recently 
discovered energetic MSPs within NGC~6624 and M28. The confrontation of the modelling of the gamma-ray emission with the observations with the present Cherenkov telescopes and the future Cherenkov Telescope Array (CTA) allows to constrain more reliably the efficiency of lepton production within the inner magnetosphere of the
MSPs and re-accelerated in their vicinity. We discuss the expected limits on this parameter in the context of expectations from the pulsar models. we conclude that deep observations of GCs, even with the present sensitivity of Cherenkov telescopes (H.E.S.S., MAGIC, VERITAS), should start to constrain the models for the acceleration and radiation processes of leptons within the inner pulsar magnetosphere and its surrounding. 
\end{abstract}

\section{Introduction}

GeV $\gamma$-ray emission has been recently discovered by the {\it Fermi}-LAT telescope from the direction of several globular clusters (GCs) (Abdo et al.~2009a, Abdo et al.~2010, Kong et al.~2010, Tam et al.~2011). In the case of two GCs, M28 and NGC~6624, exceptionally energetic pulsars have been also discovered, B1821-24 and J1823-3021A, respectively. The GeV $\gamma$-ray signals from these two GCs show periodicities with the periods of these pulsars (Freire et al.~2011, Johnson et al.~2013). Therefore, at present it is clear that the GeV emission from GCs is very likely due to a cumulative emission of the MSP population, as proposed already before this discovery (Harding et al.~2005, Venter et al.~2008, Venter et al.~2009). In fact, the level of GeV $\gamma$-ray emission allows to estimate the number of the $\gamma$-ray emitting MSPs within a specific GC (Abdo et al.~2010, Hui et al.~2011). Since MSPs are expected to inject energetic leptons from their inner magnetosphere, 
these results have triggered more attention of the modern Cherenkov telescopes. However, in most cases only upper limits on the TeV $\gamma$-ray flux have been reported, e.g. from Omega Centauri (Kabuki et al.~2007), 47 Tuc (Aharonian et al.~2009), M 13 (Anderhub et al.~2009), and  M15, M13, and M5 (McCutcheon et al.~2009). In the most complete study presented in Abramowski et al.~(2013), the upper limits of the TeV flux for several individual GCs (and also stacked upper limits) have been reported for the case of a point like and extended sources mostly for the GCs not detected by {\it Fermi}-LAT in GeV $\gamma$-rays. In the case of GC Ter 5, the longer observations ($\sim$ 90 hrs) resulted in the discovery of an extended TeV $\gamma$-ray source in the direction of this GC (Abramowski et al.~2011). Surprisingly, the centre of the TeV source is shifted from the centre of the GC by the distance corresponding to the dimension of the GC (Abramowski et al.~2011). This puts some doubts on the relation of this TeV $\gamma$-ray source to Ter 5.

GCs have been proposed to be potentially a new type of TeV $\gamma$-ray sources by Bednarek \& Sitarek~(2007). Specific models for the TeV emission in the pulsar scenario have been considered in a few papers (e.g. Bednarek \& Sitarek~2007, Venter et al.~2009, Cheng et al.~2010, Kopp et al.~2013, Zajczyk et al. ~2013). These works follow the standard scenario for the $\gamma$-ray production in which TeV $\gamma$-rays originate in the Inverse Compton (IC) scattering process of low energy radiation (optical from the GC, Microwave Background Radiation (MBR), or the infra-red and optical radiation from the galactic disk) by leptons accelerated by MSPs. The IC model also predicts synchrotron emission from the same population of leptons which might be observable in some favourable conditions between the radio and soft X-rays.
In fact, in the {\it Chandra} observations of GC Ter 5, the existence of an extended, non-thermal X-ray source centred on the core has been reported (Eger et al.~2010, Clapson et al.~2011). Similar result has been reported in the case of 47 Tuc (Wu et al.~2014).
Earlier observations have also reported evidences of an extended X-ray emission from some GCs which has been interpreted as a result of the interaction of the wind from the GC with the surrounding medium (Hartwick et al.~1982, Okada et al.~2007).
However, such X-ray sources have not been detected in the direction of a few other GCs (Eger \& Domainko~2012).
Other models for the TeV $\gamma$-ray emission from GCs (or significant modifications of that general scenario) have been also proposed (see e.g. Cheng et al.~2010, Domainko~2011, Bednarek~2012).

In this paper we re-consider the IC model for the TeV $\gamma$-ray production in GCs. In respect to the earlier version of the model, proposed by Bednarek \& Sitarek~(2007), we consider more realistically the process of propagation of leptons within GC by adding their advection from the GC with the velocity of the GC wind. We also take into consideration non-central location of MSP within the GC and calculate the synchrotron spectra produced by leptons. Such improved model has been applied to predict the non-thermal emission from a few globular clusters. As an example, we show results of calculations for M15, NGC~6624 and M28.

\section{General model for gamma-ray production in globular clusters}  

We significantly develop the basic model for TeV $\gamma$-ray emission from GCs originally proposed by Bednarek \& Sitarek~(2007) and considered with different modifications in a few papers (mentioned in the Introduction). In this model TeV $\gamma$-ray emission originates
in the comptonization process of soft radiation within the GC (stellar produced by the huge population of stars, the Microwave Background Radiation, and also the infra-red and optical radiation from the nearby Galactic disk) by energetic leptons injected directly from the inner 
magnetosphere of the MSPs or re-accelerated in the collision regions between the MSP winds or the stellar winds. These leptons diffuse within the GC (and its surrounding) up-scattering from time to time soft photons to TeV energies. Here we develop this basic scenario by assuming that leptons injected into GC from MSPs not only diffuse within the cluster but they are also advected from the cluster with the mixed pulsar wind/stellar wind. The mixed pulsar/stellar wind is expected to form due to the mixture of the matter from winds of the red giant population within the cluster with the relativistic plasma injected from the population of MSPs. The velocity of this wind can be estimated by assuming that a significant part, $\kappa$, of the MSP spin down energy, $L_{\rm MSP}$, is taken by the matter. Then, the velocity of the mixed wind can be estimated from
(see also Bednarek \& Sobczak 2014),
\begin{eqnarray}
v_{\rm w} = \sqrt{{{2\kappa L_{\rm MSP}}\over{\dot{M}_{\rm RG}}}} \approx 1.7\times 10^8 \left({{\kappa L_{35}}\over{\dot{M}_{-6}\xi_{-1}}}\right)^{1/2}~~~{\rm cm~s^{-1}},
\label{eq1}
\end{eqnarray}
\noindent
where the total  mass loss rate of the matter from all red giant stars within GC is $\dot{M}_{\rm RG} = 10^{-6}\dot{M}_{-6}$ M$_\odot$
yr$^{-1}$, the spin down power of the MSP population within the GC, $L_{\rm MSP} = L_\gamma/\xi$, is estimated from the 
observed $\gamma$-ray power of all MSPs within the GC, $L_\gamma = 10^{35}L_{35}$ erg s$^{-1}$, $\xi = 0.1\xi_{-1}$ is the efficiency of the GeV $\gamma$-ray emission by MSPs normalized to their total spin-down power, and the mixing factor fulfils the condition, $\kappa\le 1$. 
Note that the red giants are expected to be the dominant suppliers of the mass into GCs. However, the mass loss rates of red giants in globular clusters are not well known.
Different estimates, from observations of individual stars, ranges between $10^{-9}$ M$_\odot$ yr$^{-1}$ to $3\times 10^{-7}$ M$_\odot$ yr$^{-1}$ (Boyer et al.~2008, Meszaros et al. 2009). Since usually about 100 red giants can be found in specific massive globular cluster, the mass loss in the range of $10^{-7}$ M$_\odot$ yr$^{-1}$ to $3\times 10^{-5}$ M$_\odot$ yr$^{-1}$ is expected.

\begin{figure}
\vskip 7.5truecm
\includegraphics{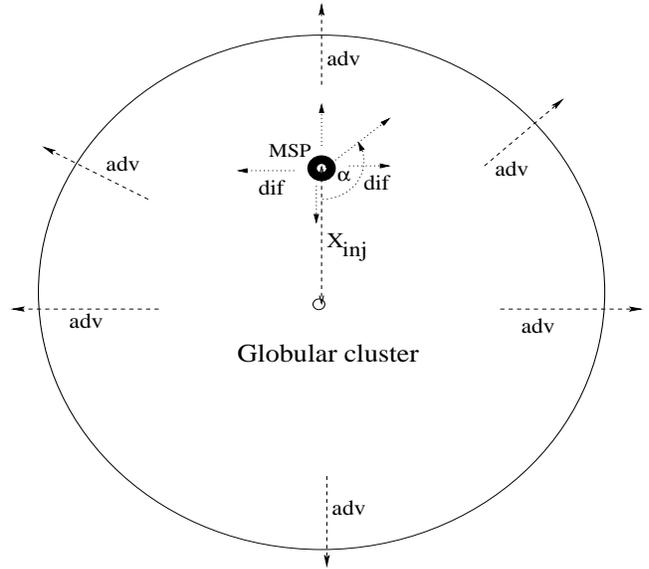}
\caption{Schematic representation of propagation of leptons injected from a MSP, at the distance $X_{\rm inj}$ from the centre of the globular cluster (GC) into the radiation field of the GC. Leptons, frozen within the plasma of the mixed stellar and the MSP winds, are advected from the GC with a specific velocity $v_{\rm adv}$. Additionally, they diffuse isotropically from the injection place in this wind. 
Their propagation within the cluster depends on the direction of diffusion defined by the angle $\alpha$ which is measured from the direction defined by the location of the MSP and the centre of the GC. The diffusion process is symmetric in the rest frame of the plasma but the advection process is generally oriented in the  outward direction from the centre of GC.}
\label{fig1}
\end{figure}

In the present work, we also consider the role of a single powerful MSP within the cluster. We assume that the pulsar is 
displaced from the centre of the GC. Since only the projected distance of the pulsar from the centre of GC can be known from observations, but not the real displacement, we consider the range of distances of the pulsar from the centre of GC. With such assumption, the propagation of leptons becomes quite complicated since the diffusion process is spherically symmetric in respect to the MSP but the advection process is assumed to be spherically symmetric in respect to the centre of the GC (see Fig.~1 for schematic illustration). 
As a result, e.g. leptons diffusing towards the centre of the GC (against the advection direction) can have relatively small drift velocities in respect to the centre of the GC in contrary to leptons diffusing outwards the centre of the GC. The leptons diffusing slowly towards towards the centre of GC lose energy on radiation processes more efficiently than leptons diffusing outwards the GC.  

Leptons, injected from the MSPs, diffuse radially from the pulsar. The distance scale for the diffusion process can be then expressed by,
\begin{eqnarray}
R_{\rm dif}(t) = \sqrt{2D_{\rm B}t}\approx 7.8\times 10^{12}(E_{\rm TeV}t/B_{-6})^{1/2}~~~{\rm cm},
\label{eq2}
\end{eqnarray} 
\noindent
where the Bohm diffusion coefficient can be expressed by 
$D_{\rm B} = R_{\rm L}c/3\approx 3\times 10^{25}E_{\rm TeV}/B_{-6}$ cm$^2$ s$^{-1}$, 
$R_{\rm L}$ is the Larmor radius of the leptons, $c$ is the speed of light, $E = 1E_{\rm TeV}$ TeV is the energy of the leptons, $B = 10^{-6}B_{-6}$ G is the strength of the magnetic field, and $t$ is the time measured from the moment of the injection of the leptons. The average diffusion velocity can be calculated by differentiating Eq.~2, 
\begin{eqnarray}
v_{\rm dif} = \left({{D_{\rm B}}\over{2t}}\right)^{1/2}\approx 3.9\times 10^{12}\left({{E_{\rm TeV}}\over{t B_{-6}}}\right)^{1/2}~~~{\rm {{cm}\over{s}}}.
\label{eq3}
\end{eqnarray} 
\noindent
Moreover, the leptons are also advected from the GC with the velocity $v_{\rm adv} = v_{\rm w}$ (see Eq.~1),
passing the distance $R_{\rm adv} = v_{\rm adv}t$. By comparing $v_{\rm dif}$ with $v_{\rm adv}$, we estimate the moment,
\begin{eqnarray}
t_{\rm adv/dif}\approx 5.3\times 10^{8}E_{\rm TeV}\dot{M}_{-6}\xi_{-1}/(L_{35}B_{-6})~~~{\rm ~s}.
\label{eq4}
\end{eqnarray} 
\noindent
at which advection process becomes dominant.
In the reference frame of the pulsar, diffusion of leptons is assumed to be spherically symmetric under the assumption of the homogeneous and isotropic magnetic field. If the magnetic field is inhomogeneous within GC, then the diffusion velocity will also depend on the direction.  We assume that the leptons are injected isotropically by the pulsar. The advection process is radially symmetric in respect to the centre of the GC as well as the radiation field produced by stars which drops with the distance from the centre of GC (Bednarek \& Sitarek~2007). Leptons diffusing in different directions, defined by the angle $\alpha$ (see Fig.~1), can produce $\gamma$-rays with various efficiency. 

The propagation of leptons is followed numerically by using the time step method. After every time step, $dt$, we determine the distance of the test lepton from the MSP and from the centre of GC by multiplying the velocity vectors defined above by the step time 'dt'. 
It is assumed that 
the average diffusion velocity vector (which is time dependent) is always directed away from the MSP and the advection velocity vector is directed from the centre of GC. The knowledge of the distance from the MSP is needed in order to take into account the change of the average diffusion velocity of these test leptons in time. The distance from the centre of GC is needed in order to correctly determine the density of the radiation field produced by the stellar population in GC. By using the time step method, we are also able to determine the location, $x_{\rm int}$, at which relativistic lepton interacts with a stellar photon or with a MBR in the IC process. This location is simulated by summing up 
contributions from every time step, $-ln(P_1) = \int [1/\lambda^{IC}_{\rm MBR} + 1/\lambda^{\rm IC}_{\star}(x)]c dt$, where $\lambda^{\rm IC}_{\star}(x)$ and $\lambda^{IC}_{\rm MBR}$ 
are the mean scattering lengths for the IC scattering in the stellar radiation and the MBR respectively at the distance 'x' from the centre of GC, and $P_1$ is a random number from (0,1). 
They are calculated by integrating the IC spectrum (in the general Klein-Nishina case, see Eq.~2.49 in Blumenthal \& Gould~1970) over energies of produced 
$\gamma$-rays. When the interaction place is determined, we simulate the type of the soft photon with which the test lepton interacts (stellar or MBR) by checking the condition, $P_2 = \dot{N}_{\star}/(\dot{N}_{\star} + \dot{N}_{\rm MBR})$, where $\dot{N}_\star = (\lambda^{\rm IC}_{\star})^{-1}$ and  $\dot{N}_{\rm MBR} = (\lambda^{\rm IC}_{\rm MBR})^{-1}$ are the scattering rates of leptons with specific energies in the stellar radiation field and in the MBR mentioned above, and $P_2$ is a random number from (0,1).
Finally, we simulate the energy of the gamma-ray photon produced in this inverse Compton scattering process on the type of radiation field determined above by applying the spectrum given by general Klein Nishina formula (Blumenthal \& Gould~1970). 
The energy of the gamma-ray photon, produced in the IC process, is subtracted from the energy of the parent lepton. Lepton also loses continuously energy on the synchrotron process. This synchrotron energy losses are subtracted from the energy budget of the test lepton after every time step. 
The cooling process of the test lepton, during its propagation in GC, is followed up to the moment when it moves outside the maximum distance from the centre of the GC, $R_{\rm int}$. Depending on the parameters of the considered scenario (e.g. such as the location of the MSP with GC, the energy of lepton, ...), we inject up to several thousand test leptons in specific direction in order to
obtain good quality spectrum of produced $\gamma$-rays. Described above numerical procedure is repeated for the test leptons injected from the MSP at different directions (defined by the angle $\alpha$) and at different locations of the MSP within GC. 
Note that the density of synchrotron photons within GC is typically $\sim$(4-5) orders of magnitude lower in the optical range than the density of the stellar radiation. Therefore, we neglect the contribution of the synchrotron radiation to the radiation field within the GC when calculating the $\gamma$-ray spectra from the IC process.

Two models for the injection spectrum of leptons into the GC are considered. In the first one, leptons are assumed to be injected from the pulsar inner magnetosphere with a close to mono-energetic spectrum (see e.g. Venter et al.~2009, Zajczyk~et al.~2013). Energies of these leptons are not well known but the values in the TeV energy range are usually expected. In the second model, we assume that the leptons are additionally accelerated in collisions of the pulsar wind with the winds from the GC stars or with the winds from other pulsars. In such case, the power law spectrum is expected due to the acceleration process occurring on the shocks (e.g. Bednarek \& Sitarek~2007, Kopp et al.~2013). The spectral index of such injected spectrum of leptons is expected to be close to -2 in the energy range between 100 GeV and the maximum energies allowed by the acceleration process and the energy losses or the escape from the acceleration region. These maximum energies are expected to be in the range between a few to a few tens of TeV (e.g. see discussion in Bednarek \& Sitarek~2007). 

We consider the case of lepton injection from the whole population of MSPs within the GC with the radial distribution
similar to that observed in normal stars within the GC. In another case, the propagation of leptons from a single non-centrally located, energetically dominating MSP is considered. In our model,
the wind cavity dominated by pulsar wind around specific MSP within GC is determined by the balance between pressure of the total GC wind and the specific pulsar wind.
The pressure of the GC wind is estimated from $P_{\rm GC} = \dot{M}_{\rm GC}v_{\rm adv}/(4\pi X_{\rm inj}^2)$, where $X_{\rm inj}$ is the injection distance from the centre of the GC. The pressure of the pulsar wind is $P_{\rm MSP} = L_{\rm MSP}/(4\pi R_{\rm MSP}^2c)$, where $R_{\rm MSP}$ is the distance from the MSP. By comparing these pressures, we estimate the radius of the wind cavity (in units of the distance of MSP from the centre of the GC) on $R_{\rm MSP}/X_{\rm inj}\approx 0.02(L_{35}/\dot{M}_{-6}\xi_{-1}v_8)^{1/2}$, where $v_{\rm adv} = 10^8v_8$ cm s$^{-1}$. The wind cavities around specific MSPs are expected to be rather small in comparison to the distance scale within the GC, even around such dominating MSP as observed in NGC~6624 and M28. Therefore, we can assume that leptons are injected from a point-like source centred on the MSP.

\begin{figure*}
\vskip 7.truecm
\includegraphics{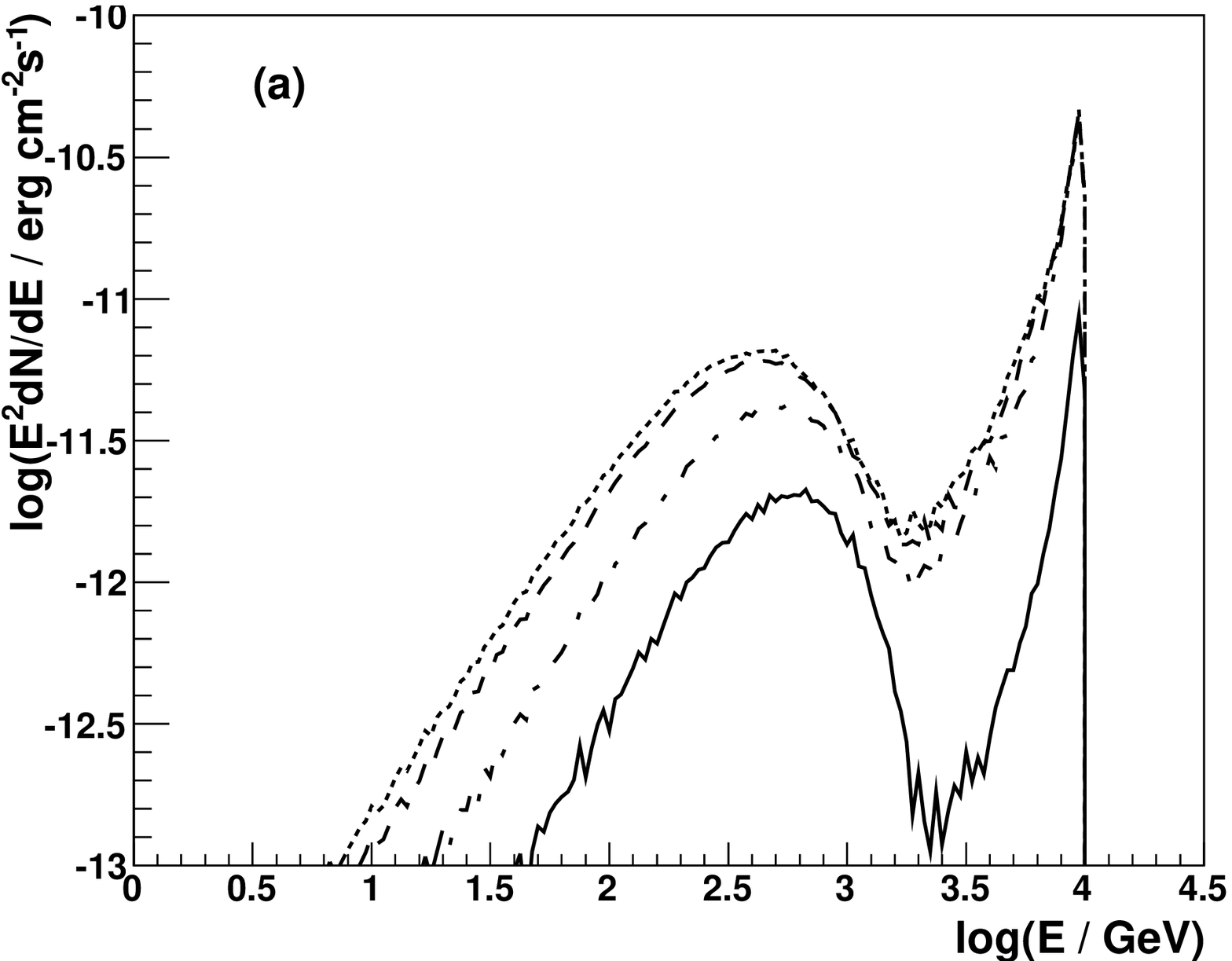}
\includegraphics{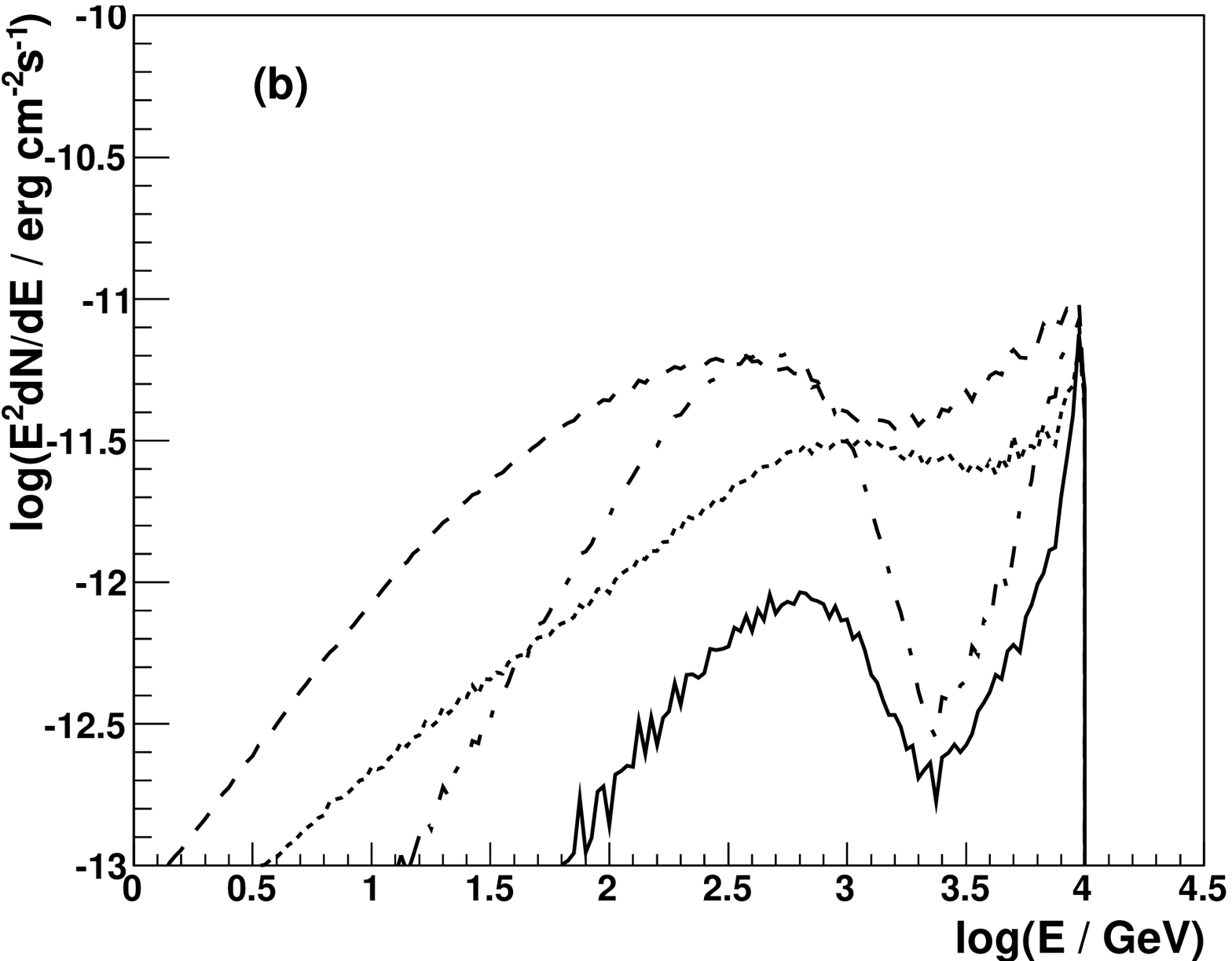}
\includegraphics{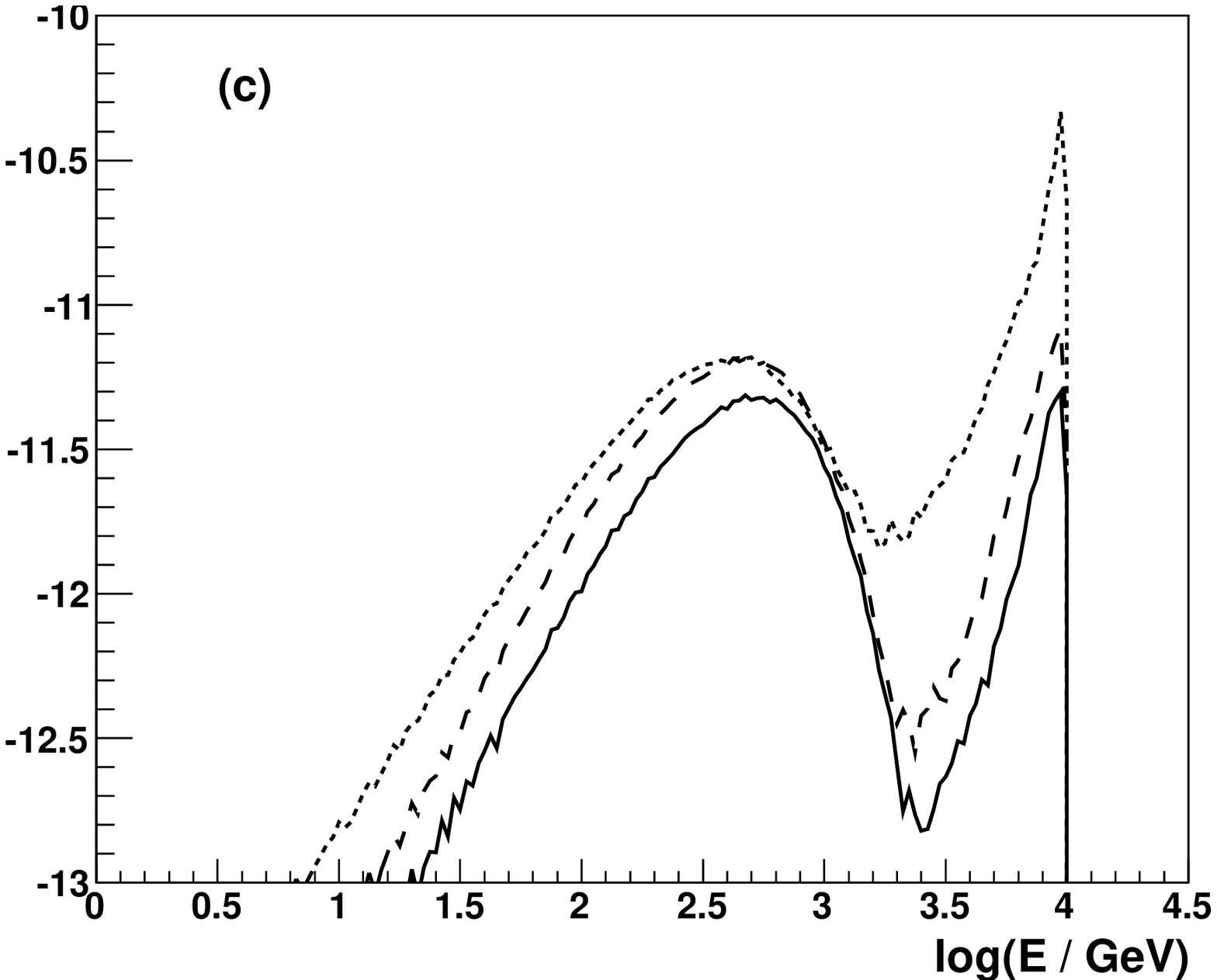}
\includegraphics{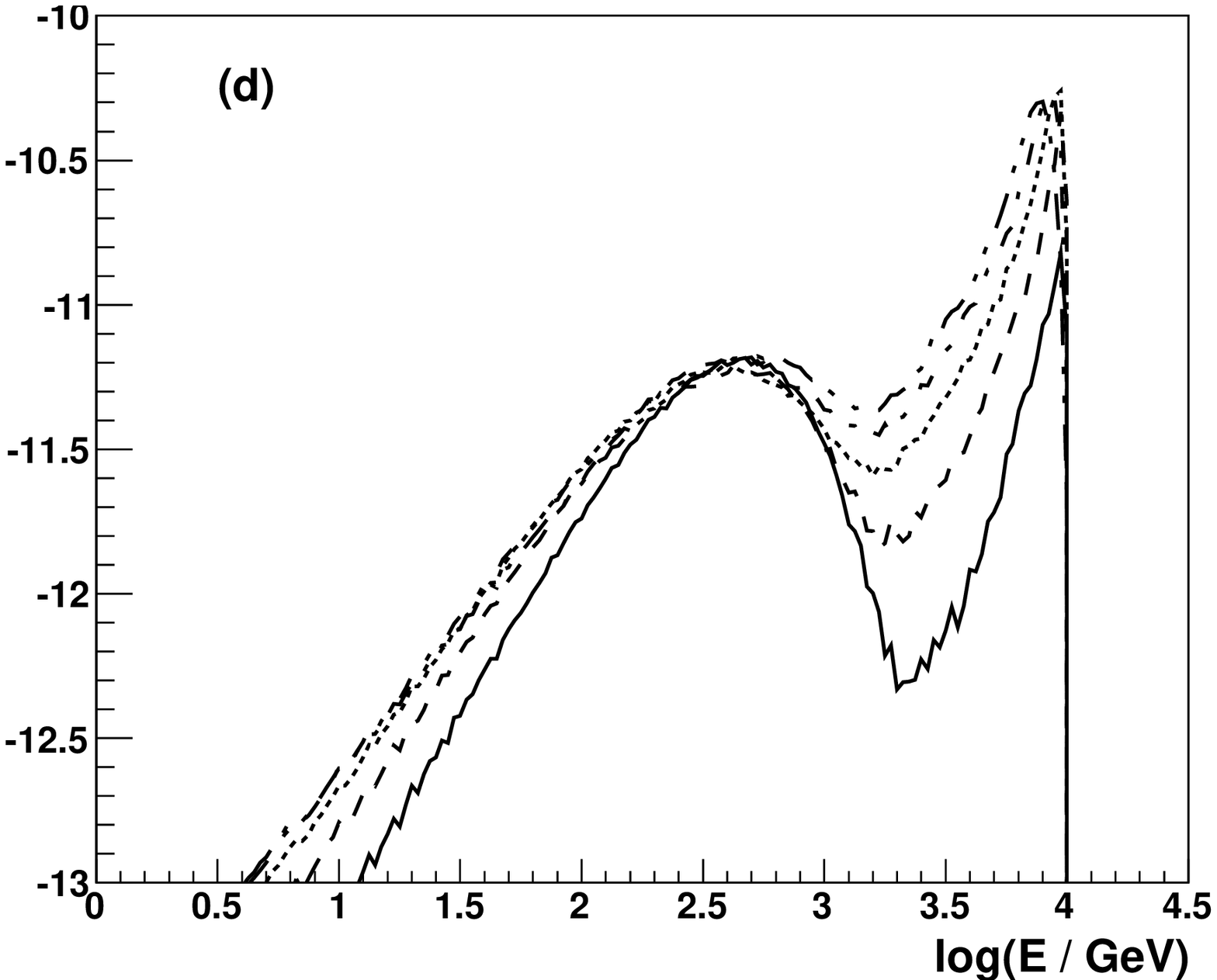}
\includegraphics{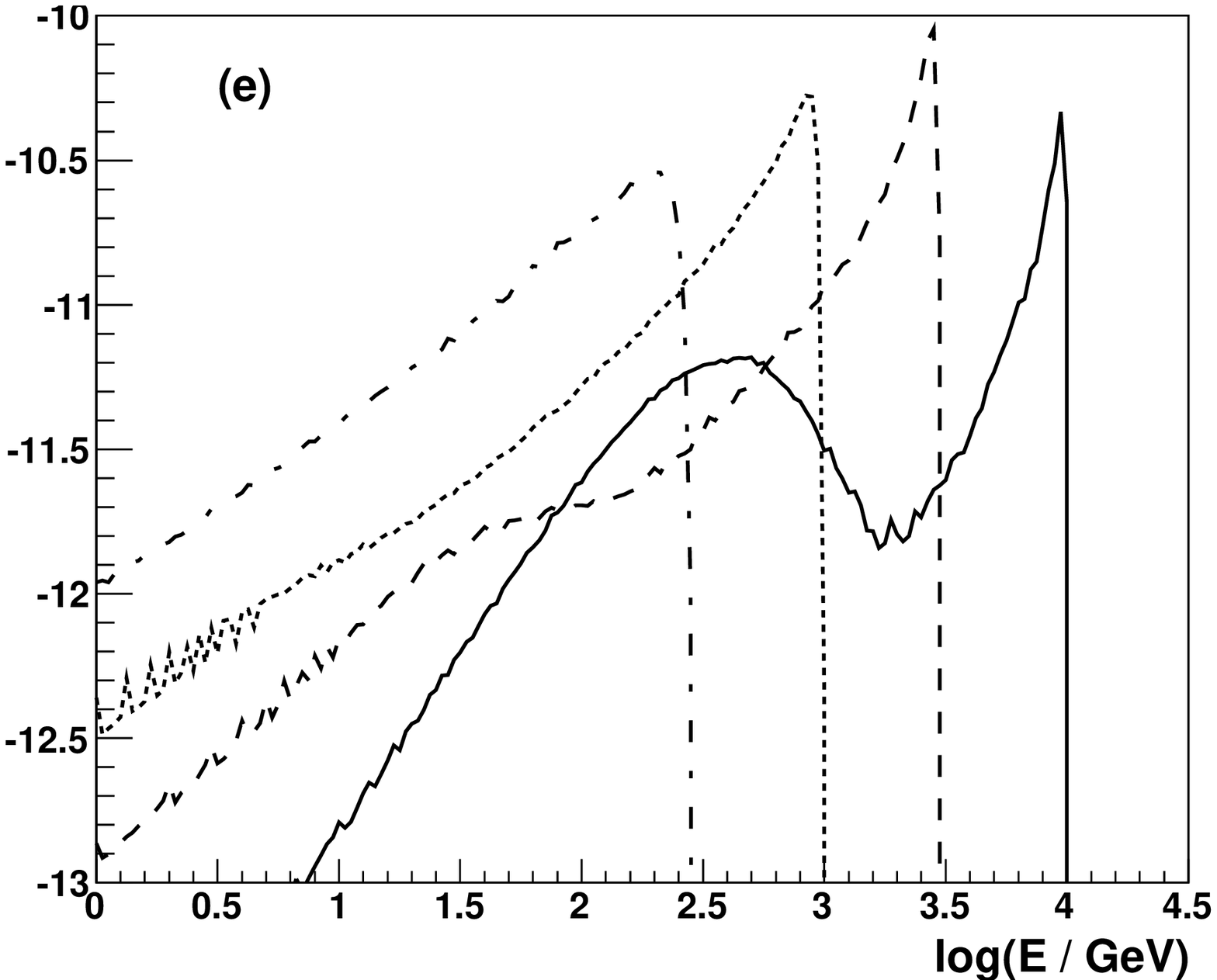}
\includegraphics{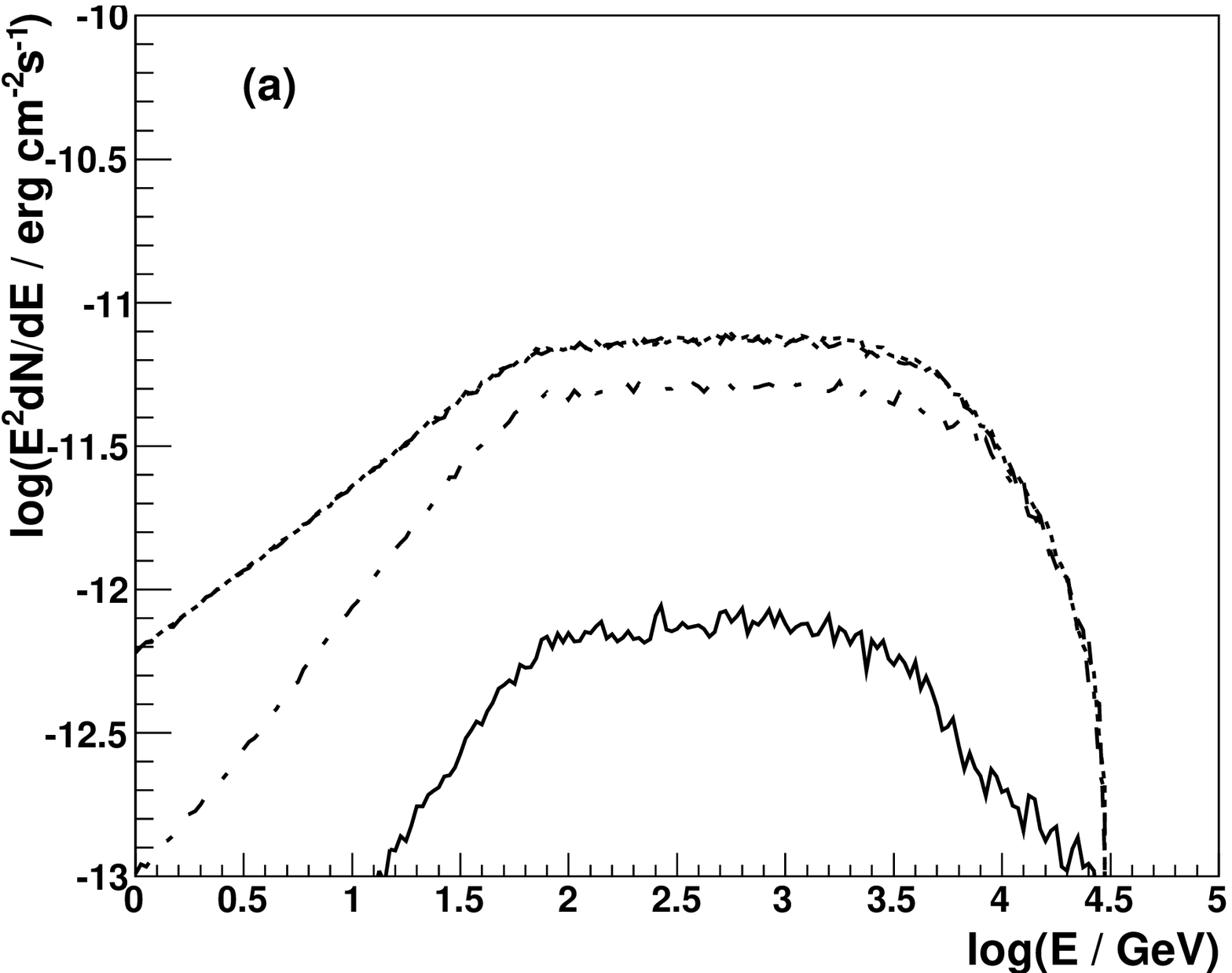}
\includegraphics{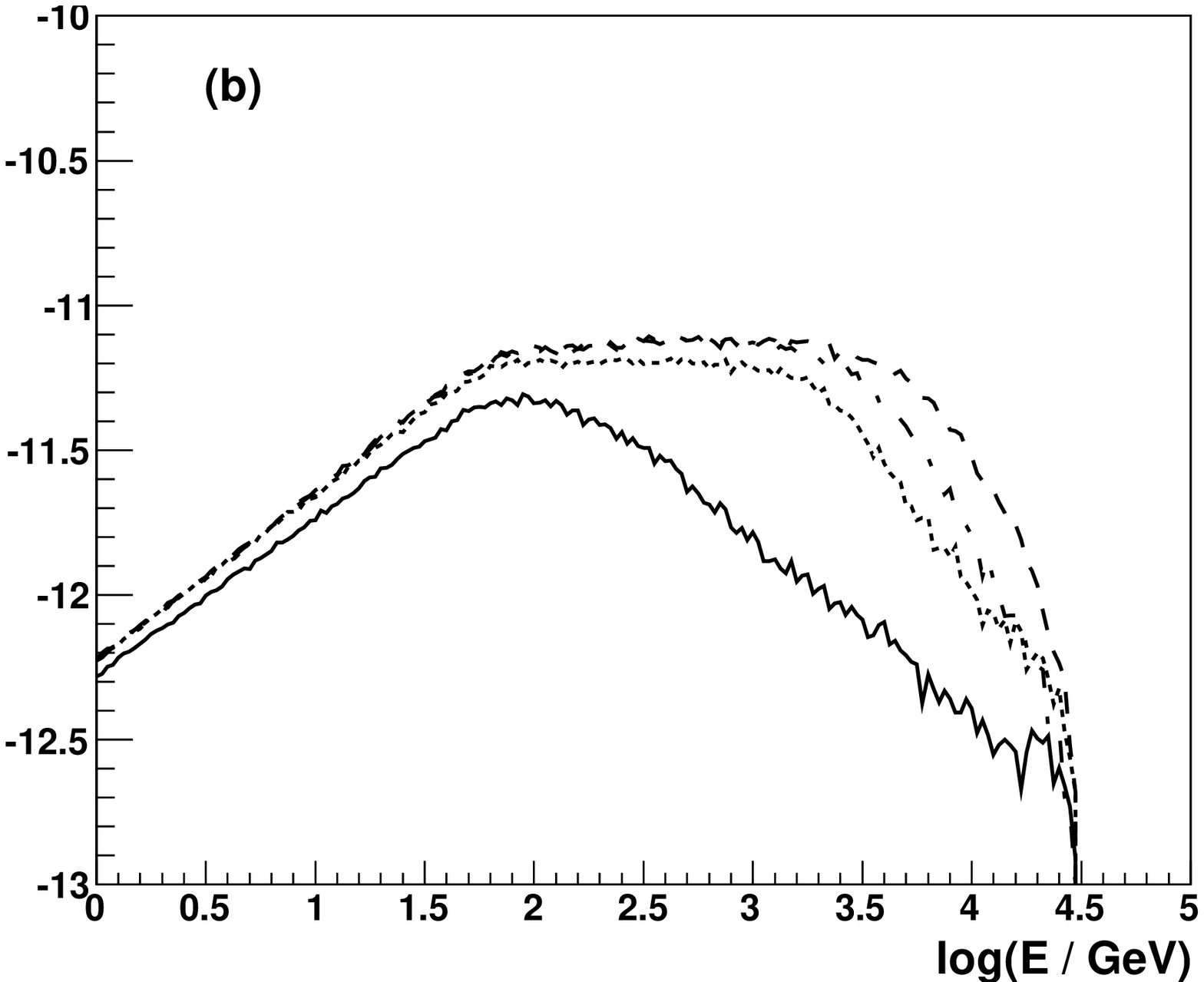}
\includegraphics{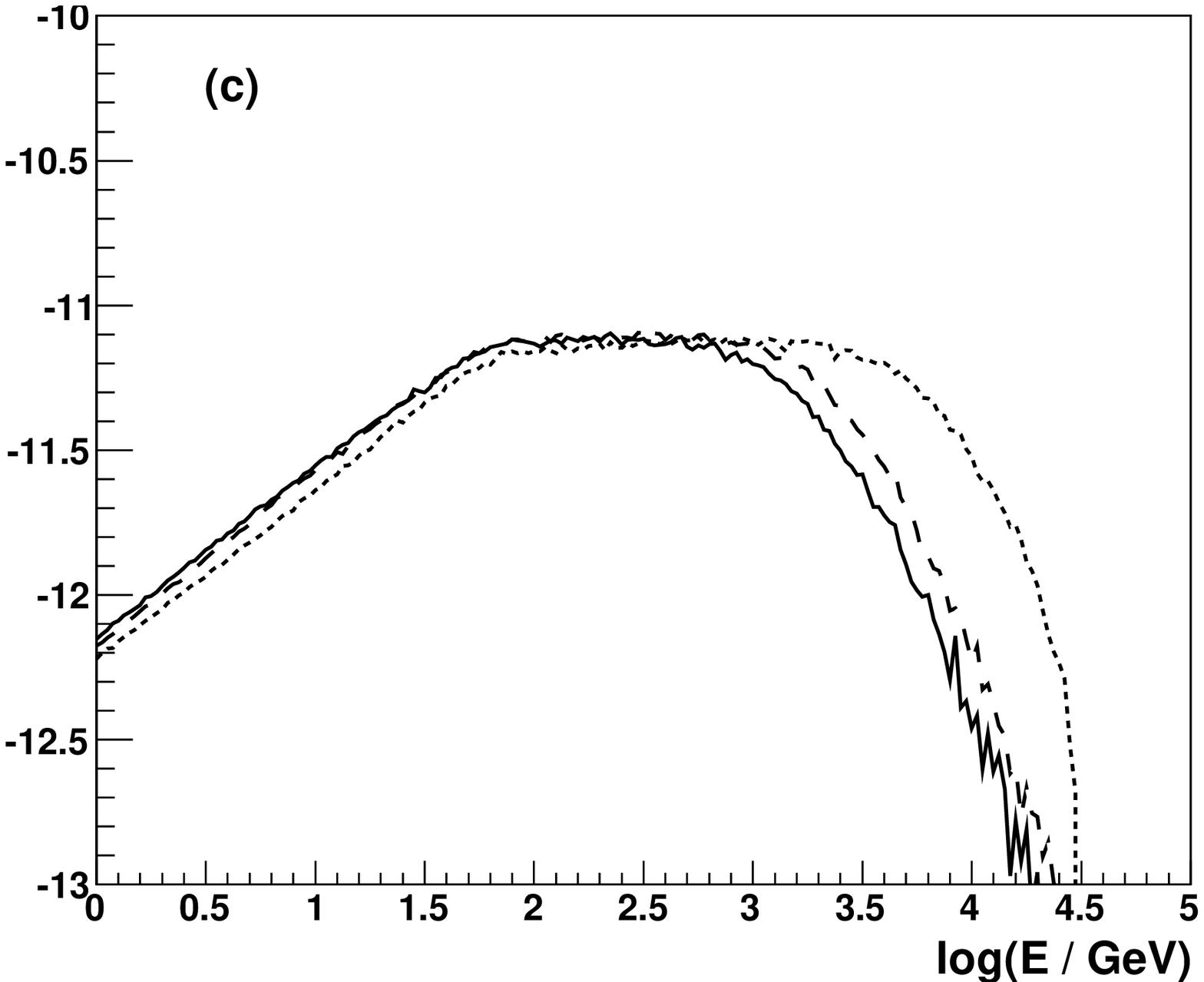}
\includegraphics{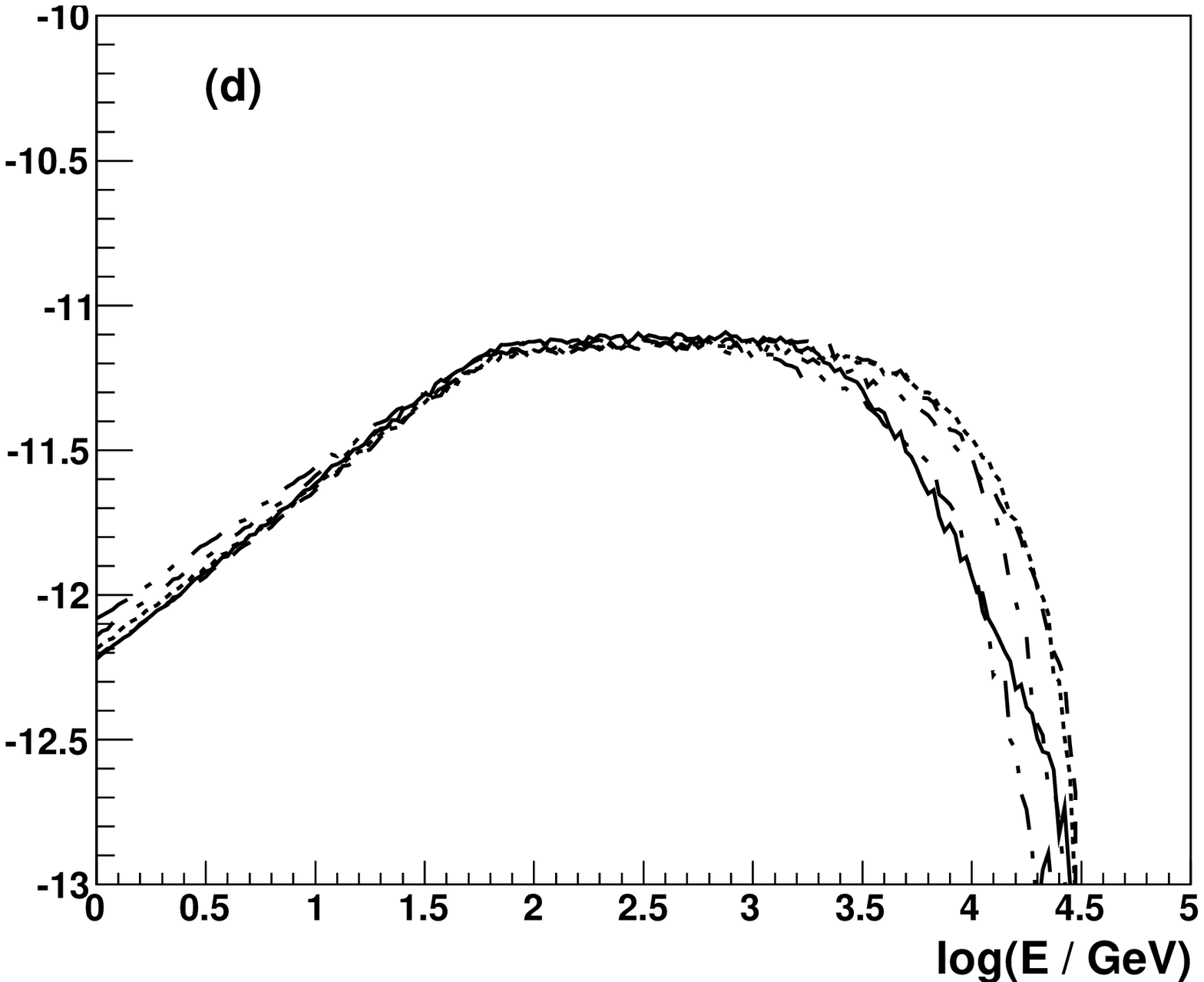}
\includegraphics{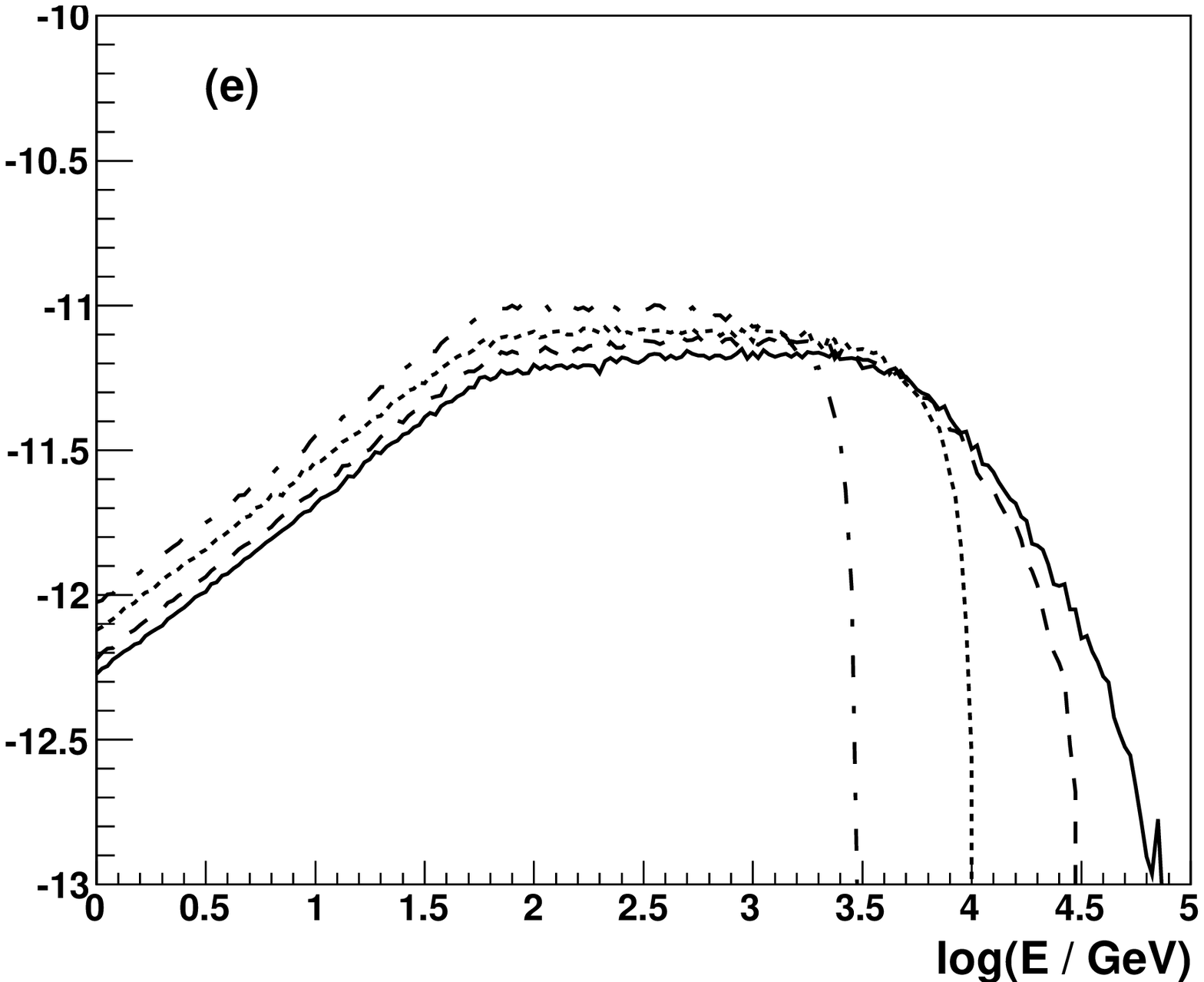}
\caption{Gamma-ray spectral energy distribution (SED) produced by mono-energetic leptons (upper panel) and leptons with the power law spectrum (bottom panel). These leptons comptonize stellar radiation field within the GC and the MBR. Specific figures show the spectra 
as a function of  the advection velocity $v_{\rm adv}$ equal to $10^8$ cm s$^{-1}$ (solid curve), $10^7$ cm s$^{-1}$ (dot-dashed), $10^6$ cm s$^{-1}$ (dashed), no advection (dotted) (figure (a)). The other parameters of the model are the following, magnetic field strength $B = 10^{-6}$ G, diffusion direction of leptons at the angle $\alpha = 1^\circ$, lepton energy $E_{\rm e} = 10$ TeV, the distance from the centre of the GC $X_{\rm inj} = 2$ pc. Dependence on the magnetic field strength for $B = 10^{-6}$ G (dot-dashed), $3\times 10^{-6}$ G (dashed), $10^{-5}$ G (dotted), $3\times 10^{-5}$ G (solid) (b). The other parameters are:  $\alpha = 90^\circ$, $X_{\rm inj} = 2$ pc and no advection included. Dependence on the direction of the advection defined by the angle $\alpha = 1^\circ$ (dotted), $90^\circ$ (dashed), and $179^\circ$ (solid) (c). The other parameters are: energy of leptons 10 TeV, magnetic field strength $10^{-6}$ G and not advection is included. Dependence on the distance from the centre of the GC $X_{\rm inj} = 0.1$ pc (solid), 2 pc (dashed), 4 pc (dotted), 6 pc (dot-dashed), and 8 pc (dot-dot-dashed) (d). The other parameters are: $\alpha = 1^\circ$, $B = 10^{-6}$ G, $E_{\rm e} = 10$ TeV, and no advection included. Dependence on the lepton energy $E_{\rm e} = 300$ GeV (dot-dashed), 1 TeV (dotted), 3 TeV (dashed), and 10 TeV solid) (e). The other parameters are $\alpha = 1^\circ$, 
$B = 10^{-6}$ G, $X_{\rm inj} = 2$ pc, and no advection included.
Leptons have the power law spectrum  with the spectral index equal to -2 between 100 GeV and the maximum energy equal to 30 TeV. In the bottom figure (e) the maximum  energies are equal to 3 TeV (dot-dashed), 10 TeV (dotted), 30 TeV (dashed), and 100 TeV (solid).
In these calculations the following parameters of the GC has been applied, luminosity $6.25\times 10^5$ L$_\odot$, the core radius 0.24 pc, the half mass radius 1.97 pc, the distance to GC 5 kpc, the millisecond pulsar has the period of 3 ms and the surface magnetic field of $2\times 10^9$ G, and lepton acceleration efficiency is $10\%$.}
\label{fig2}
\end{figure*}
\section{Gamma-rays from a single source of leptons in a GC}  

In order to have an impression about the dependence of $\gamma$-ray spectra on the parameters describing the source of leptons and their propagation, we perform numerical calculations in the case of injection of the leptons from a point-like source located at a fixed distance from the centre of the GC. Leptons are expected to diffuse in specific directions defined by the angle $\alpha$ which is the angle between the direction  of lepton diffusion and the direction towards the centre of the GC.
At first, the case of the mono-energetic leptons is considered from a single pulsar located at the distance, $X_{\rm inj}$, from the centre of the GC. 
Injection of quasi mono-energetic leptons is predicted based on the modelling of the radiation processes in the inner MSP magnetosphere. For example, in terms of the extended polar cap model, three-dimensional numerical modelling (see Dyks~2002, Zajczyk~2012) shows that leptons with Lorentz factors in the range $10^{5.5}-10^7$ should escape through the light cylinder radius (see also Zajczyk et al.~2013). Also leptons with a close to the mono-energetic spectrum, peaked at several TeV, are expected in the case of the polar cap, distorted magnetic field dipole model (Harding \& Muslinov~2011, Venter et al.~2015). Mono-energetic leptons escaping from the inner pulsar magnetosphere are expected to take a few percent of the spin-down energy of the MSPs (e.g. Venter \& de Jager~2005).

\begin{figure*}
\vskip 4.truecm
\includegraphics{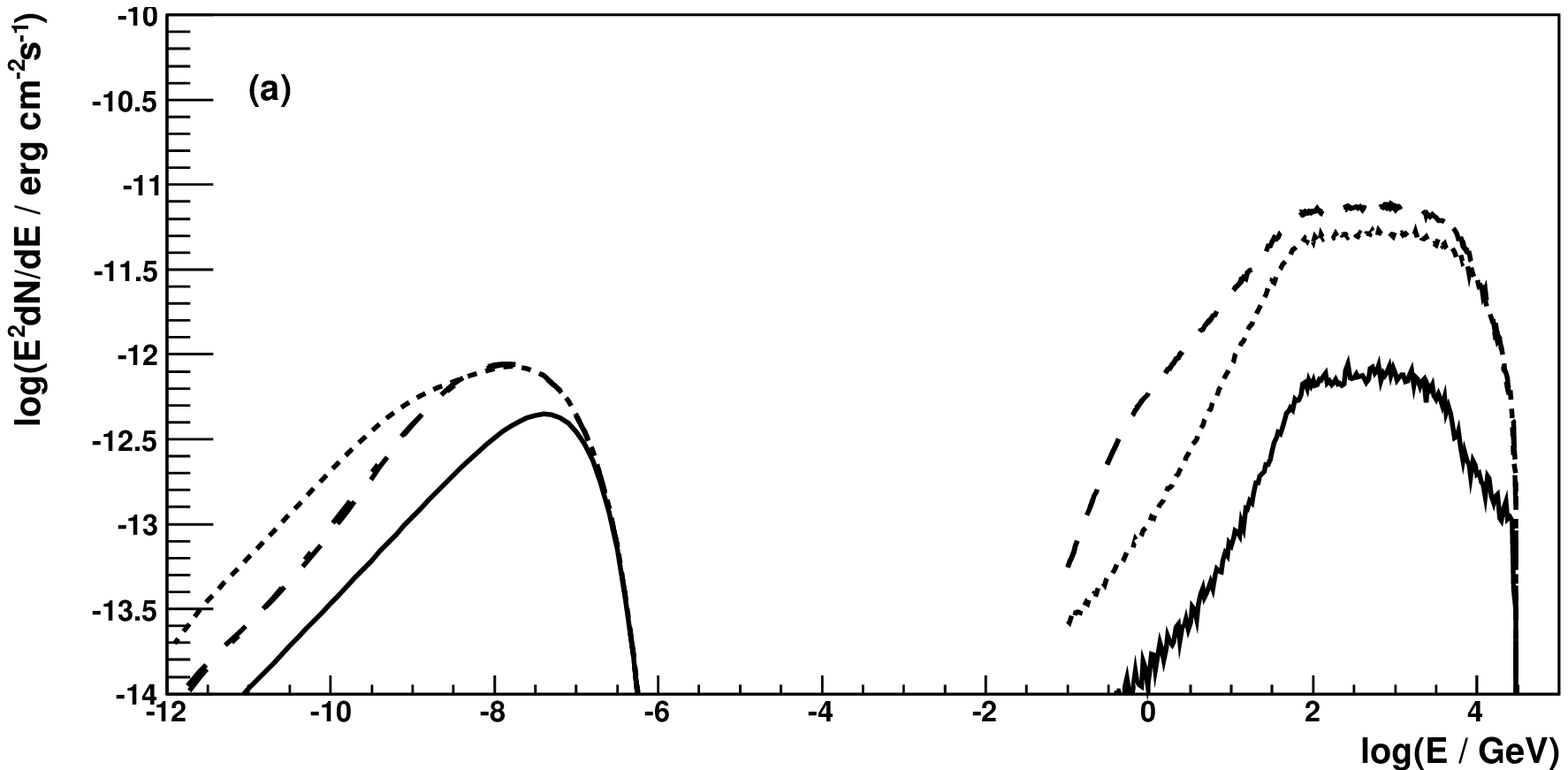}
\includegraphics{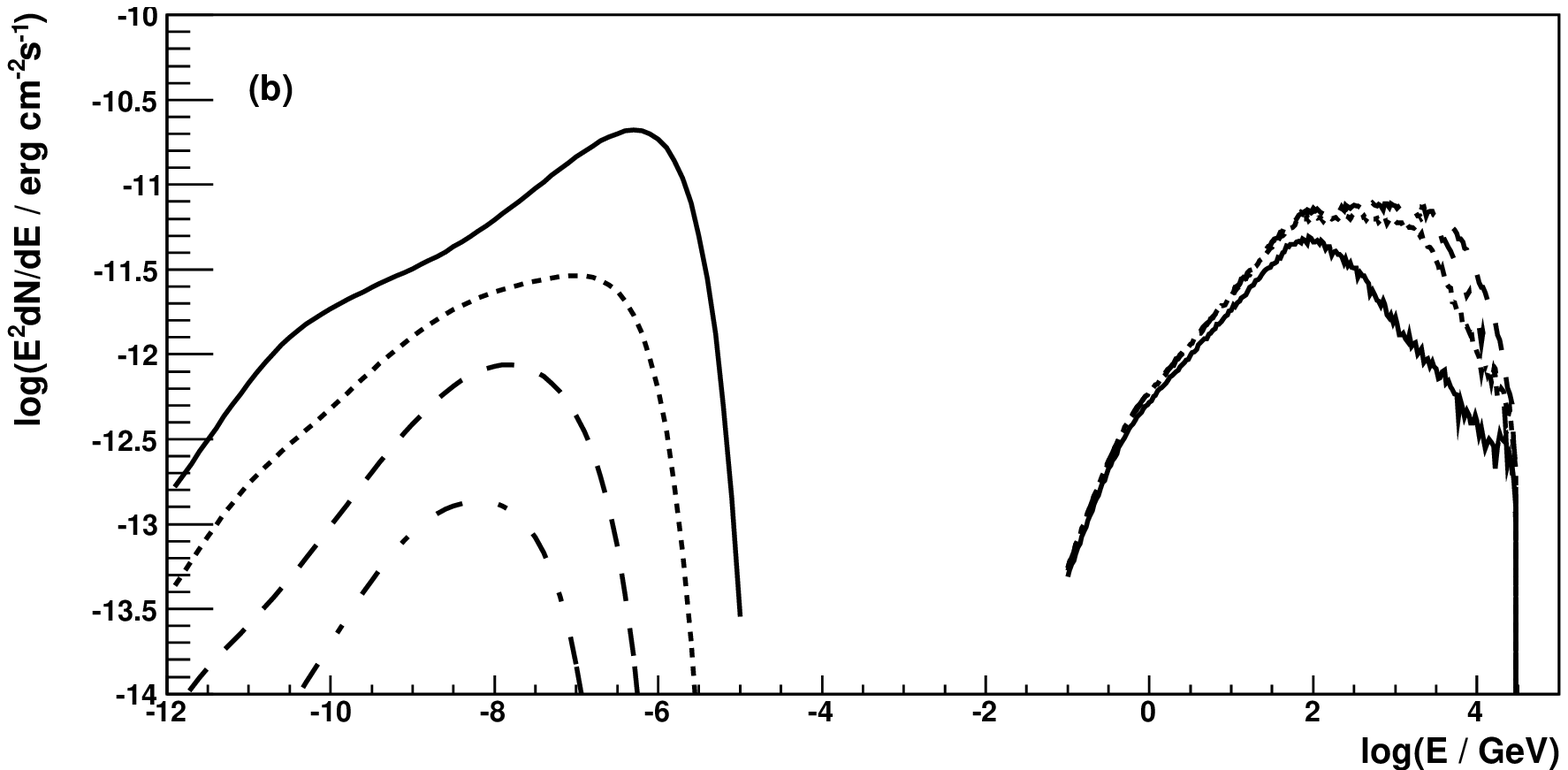}
\includegraphics{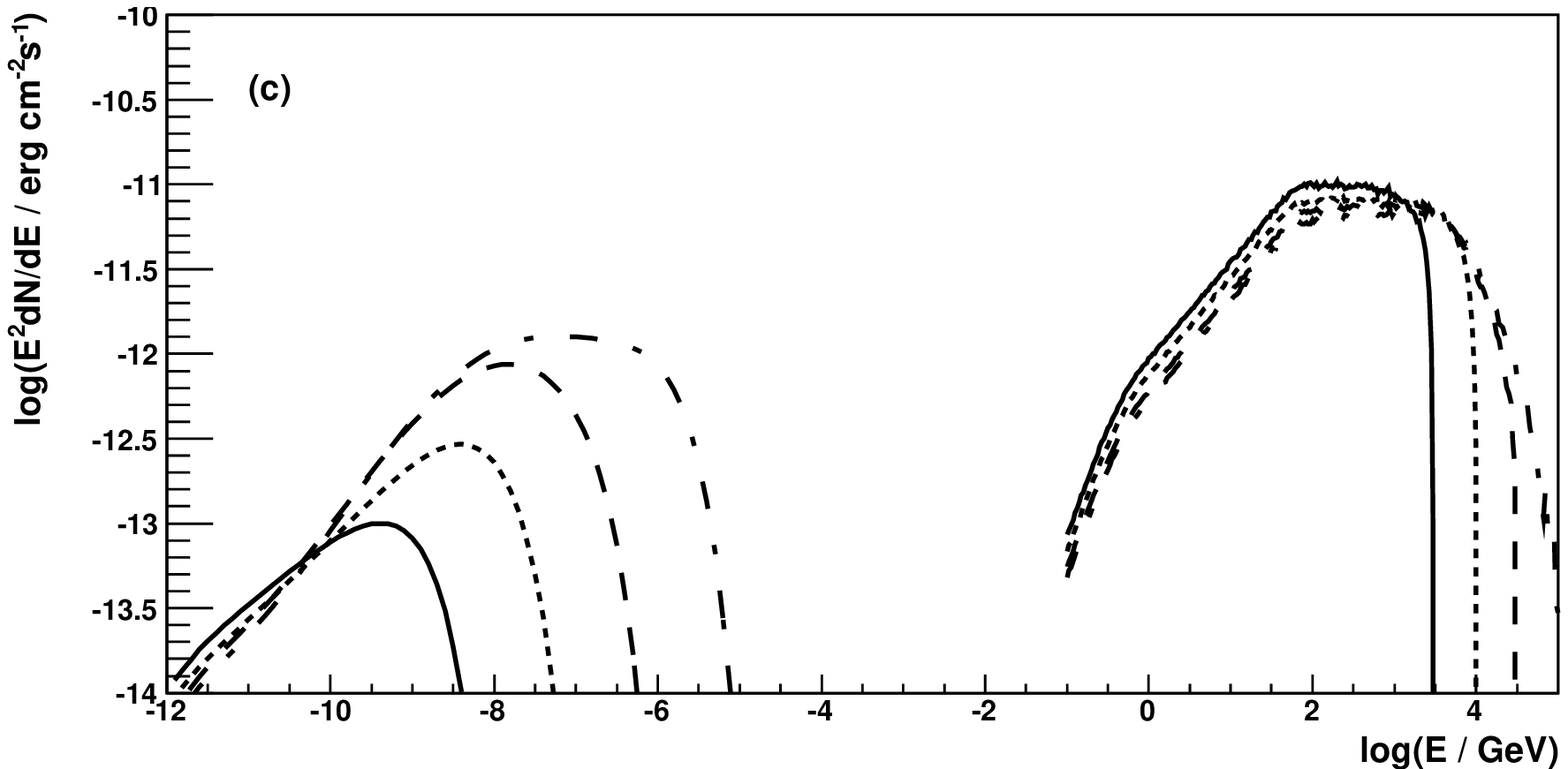}
\caption{The SED of synchrotron and IC $\gamma$-ray spectra produced by leptons, injected with a power law spectrum,
within the GC with parameters considered in Fig.~2. The dependence of the synchrotron spectra on the 
advection velocity of the GC wind is shown for $v_{\rm adv} = 0$ (dashed), $10^7$ cm s$^{-1}$ (dotted) and $10^8$ cm s$^{-1}$ (solid) (left figure). The spectra for different strength of the magnetic field  $B = 1\mu$G (dot-dashed), $3\mu$G (dashed), $10\mu$G (dotted) and $30\mu$G (solid) are shown in middle figure and as a function 
of the maximum energy of leptons  $E_{\rm max} = 100$ TeV (dot-dashed), 30 TeV (dashed), 10 TeV (dotted) and 3 TeV (solid)
in the right figure. The diffusion direction of leptons is equal to $1^\circ$ (inwards), the spectrum of leptons with the spectral index of -2 and the low energy cut-off at 100 GeV.  In the left and right figures the magnetic field has been fixed on $3\mu$G.}
\label{fig3}
\end{figure*}

Leptons are advected and diffuse through the region of the inhomogeneous radiation field of the GC comptonizing stellar radiation and the MBR.
In the upper panel of Fig.~2, the $\gamma$-ray spectra are shown for different advection velocities of the GC wind $v_{\rm adv}$, strengths of the magnetic field $B$  within the GC, diffusion directions of leptons defined by the angle $\alpha$, distances, $X_{\rm inj}$, of the injection place from the centre of the GC, and lepton energy $E_{\rm e}$. 
Although the mono-directional and mono-energetic injection of leptons represents the simplest possible example of our model, 
the dependence of $\gamma$-ray spectra on individual parameters does not have to be necessary monotonic.
In fact, we are considering here three radiation processes (the inverse Compton scattering of optical photons in the Klein-Nishina (KN) regime, the inverse Compton scattering of the MBR in the Thomson (T) regime and the synchrotron process). These three processes compete with each other for the energy of the same lepton and also with their escape processes, advection from GC and diffusion. This last process is determined by the magnetic field strength. Note moreover, that the densities of the stellar radiation and the MBR photons are comparable in the central region of GC but not in its outer parts. This introduces farther complication to this scenario. Also, the inverse Compton cross section, for scattering of optical photons by leptons with energies $\sim$10 TeV, becomes lower by about an order of magnitude in respect to the scattering of the MBR. Therefore, such complicated system is difficult to analyse analytically and numerical methods had to be applied.
 
Our numerical simulations show that the general $\gamma$-ray spectra are characterised by two features, i.e. a broad bump at sub-TeV energies (due to the comptonization of the MBR in the Thomson regime) and the Compton peak due to comptonization of the stellar radiation in the Klein-Nishina regime. The spectra look similar in the case of different advection velocities but their intensity drops significantly for fast winds from GC since leptons do not have enough time to comptonize (mainly stellar) photons to the $\gamma$-ray energies (see Fig 2a). The magnetic field strength has a non-trivial effect on the 
spectra in the region of the sub-TeV $\gamma$-ray broad peak (due to the comptonization of the MBR) in respect to the close to mono-energetic peak at TeV energies due to comptonization of the stellar radiation (see Fig 2b). For weak magnetic field ($B = 10^{-6}$ G), the diffusion process is fast. Leptons escape from the GC with relatively low energy losses and the $\gamma$-ray spectrum is dominated by the broad IC peak due to the scattering of the MBR. For stronger magnetic field, the synchrotron energy losses become comparable to the IC losses on the MBR. Therefore, the IC spectrum from scattering of the MBR does not increase proportionally to the magnetic field strength but becomes broader due to more effective energy losses of the leptons. The enhanced energy losses of the leptons (on the synchrotron and the IC scattering of the MBR) are also responsible for the broadening of the Compton peak due to the scattering of the optical radiation by these leptons.  For strong magnetic fields (e.g equal to $10^{-5}$ G), the synchrotron process dominates the energy losses of the leptons. The inverse Compton scattering of the MBR becomes inefficient but the Compton peak, due to the IC scattering of the optical radiation, is still relatively strong due to the slow diffusion of leptons through the central region of the GC where the optical radiation field is dense.
We also show how the spectra depend on the angle $\alpha$, describing the direction of diffusion of the leptons in the case of their mono-directional injection (see Fig~3c). We observe significant drop of the $\gamma$-ray emission, produced in the comptonization of the stellar radiation for larger diffusion angles since the radiation field strongly depend on this angle. 
On the other hand, the power in the higher energy Compton peak also strongly depends on the injection distance of leptons from the centre of the GC but the broad peak due to comptonization of the MBR is almost not influenced (see Fig~2d). This is due to the fact that the injection distance (for small injection angles) strongly effects the stellar radiation field seen by the diffusing leptons. We also investigate dependence of the $\gamma$-ray spectra on the energy of the injected leptons (Fig~2e). Note that the location and the strength of the broad bump, due to the scattering of the MBR, strongly depends on the lepton energy. It clearly appears only in the case on leptons with multi-TeV energies but disappears in the case of leptons injected with sub-TeV energies. It is due to the KN suppression in the IC cross section for the multi-TeV electrons scattering the stellar radiation. These electrons can then comptonize efficiently the MBR in the Thomson regime.
 
We have also calculated the $\gamma$-ray spectra assuming that leptons are injected into the radiation field of the GC
with a power law spectrum (see bottom panel in Figs.~2). The lower energy cut-off in the leptons' spectrum is expected to correspond to the energies of the leptons injected from the MSP into the GC radiation field. These leptons are additionally accelerated in turbulent collision regions at the shocks between the pulsar winds or between the pulsar wind and the stellar winds. In Fig.~2 (bottom panel) we show the $\gamma$-ray spectra calculated for the same parameters as assumed in the case of injection of the mono-energetic leptons. The effects of specific parameters on the $\gamma$-ray spectra in the case of leptons with the power law spectra are generally similar to the previously considered case of the mono-energetic injection 
of the leptons. The $\gamma$-ray flux drops drastically for fast GC winds and the spectra show evidences of steepening at the highest energies (bottom figure (a) in Fig.~2). Also the strong magnetic field in the GC results in additional steepening of the power law $\gamma$-ray spectrum at the highest energies due to the additional extraction of energy from the leptons by the synchrotron process (see bottom panel in Fig. 2b). The $\gamma$-ray spectra do not depend strongly on the diffusion angle of the leptons
at $\sim$TeV energies. However, these spectra cuts-off at lower energies for diffusion directions of the leptons at larger angles since the leptons are immersed for a shorter time in the stellar radiation of the GC (bottom Fig.2c). Similar effect is observed in the case of different injection distance from the centre of the GC for diffusion directions at small angles. The power law $\gamma$-ray spectra cuts-off at lower energies for larger injection distances (bottom Fig. 2d). As expected, the maximum energies of leptons determine the maximum energies of produced $\gamma$-rays. The cut-off in the $\gamma$-ray spectrum is observed close to the maximum energies of the injected leptons (bottom Fig. 2e). Moreover, the $\gamma$-ray spectra clearly steepen at the highest energies for large maximum energies of injected leptons (above $\sim$10 TeV). This effect is due to the strong synchrotron energy losses of the leptons with the highest energies. 

Our numerical code calculates also the synchrotron spectra produced by the leptons during their advection and diffusion processes through the GC. For typical parameters of the considered model, the synchrotron emission extends through the optical and the  
X-ray energy range. Therefore, it is expected that our model might be additionally tested with the observations of the diffusive emission from the GC in the X-rays. Because of that, we investigate the synchrotron emission from leptons
injected within the GC. From Fig.~3, it is clear that the synchrotron X-ray emission is expected from GCs only in the case of extreme
parameters. The magnetic field strength should be above $\sim$10 $\mu$G in order to produce  
soft X-ray emission by leptons with energies of 30 TeV (Fig.~3b,c). If the magnetic fields within GCs are rather weak ($\sim 3$ $\mu$G), leptons should be accelerated above $\sim$100 TeV in order to produce observable X-ray emission. In this case, the level of the synchrotron emission should even dominate over the level of the IC $\gamma$-ray emission. Fast advection of the leptons from the GC causes comparable drop in the flux of the synchrotron and the IC component (Fig.~3a). However, the synchrotron emission, produced by the leptons with more typical parameters, is expected to peak at the optical/UV energy range. This synchrotron emission will be difficult to extract from the dominating radiation field produced by the huge population of classical stars within the GC.

\section{Injection of leptons from the pulsar population in M15}

\begin{figure*}
\vskip 6.5truecm
\includegraphics{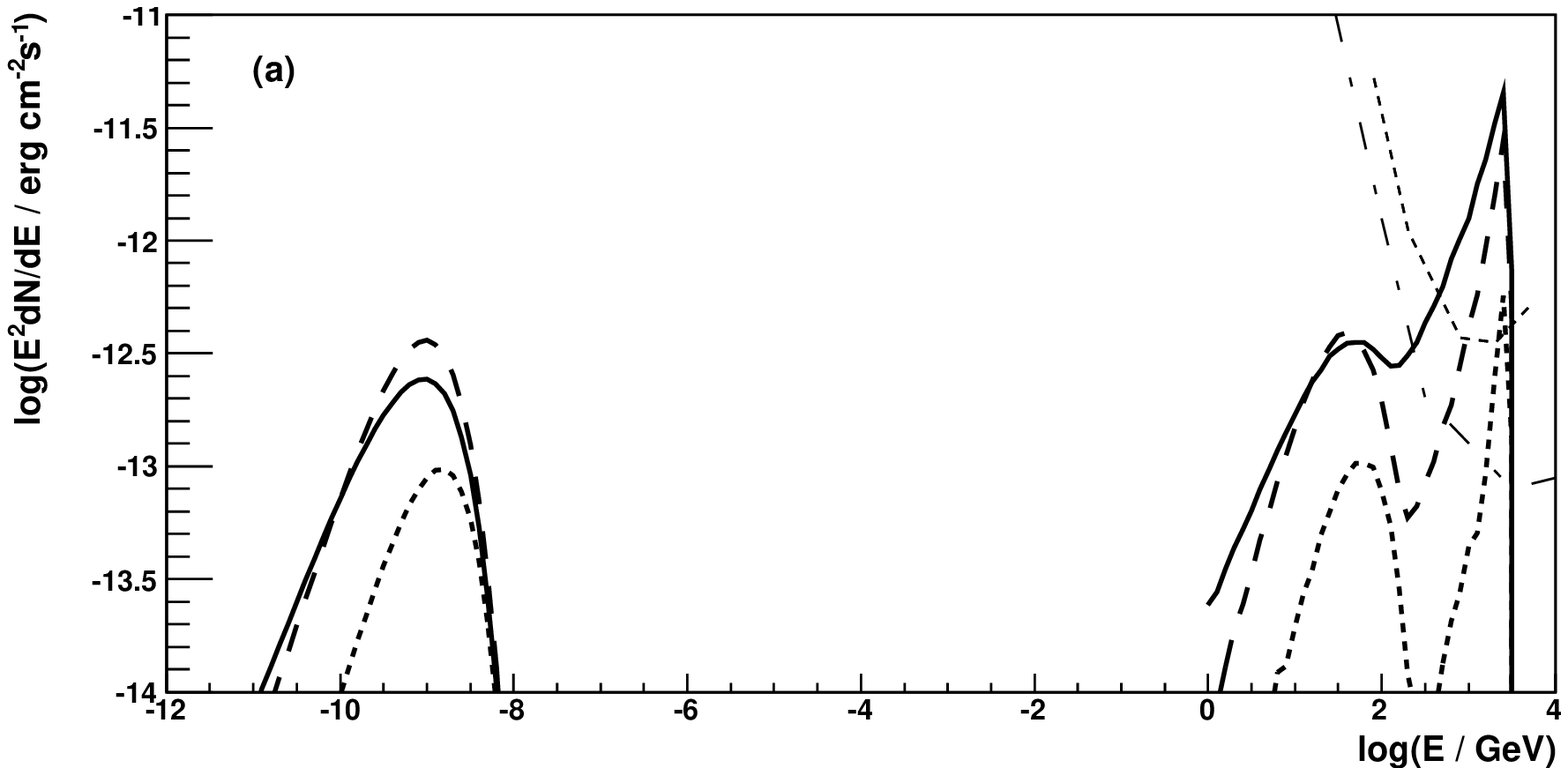}
\includegraphics{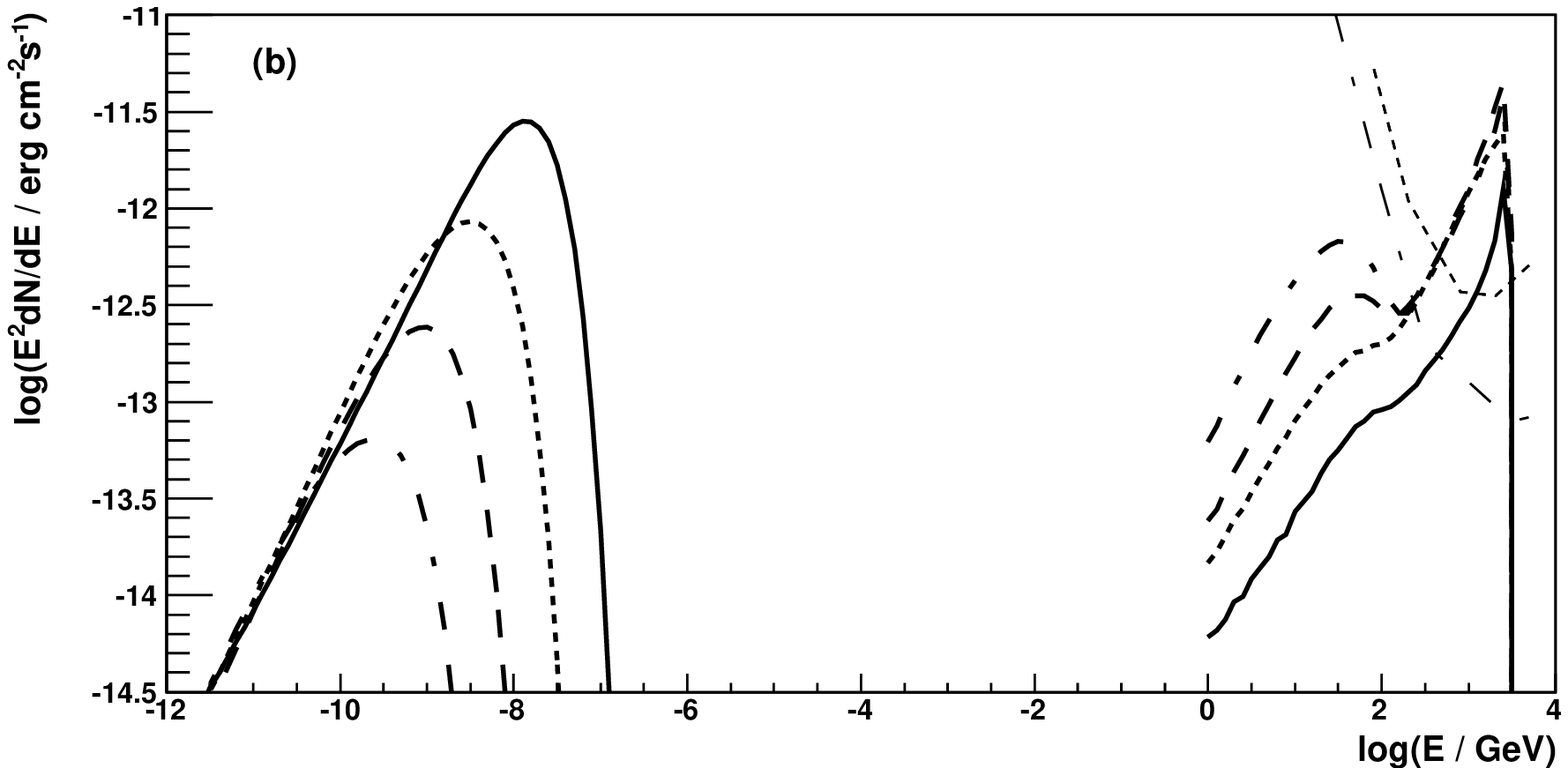}
\includegraphics{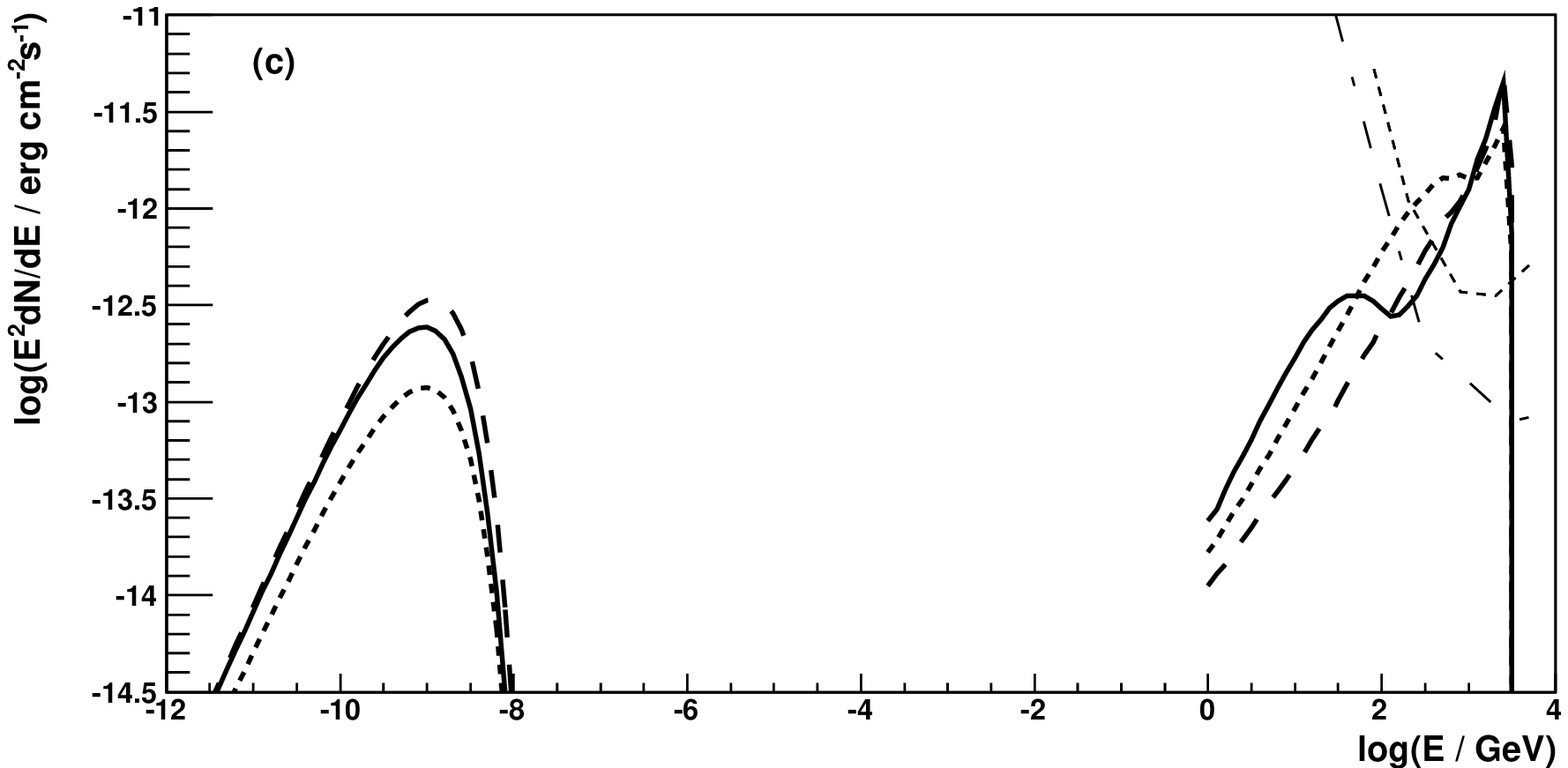}
\includegraphics{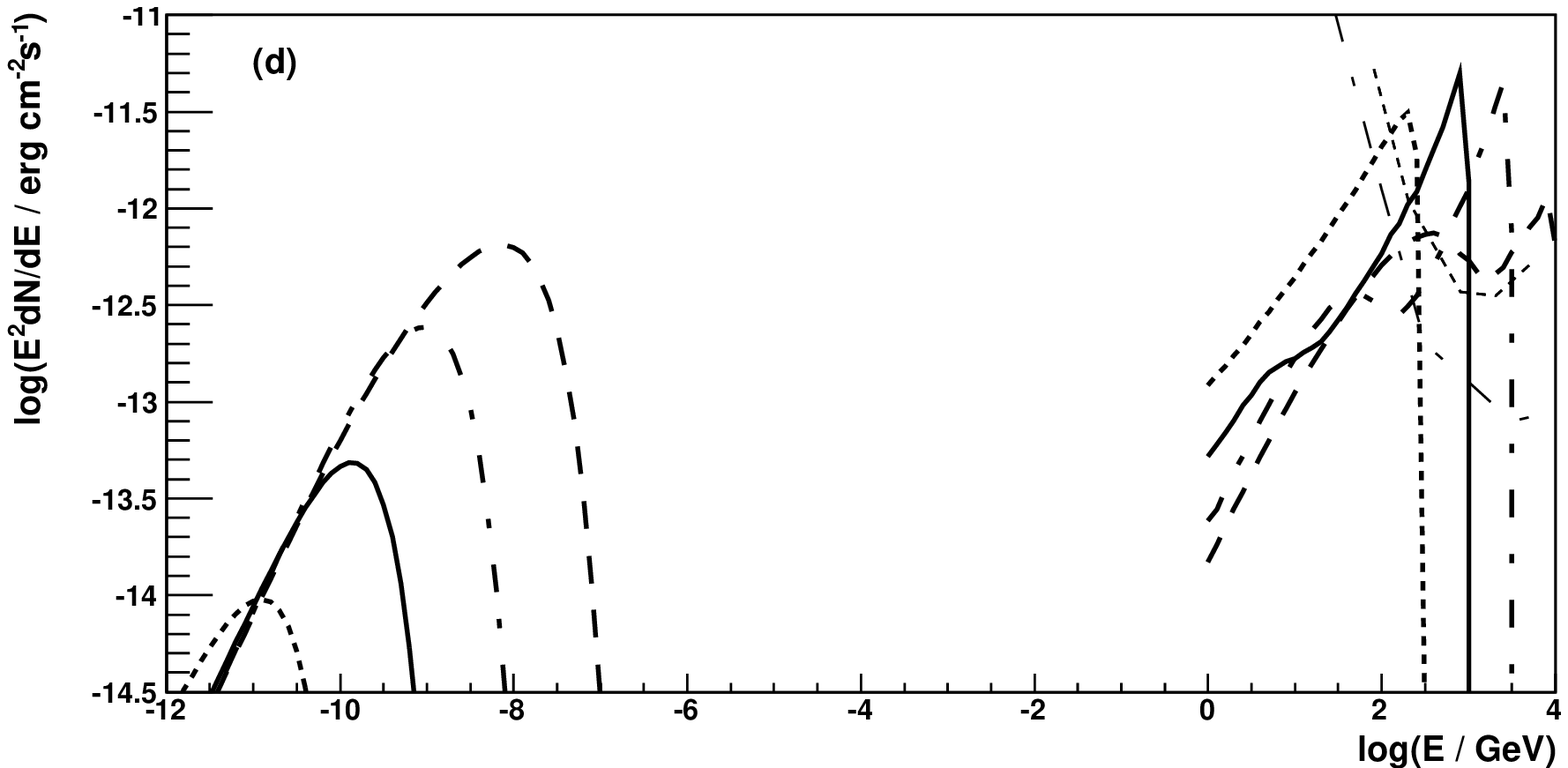}
\includegraphics{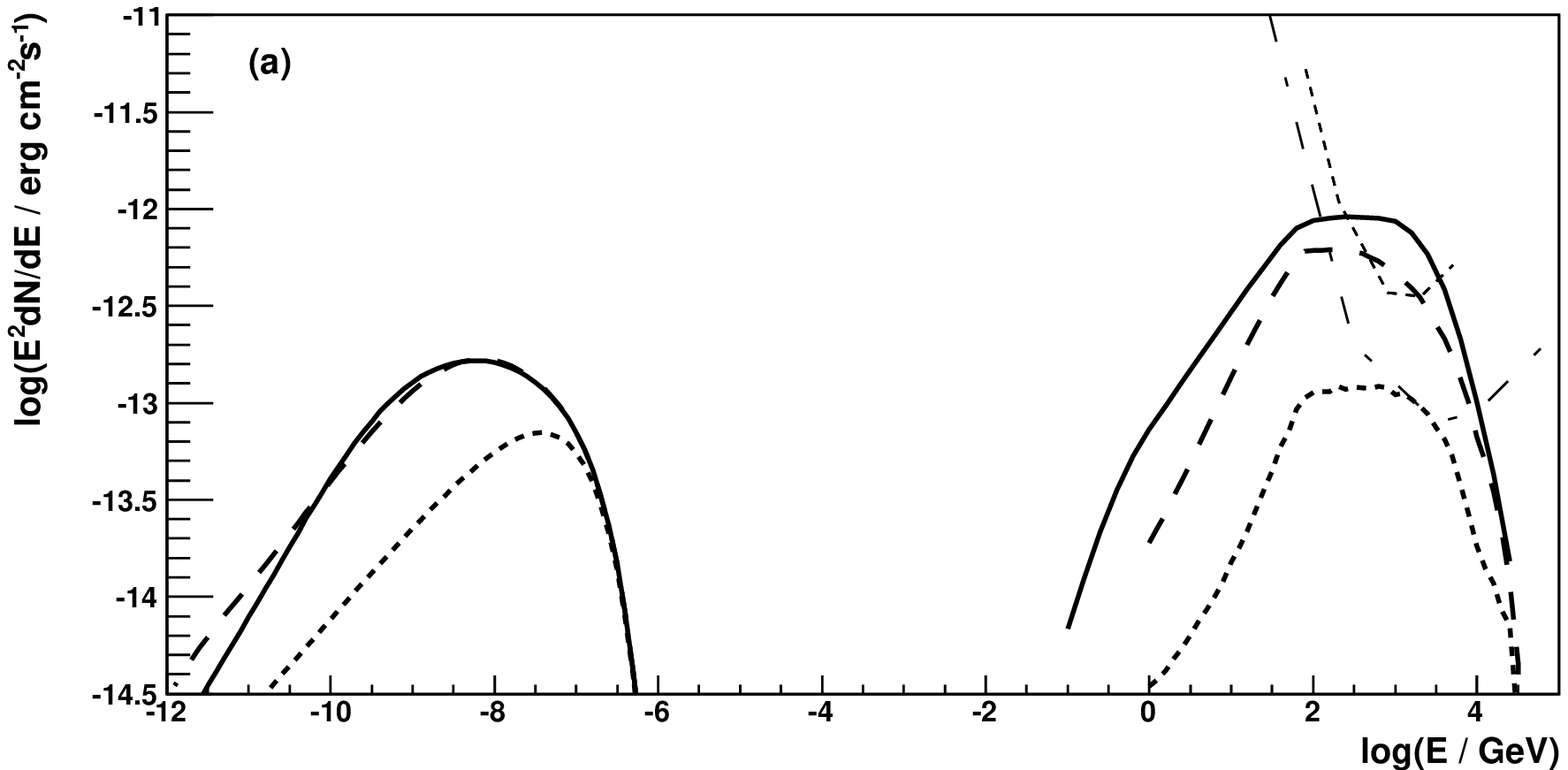}
\includegraphics{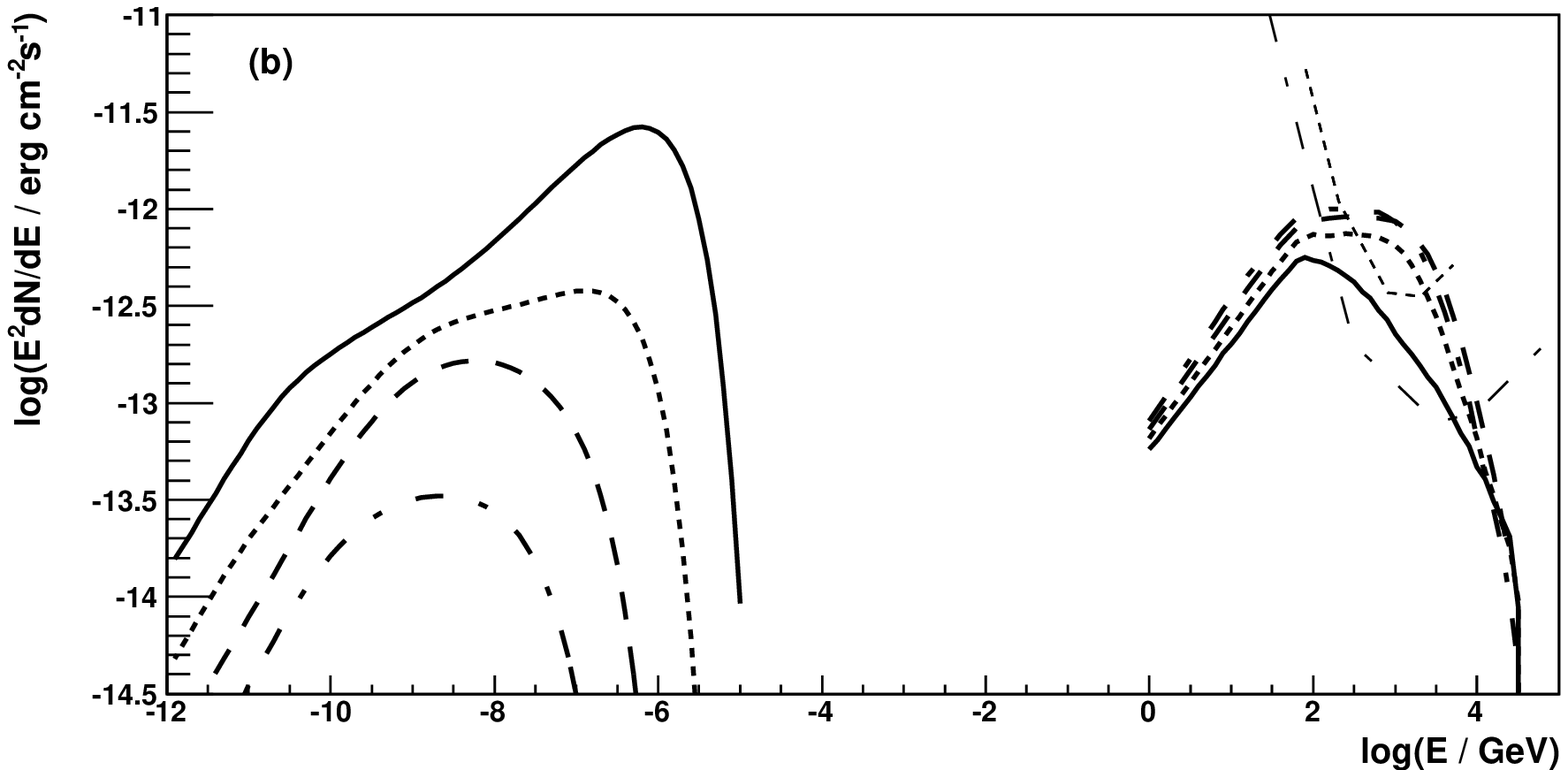}
\includegraphics{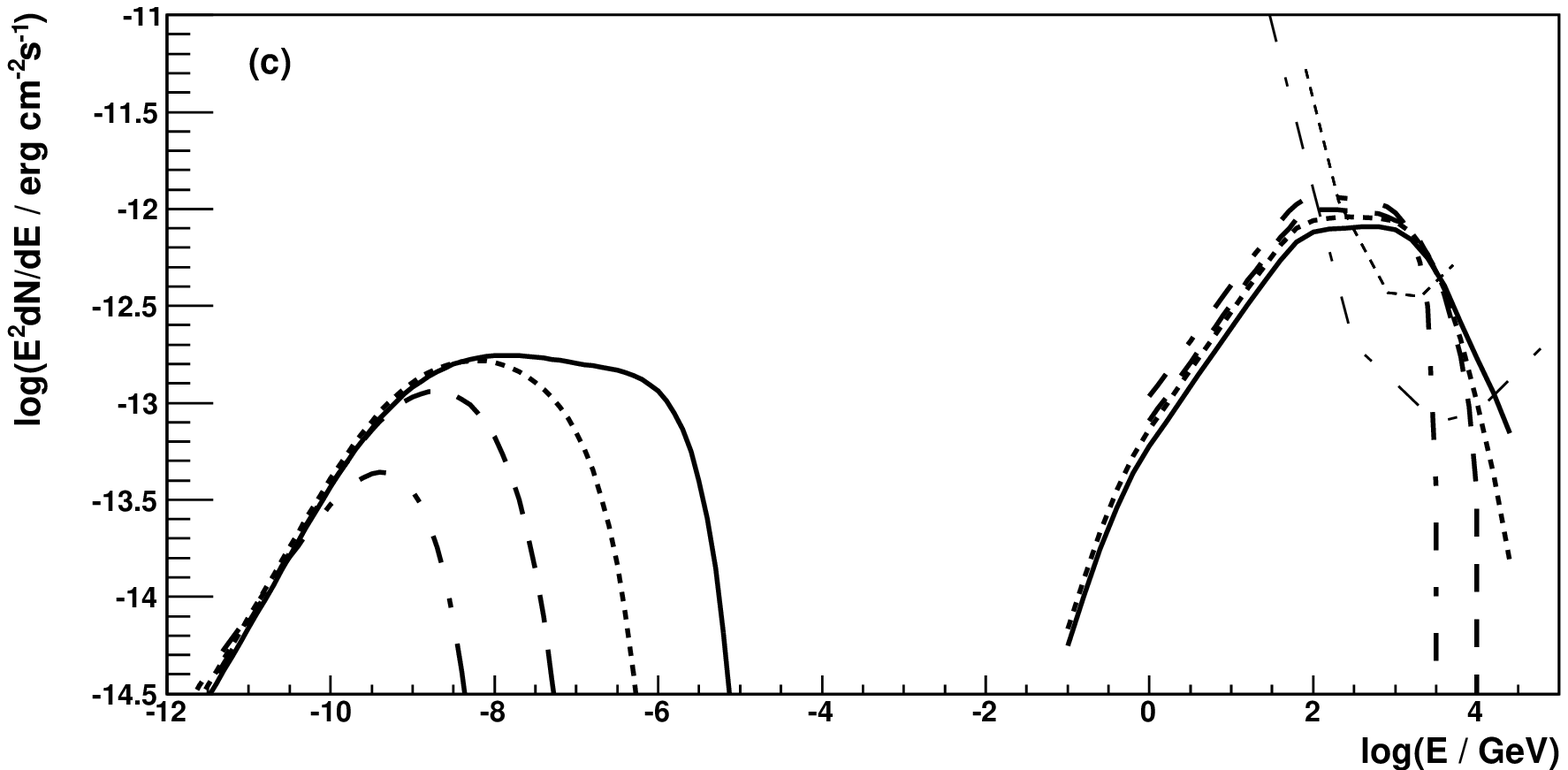}
\includegraphics{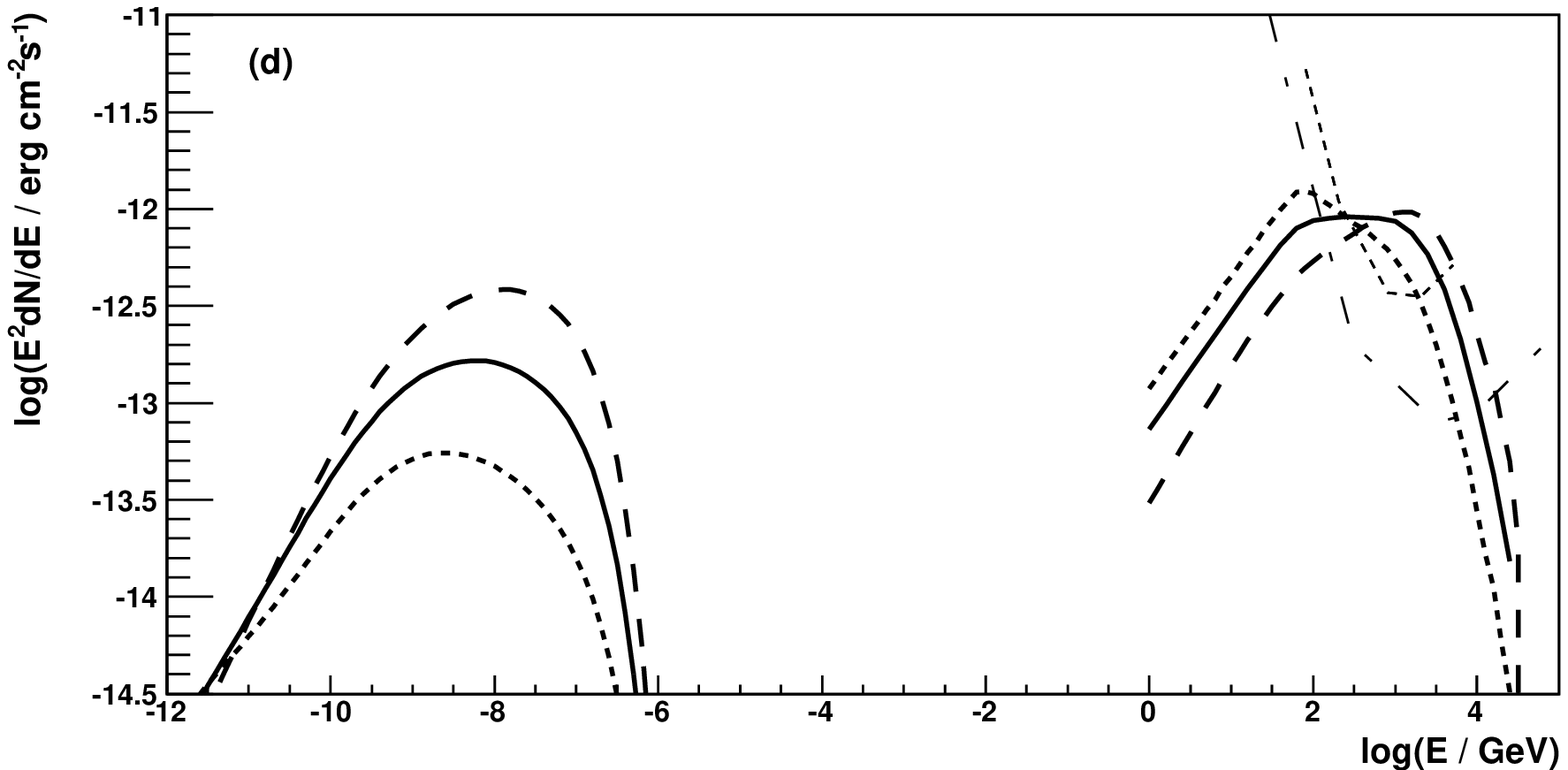}
\caption{SED produced in the case of isotropic injection of mono-energetic leptons (upper panel) and leptons with a power law spectrum (bottom panel) from a MSP population within M15. The leptons are injected from the MSPs which positions in the globular cluster M15 are randomly simulated from the distribution similar to the distribution of normal stars. We investigate the dependence of the IC and synchrotron spectra on the advection velocity from the GC, $v_{\rm adv} = 10^7$ cm s$^{-1}$ (dashed curve),  $10^8$ cm s$^{-1}$ (dotted), and no advection (solid) (figure (a)), on the magnetic field strength within the cluster, $B = 1\mu$G (dot-dashed), $3\mu$G (dashed), $10\mu$G (dotted), and $30\mu$G (solid) (figure (b)), on the infra-red radiation field within the GC with the characteristic temperature of 40 K and the energy density equal to the energy density of the MBR and the optical radiation of M15 (dashed curve), the optical radiation and 10 times of the MBR (dotted curve), and with only with density of the MBR and the optical radiation (solid curve) (figure (c)), and on the energy of injected leptons, $E_{\rm e} = 300$ GeV (dotted), 1 TeV (solid), 3 TeV (dot-dashed), 10 TeV (dashed) (figure (d)). The basic parameters for these calculations are $E_{\rm e} = 3$ TeV, $B = 3\mu$G, and no advection. In the case of the leptons with the power law spectrum, the spectral index equal to -2.05 above 100 GeV is assumed in all cases. 
Dependence of the spectra on the advection velocity for the parameters as above (see bottom panel, figure (a)), the magnetic field strength (b), and the maximum energy of injected leptons for  $E_{\rm e} = 3$ TeV (dot-dashed), 10 TeV (dashed), 30 TeV (dotted), 100 TeV (solid) (c).
The dependence of the spectra on the spectral index of leptons, equal to -1.5 (dashed curve), -2.05 (solid), and -2.5 (dotted), are shown in the bottom figure (d). The parameters of the GC M15 are assumed. Sensitivity of the present Cherenkov telescopes (e.g. MAGIC, Aleksi\'c et al.~2016) has been shown by the thin-dotted curve and of the CTA (Acharaya et al.~2013) by the thin-dot-dashed curve.}
\label{fig4}
\end{figure*}

The more complete scenario for the radiation processes in GCs defined in this paper is considered for the specific GCs. It is assumed that a certain number of MSPs is present within the GC. Their positions within the GC are randomly simulated from their distribution within the cluster  which is similar to the distribution of normal stars described by the Michie-King model (Michie 1963, see also Kuranov \& Postnov~2006). These MSPs are responsible for the injection of relativistic leptons within the GC either from their inner magnetosphere (close to mono-energetic leptons) or re-accelerated within the cluster (power law spectrum of the leptons). The leptons are assumed 
to be injected isotropically by the MSPs. The example calculations have been performed for the parameters of M15. The basic parameters of this cluster are the following, the distance to the cluster 10.4 kpc, the stellar luminosity $7\times 10^5$ L$_\odot$, the core radius 0.43 pc, and the half mass radius 3.04 pc. The injection rate of the leptons from the MSPs is normalized to the power of $2\times 10^{34}$ erg~s$^{-1}$, which might be for example provided by the population of 20 MSPs with a typical period of 4 ms and surface magnetic field strength of $3\times 10^8$ G, assuming $10\%$ efficiency of the rotational energy loss rate transferred to the leptons. Note that 8 MSPs have been discovered up to now within M15, three of them have the periods close to 4 ms (http://www.naic.edu/pfreire/GCpsr.html).  We consider the propagation of the leptons up to the distance of 20 pc from the centre of M15. This distance corresponds to the angular dimension of the source of the order of $\sim$6' which is of the order of typical point spread function of the Cherenkov telescopes. 

\begin{figure*}
\vskip 3.5truecm
\includegraphics{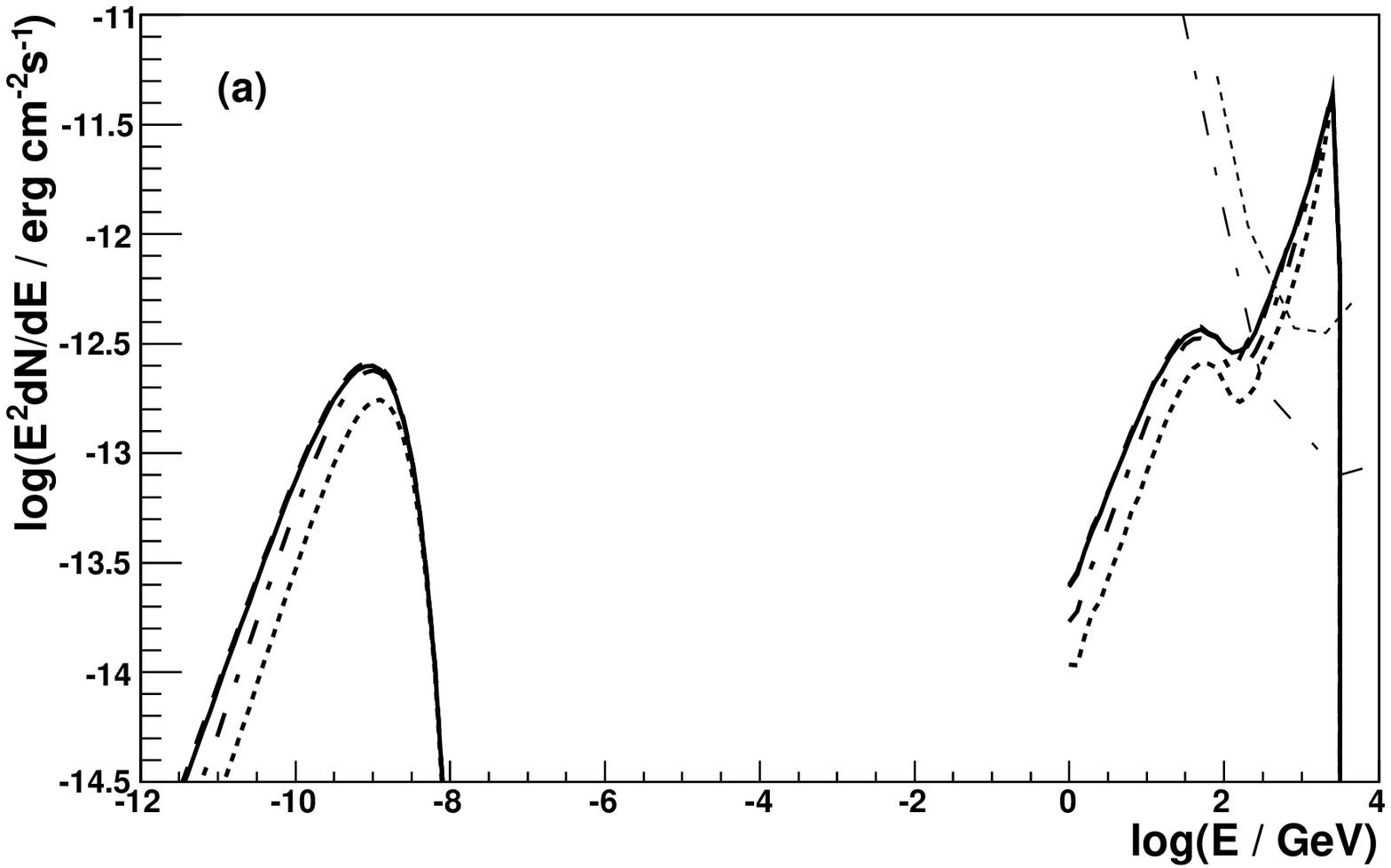}
\includegraphics{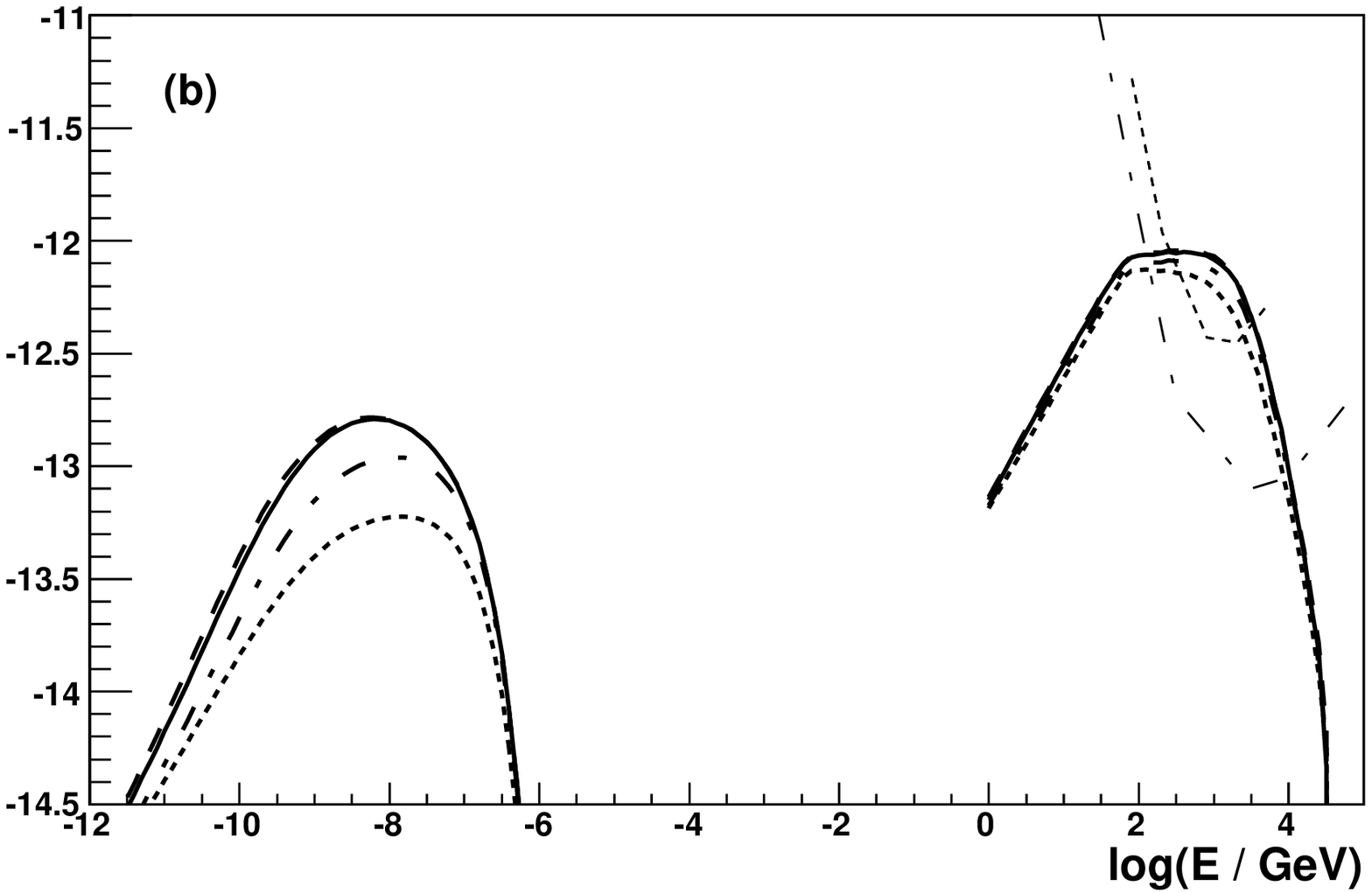}
\includegraphics{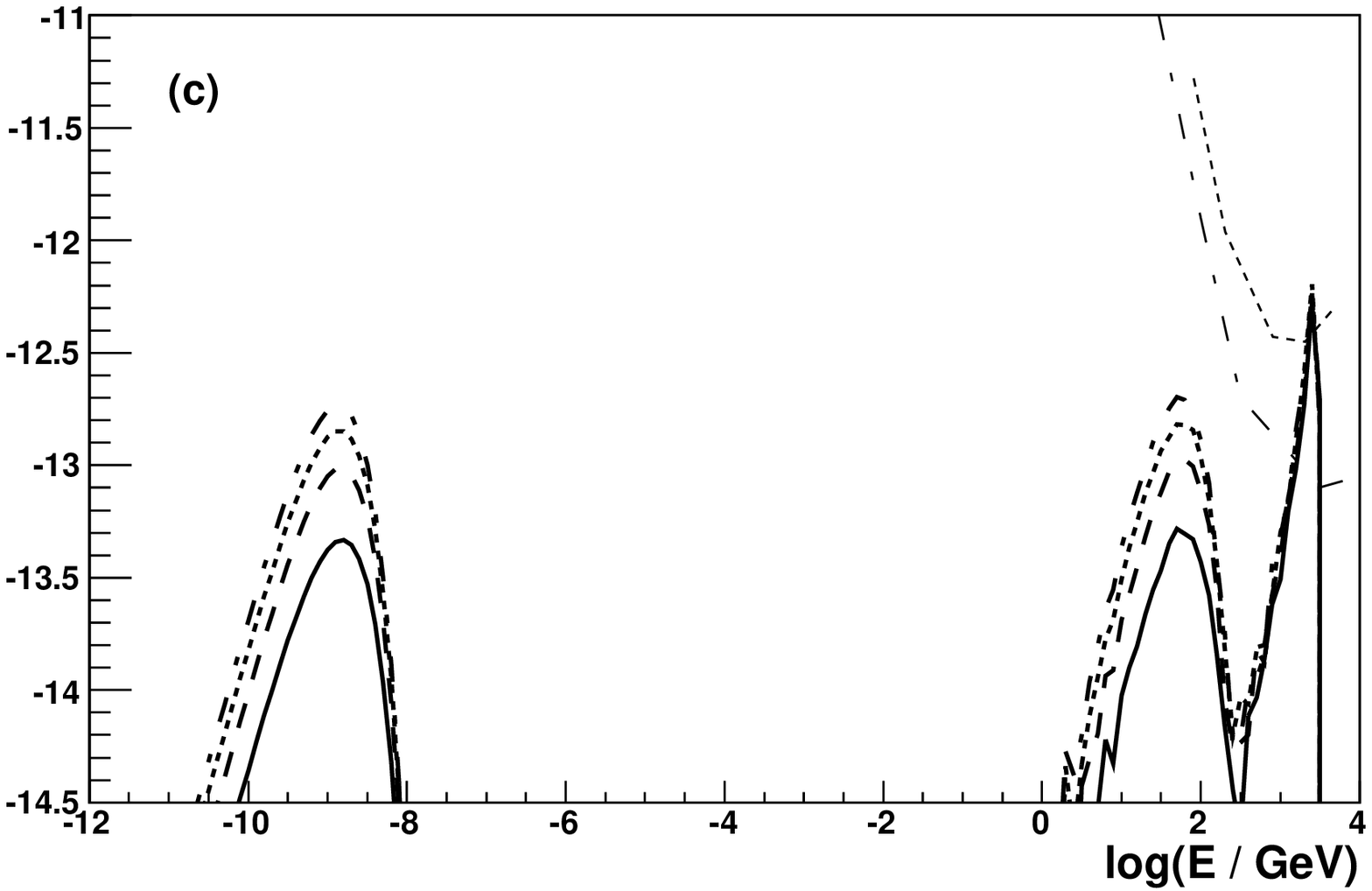}
\includegraphics{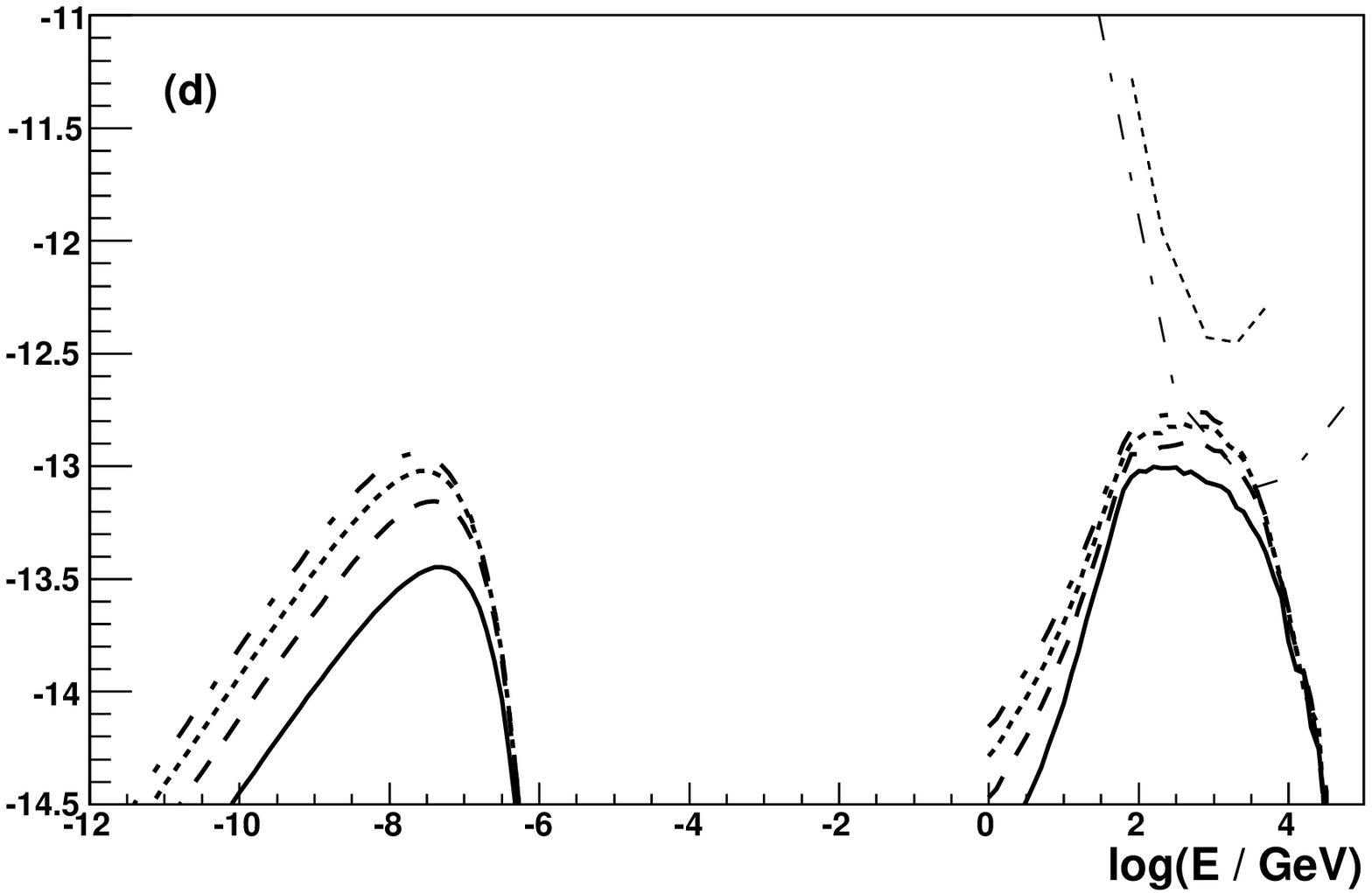}
\caption{SED produced by the mono-energetic leptons (figures (a) and (c)) and leptons with the power law spectrum ((b) and (d)) within spherical regions with different radii, $R_{\rm int}$, from the centre of the globular cluster M15. The cases with no advection are shown in (a) and (b) and with the advection with the wind velocity $10^8$ cm s$^{-1}$ (c) and (d). (a) and (b) are shown for $R_{\rm int} = 4$pc (dotted), 6pc (dot-dashed), 10pc (solid), and 20pc (dashed) and (c) and (d) for $R_{\rm int} = 10$pc (solid), 20pc (dashed), 30pc (dotted), and 40pc (dot-dashed). The magnetic field strength is equal to 3$\mu$G. Leptons are injected isotropically from the MSPs with the distribution as in Fig.~4. The energy of mono-energetic leptons is fixed on 3 TeV, and the power law spectrum with spectral index -2.05 extends between 100 GeV and 30 TeV.}
\label{fig5}
\end{figure*}

The results of calculations for the case of injection of the mono-energetic leptons (in most cases energy has been fixed on 3 TeV) are shown in the upper panel of Fig.~4. We investigate the synchrotron and the IC spectra produced by these leptons as a function of the parameters describing the initial spectrum of the leptons and defining their propagation (diffusion and advection processes) within and around the GC. As in the previous section, we assume that in most cases the leptons interact with the optical radiation produced by stars and with the MBR. 
In Fig. 4a, we show the dependence of the spectra on the advection velocity with the GC wind. 
Note that the synchrotron spectra and the IC spectra, due to the scattering of the MBR, are on a slightly higher level for the advection velocity $10^7$ cm s$^{-1}$ than for the case with no advection. In fact, such situation can be understood since a part of the isotropically injected leptons diffuse towards the centre of the GC, i.e. against direction of the advection. These leptons can move very slowly in respect to the centre of the GC and their cooling on the synchrotron and the IC scattering of the MBR can become more efficient than  in the case without advection.
For the advection velocities close to  $\sim$10$^8$ cm s$^{-1}$, the intensity of the synchrotron spectrum and the IC 
$\gamma$-ray spectrum (both in the region of the broad bump due to the scattering of the MBR and in the Compton peak due to the scattering of optical photons) drops by a factor of the order of three in respect to the case without the advection or the advection with the velocity $10^7$ cm s$^{-1}$. 
These spectra also strongly depend on the strength of the magnetic field within the GC, which determine the diffusion process
and the synchrotron energy losses of the leptons (see upper Fig. 4b). The synchrotron emission usually stays in the optical/UV energy range. For energies of the leptons of the order of a few TeV, these spectra do not extend to the soft X-rays. Therefore, they cannot be observed by the X-ray satellites. The level of the synchrotron emission becomes comparable to the level of the IC $\gamma$-ray emission for a rather strong magnetic field within the GC (of the order of $30\mu$G). Note that the broad IC bump, peaking in the sub-TeV energies, clearly weakens for stronger magnetic fields due to extraction of energy from the leptons by the synchrotron process already within the GC. We also investigate the dependence of the non-thermal spectra on different radiation fields within the GC. We performed calculations of the non-thermal emission assuming  the infra-red radiation from the galactic disk in addition to the stellar radiation field. The strong infra-red radiation from the disk (characteristic temperature 40 K and energy density 10 times energy density of MBR) results in the 
appearance of an additional bump in the $\gamma$-ray spectrum above $\sim$100 GeV, i.e. just below the Compton peak (see dotted curve in Fig. 4c). Therefore, the infra-red radiation from the galactic disk can help in detecting the sub-TeV $\gamma$-ray emission from the GC. 
We also investigate the spectra for different energies of injected the leptons, in the range of 0.3-10 TeV, expected from the modelling of processes in the MSP inner magnetosphere (Fig. 4d). For leptons with large energies, the synchrotron emission strengthens and shifts to larger energies but never clearly reaches the X-ray energy range. On the other hand, the $\gamma$-ray emission strengthens in the region of the broad bump and weakens in the region of the Compton peak. The most favourable conditions for the detection of $\gamma$-ray emission from GCs are expected in the case of the injection of the leptons with the intermediate energies, i.e. close to $\sim$1-3 TeV.

We also investigate the radiation produced by leptons injected with a power law spectrum (bottom panel in Fig.~4). Note that in this case we assume the lepton spectrum extending to clearly larger energies than considered in the mono-energetic case since the leptons injected from the MSPs can be additionally re-accelerated within the GC up to energies of the order of a few tens of TeV. 
As in the case on the mono-energetic injection of the leptons, the velocity of the wind from the GC and the magnetic field strength have strong effect on the synchrotron and the IC $\gamma$-ray spectra (see Fig. 4a and b in the bottom panel). However, 
the maximum energies of the accelerated leptons have only a weak effect on the expected $\gamma$-ray spectra. Although, the maximum energy of the leptons have a significant effect on the expected synchrotron spectrum (see Fig. 4c). The weak effect of the maximum lepton energies on the level of TeV $\gamma$-ray emission is due to the strong Klein-Nishina effects for the extremely relativistic leptons scattering optical radiation. 
Finally, we investigate the dependence of the non-thermal emission on the spectral index of the accelerated leptons (bottom panel in Fig. 4d). The IC $\gamma$-ray spectra are not so strongly influenced in this case since we define the power law  spectrum of leptons in a relatively narrow  energy range (between 0.1-30 TeV). But, the levels of the synchrotron spectra are significantly influenced since they are mainly produced by the leptons with the highest energies. 

In order to have an impression whether the $\gamma$-ray source towards M15 should be point like or extended for the Cherenkov telescopes, we calculate the spectra produced within a sphere with a specific radius, $R_{\rm int}$, around the centre of M15. The cases, with no advection and with strong advection, with velocity $10^8$ cm s$^{-1}$, are considered. In the first case, we consider the range of radii between 4-20 pc (see (a) and (b) in Fig.~5) and in the second case between 10-40 pc ((c) and (d) in Fig.~5). In no advection case, we show that the emission mainly comes from within $\sim$10 pc from the core of M15 which means that a point-like TeV $\gamma$-ray source should be observed by the Cherenkov telescopes. However, if the GC wind is fast, then emission on clearly lower level should emerge from the extended region around M15 with the radius of the order of $\sim$10' (corresponding to $R_{\rm int} = 30$ pc). We conclude that if the wind from the GC is fast (of the order of $10^8$ cm s$^{-1}$) then the TeV $\gamma$-ray source towards M15 will be difficult to detect by the present Cherenkov telescopes and the observations with the CTA will be needed.

\section{Non-thermal radiation from observed MSP within GC}

In the case of several GCs, GeV $\gamma$-ray emission has been discovered. This emission is likely  due to a cumulative emission from the whole population of MSPs within a specific GC. This expectation has got recently  strong support with the discovery of two energetic pulsars within two GCs (PSR B1821-24 in M28 and J1823-3021A in NGC~6624).
The pulsed $\gamma$-ray emission of these two pulsars take significant amount of the total $\gamma$-ray emission 
from their GCs. Since the parameters of these two pulsars are well known, it is worth to calculate the expected non-thermal emission produced by leptons accelerated within their inner magnetosphere and their vicinity during propagation of leptons through the environment of their parent GCs. Such emission should determine the lower level of the non-thermal emission expected from their parent GCs. This emission should be independent on some of the unknown input parameters of the considered here model.  
Below we show the results of calculations of the non-thermal synchrotron and the IC $\gamma$-ray spectra assuming that the leptons are injected into the GC environment only from these specific MSPs. The spectra of the leptons are normalized in such a way that the power in the injected leptons is equal to the GeV $\gamma$-ray power emitted by these specific MSPs, i.e. $L_{\pm} = L_\gamma^{\rm MSP}$.

\subsection{The case of J1823-3021A in NGC~6624}

The energetic millisecond pulsar, J1823-3021A, has been discovered in the $\gamma$-rays by the Fermi Collaboration within the globular cluster NGC~6624 (Freire et al.~2011). The $\gamma$-ray power, $(8.4\pm 1.6)\times 10^{34}$ erg s$^{-1}$, is the highest observed for any MSP. The pulsar has the period 5.4 ms and the surface magnetic field of $4.3\times 10^9$ Gs, being one of the most energetic objects of this type. It is offset from the centre of NGC~6624 by $\sim$0.05', corresponding the the projected distance of $\sim 0.12$ pc for the distance of NGC~6624 estimated on 8.4 kpc (Valenti et al.~2007).

We investigate the expected non-thermal radiation produced by leptons which are accelerated directly by this MSP or re-accelerated on the shocks formed during interaction of the pulsar wind with the companion star, stars within the GC or the winds from other MSPs.
As in the case of M15, it is assumed that the leptons are injected with close to mono-energetic spectrum (see e.g. Zajczyk et al.~2013). In the second model, we assume that the leptons reach a power law spectrum with the spectral index close to -2.
The synchrotron spectra, produced by the leptons in the magnetic field of the GC, and the IC $\gamma$-ray spectra, produced by the leptons which scatter the thermal radiation from stars within the GC and the MBR, are calculated in terms of the model described above. We investigate the dependence of radiation output as a function of different parameters describing the injected spectra of the leptons and their propagation within NGC~6624. 
It is assumed that the relativistic leptons take a part of the rotational energy of the MSP J1823-3021A, comparable to the energy lost on the GeV $\gamma$-ray emission. This is estimated to be of the order of 10$\%$ of the spin down energy loss rate (Freire et al.~2011).

\begin{figure*}
\vskip 3.5truecm
\includegraphics{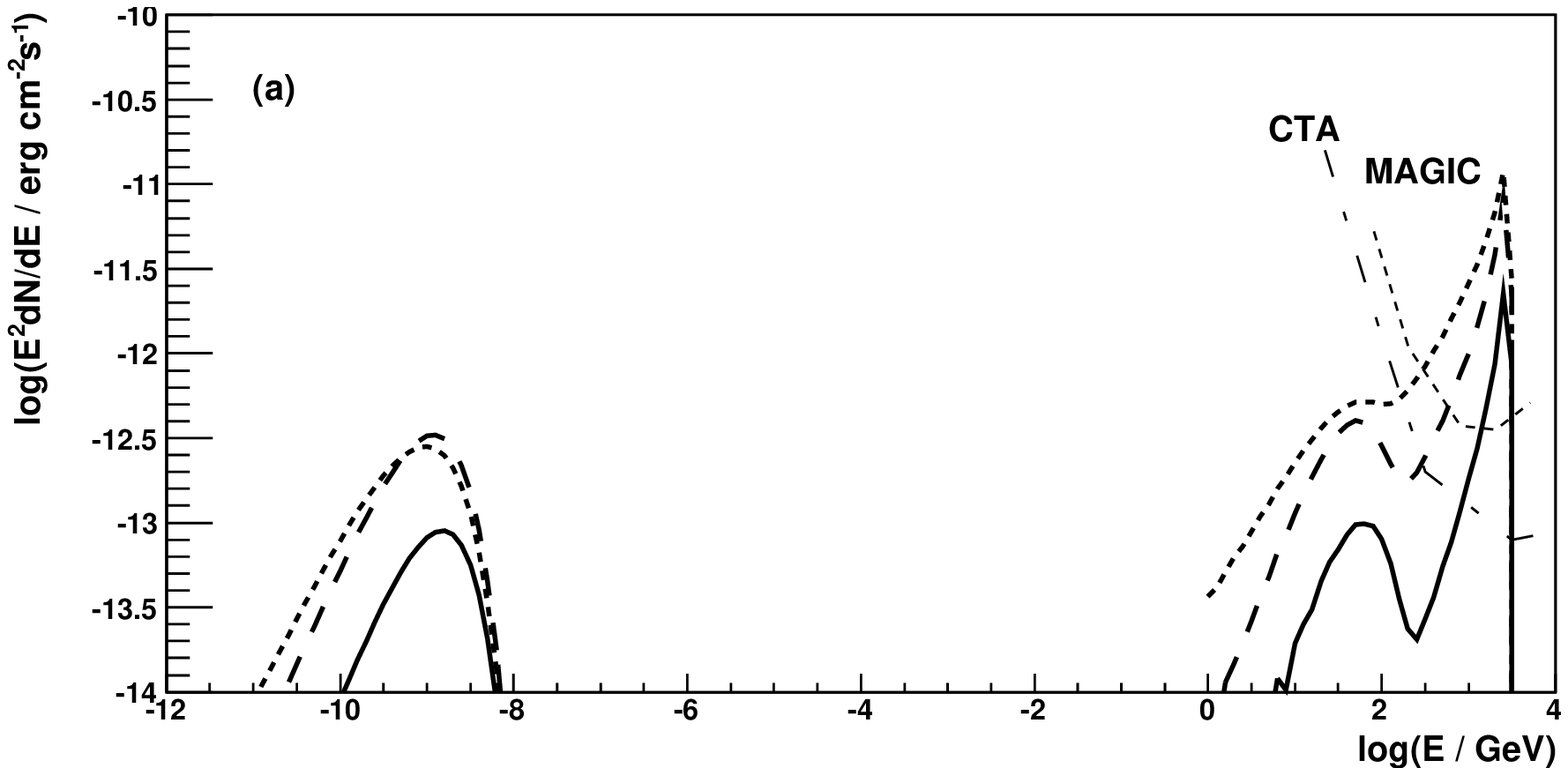}
\includegraphics{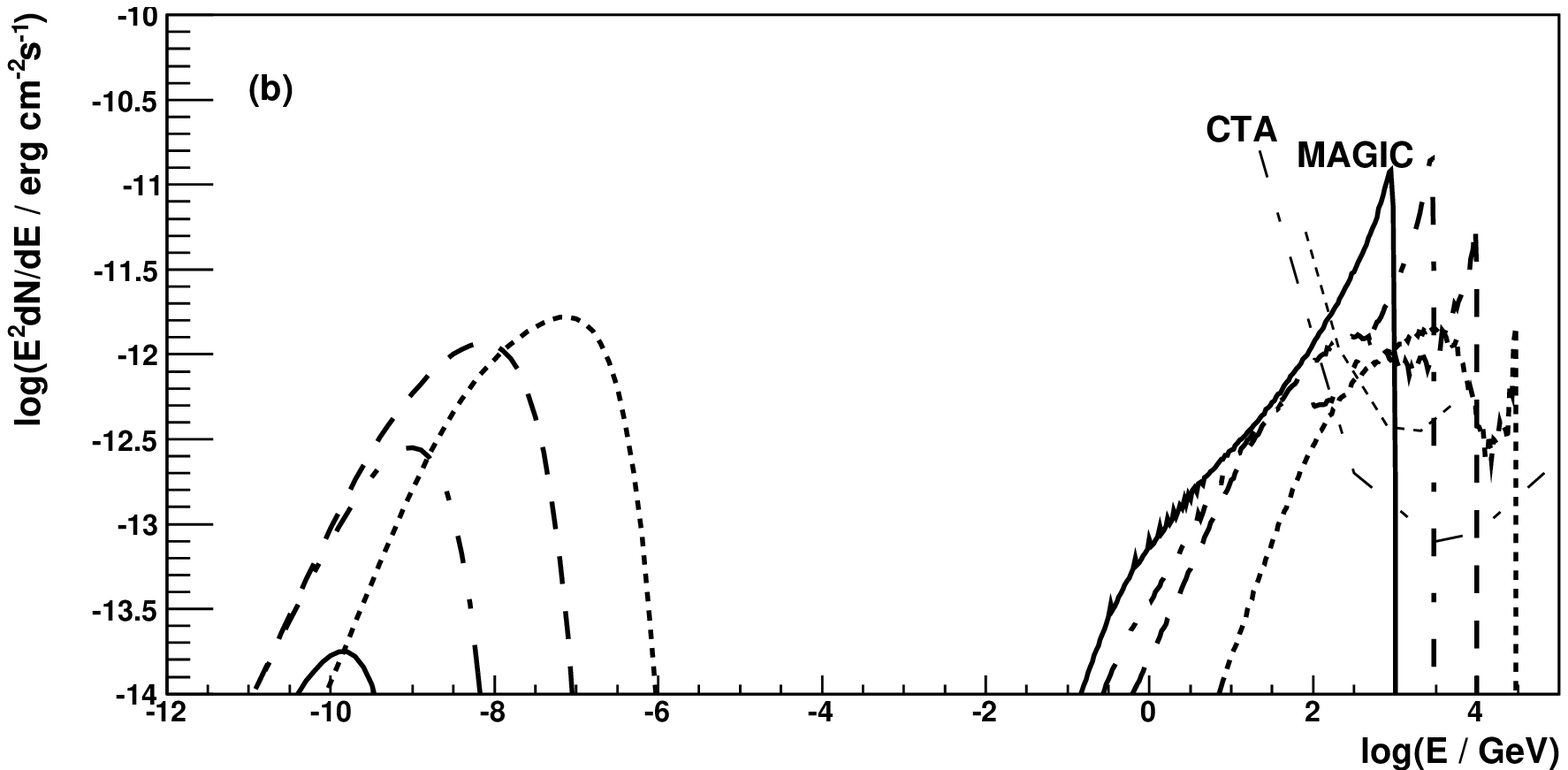}
\includegraphics{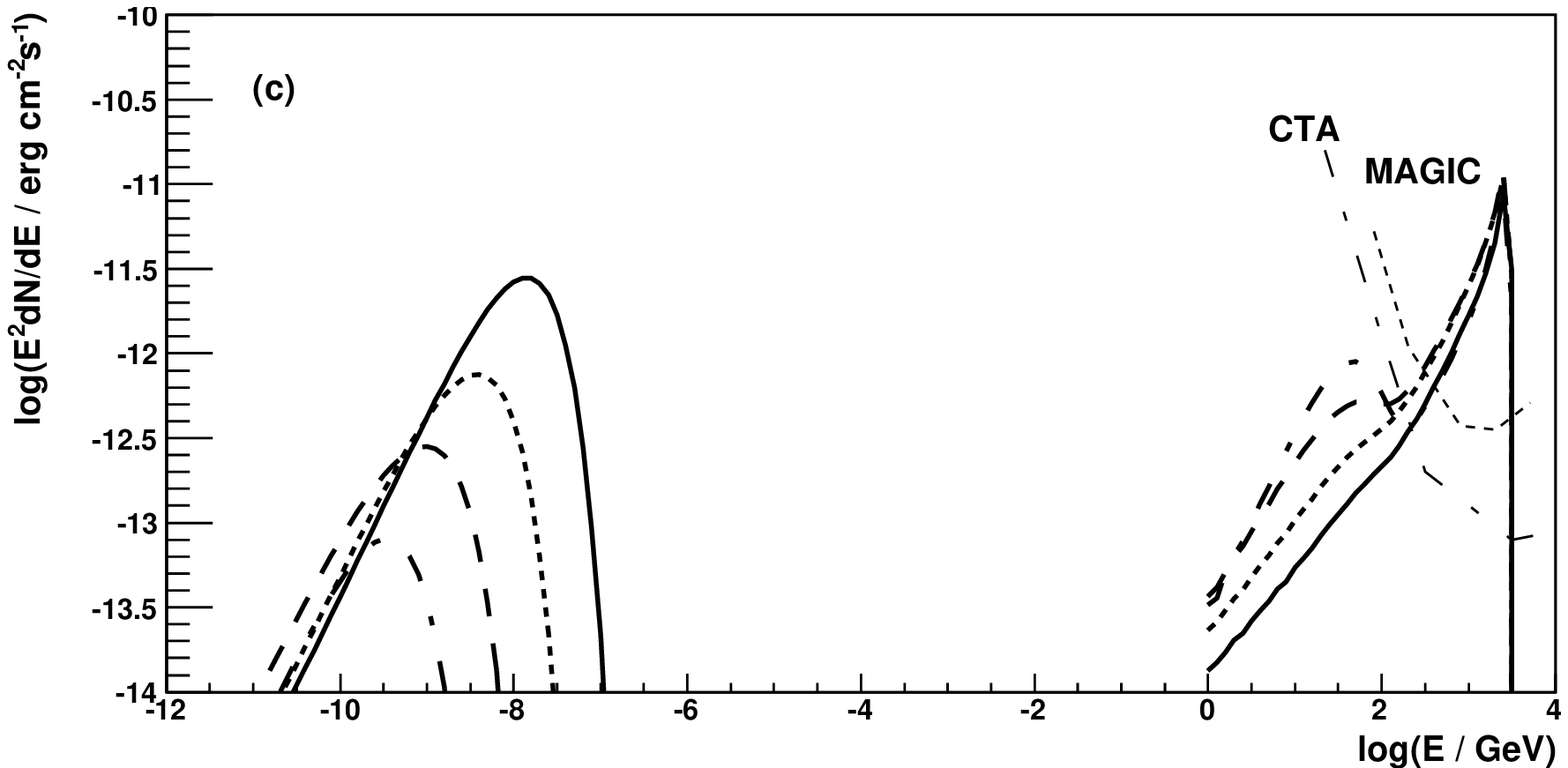}
\includegraphics{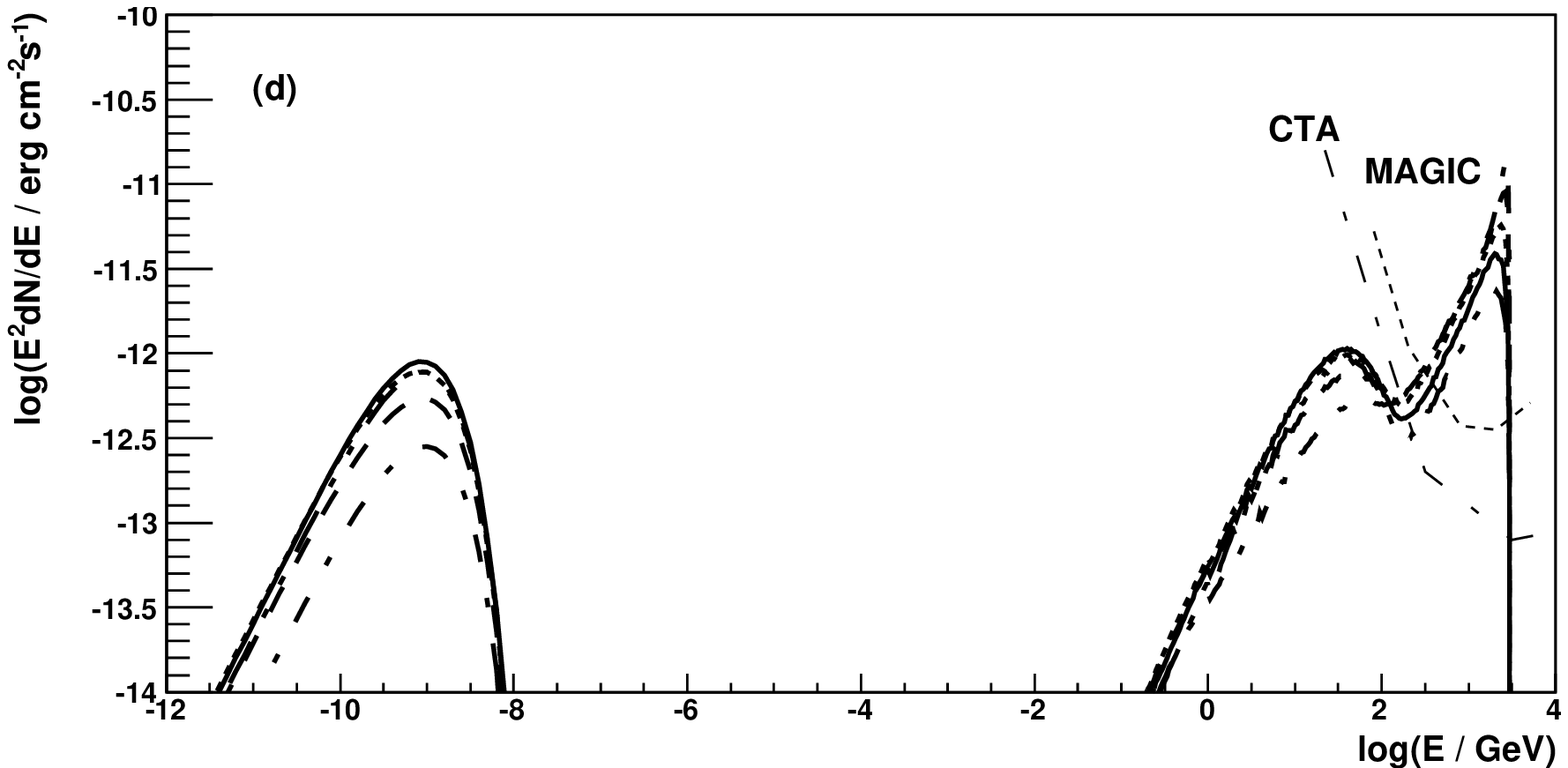}
\caption{SED produced by mono-energetic leptons injected from the millisecond pulsar J1823-3021A into the globular cluster NGC~6624. The leptons diffuse through the GC. They can be also advected from the GC with the GC wind. The spectra are investigated as a function of the velocity of the GC wind for $v_{\rm adv} = 0$ (dotted), $10^7$ cm s$^{-1}$ (dashed), and $10^8$ cm s$^{-1}$ (solid) (figure (a)). Other parameters of the model are the following, the distance from the core $X_{\rm inj} = 0.12$ pc, magnetic field strength $B = 3\mu$G, and the energy of leptons 30 TeV. Dependence on energies of the leptons is shown for $E_{\rm e} = 1$ TeV (solid), 3 TeV (dot-dashed), 10 TeV (dashed), and 30 TeV (dotted) (figure (b)), and other parameters as mentioned in (a). Dependence on the magnetic field strength is shown for $B_{\rm c} = 1 \mu$G (dot-dashed), 3$\mu$G (dashed), 10$\mu$G (dotted), and 30$\mu$G (solid) for other parameters as above and the distance of MSP from the core $0.12$ pc (c). Dependence on the real distance from the core of GC for $d = 0.12$ pc (dot-dashed), 2 pc (dashed), 4 pc (dotted), 6 pc (solid), and 8 pc (dot-dot-dashed) assuming other parameters as above and no advection (d). The MAGIC and CTA 50 hr sensitivities are marked by the thin dashed and dot-dashed curves, respectively.}
\label{fig6}
\end{figure*}
\begin{figure*}
\vskip 6.5truecm
\includegraphics{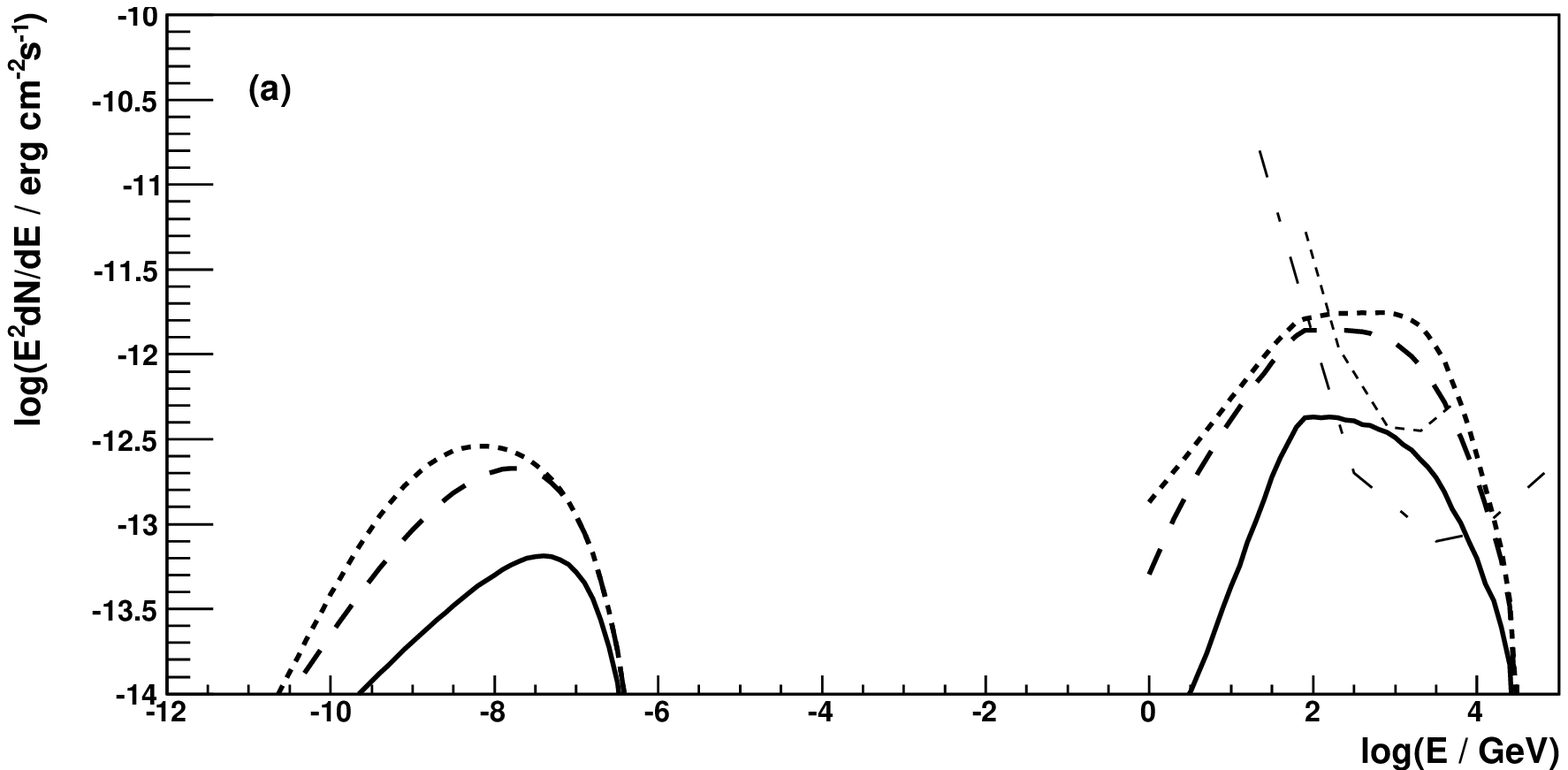}
\includegraphics{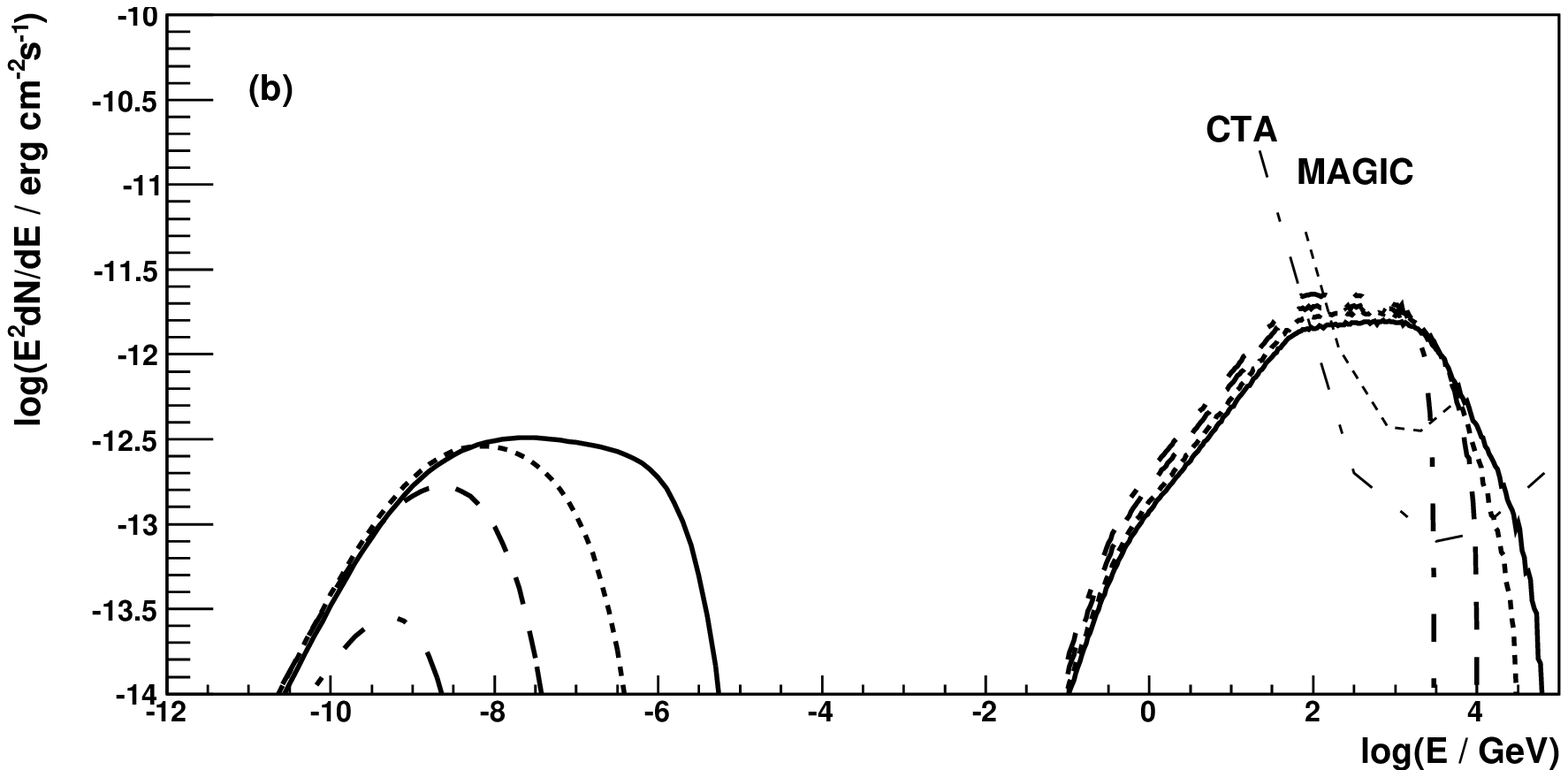}
\includegraphics{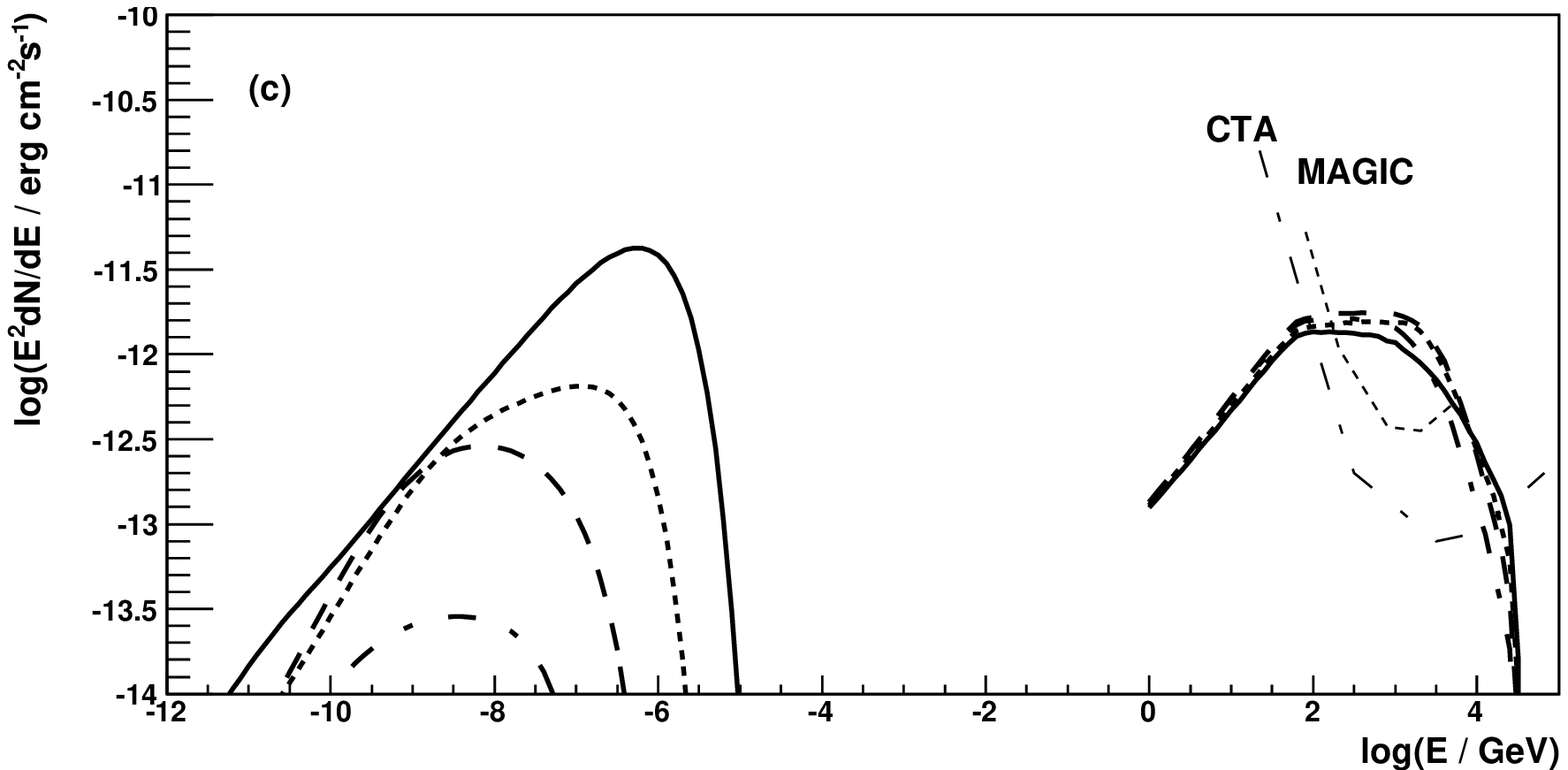}
\includegraphics{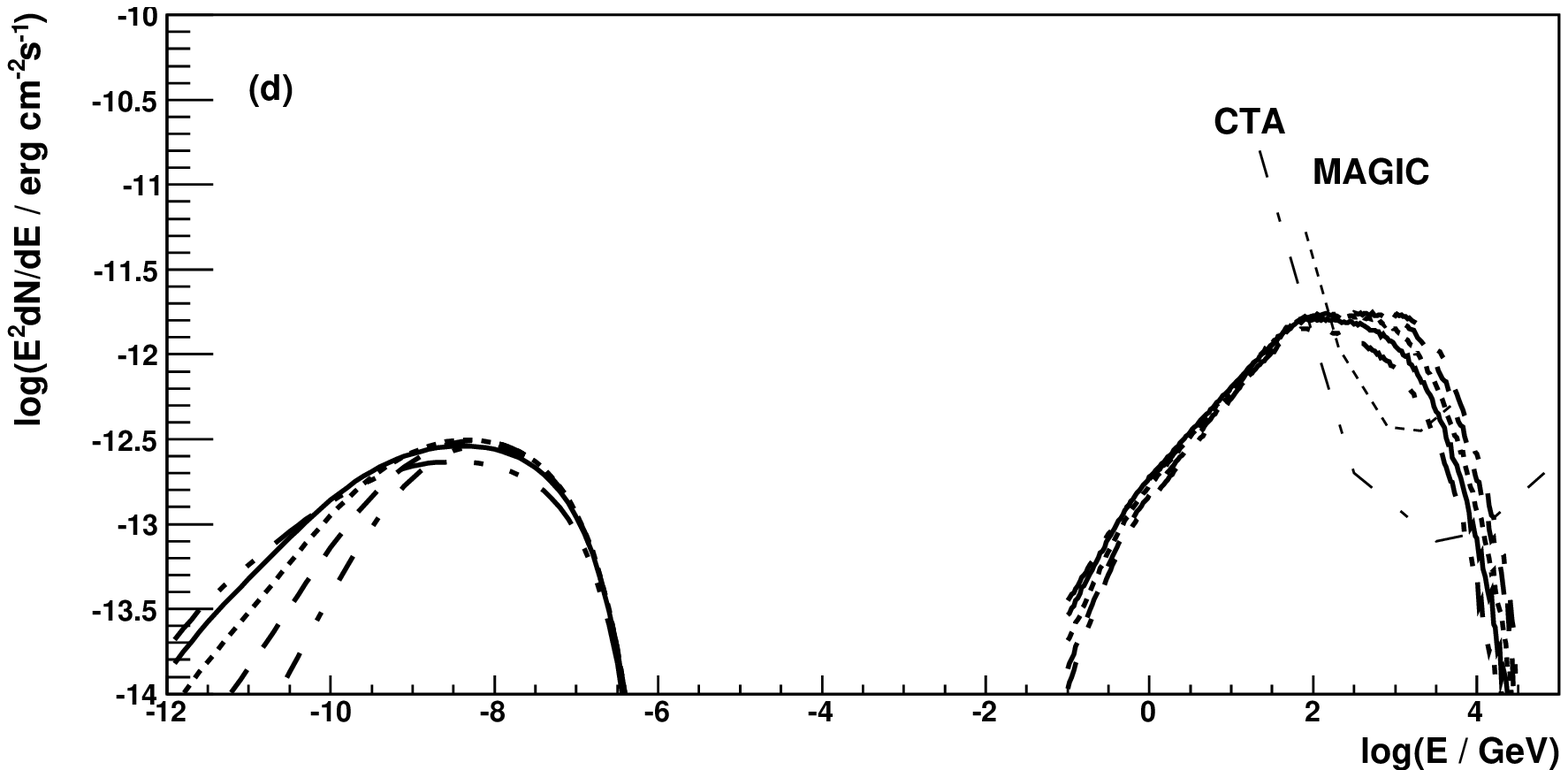}
\includegraphics{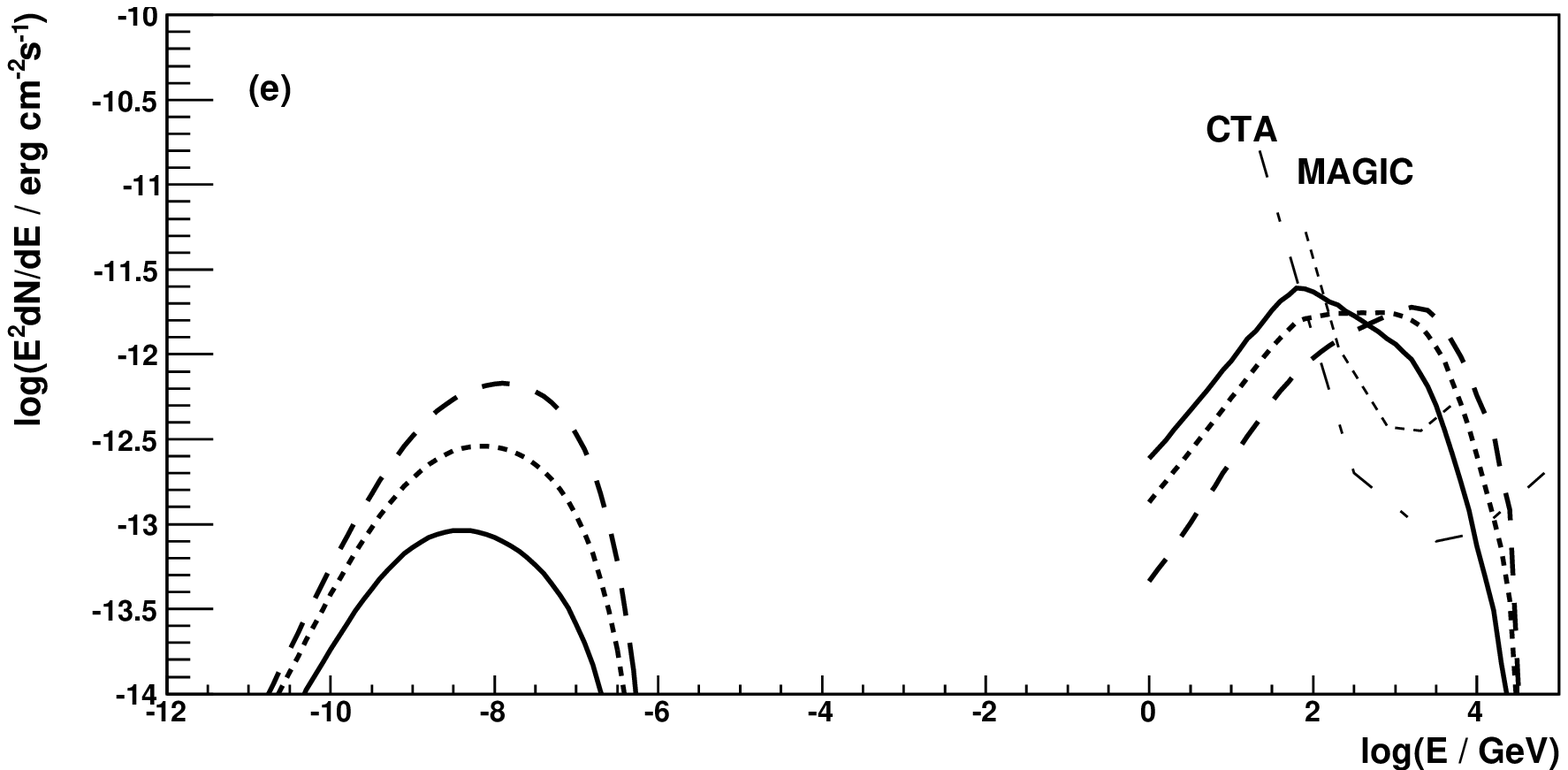}
\caption{As in Fig.~6 but for leptons with a power law spectrum and the spectral index equal to -2 between 100 GeV and 30 TeV. The dependence of the spectra on the advection velocity of the GC wind (figure (a)), on the maximum energy of injected leptons equal to $E_{\rm e} = 3$ TeV (dot-dashed), 10 TeV (dashed), 30 TeV (dotted), 100 TeV (solid) (b), on the magnetic field strength (c), on the injection distance from the centre of the GC (d). The other parameters are as in Fig.~6. Dependence on the spectral index equal to 1.5 (solid), 2.05 (dotted), and 2.5 (dashed), for $B = 3\mu$G, and no advection (e).}
\label{fig7}
\end{figure*}

At first, we investigate the dependence of the non-thermal emission on different parameters of the model in the case of mono-energetic injection of the leptons (Fig.~6). We fix the energy of the leptons on 3 TeV, apply the Bohm diffusion model, and uniform magnetic field strength in the GC equal to 3$\mu$G. As in the case of M15, the $\gamma$-ray IC spectra, calculated for different sets of input parameters, are typically composed from two features, the first one due to the comptonization of the MBR in the T regime (a broad lower energy bump) and the second one due to comptonization of the stellar radiation from the GC in the KN regime (a peak at higher energies). The synchrotron emission appears typically at energies below X-ray energy range.
Dependence of the spectra on the advection velocity of the GC wind (considered cases without advection, advection with velocity of $v_{\rm adv} = 10^7$ cm s$^{-1}$ and $10^8$ cm s$^{-1}$) are shown in Fig.~6a. Both, the synchrotron and the IC emission, are clearly weaker for faster GC winds since the leptons escape more efficiently from the GC. Moreover, the spectra becomes narrower for larger $v_{\rm adv}$, due to ineffective cooling of the leptons. We investigate the dependence of spectra on the energy of the injected leptons assuming that they are in the range from 1 TeV and to 30 TeV (Fig.~6b). Note that the synchrotron spectra cannot reach the X-ray range even for the case of the most extreme considered maximum energies of the leptons equal to 30 TeV. On the other hand, the IC spectra strongly depend on the lepton energy. For the TeV energies, the spectra are dominated by scattering process of the optical radiation in the KN regime. For leptons with the largest considered energies (i.e. 30 TeV), the scattering of the MBR in the T regime starts to dominate. The synchrotron spectra strongly depend on the strength of the magnetic field within the GC (Fig.~6c). We investigate the range of magnetic fields between 1$\mu$G and 30$\mu$G.
For the strongest fields, the synchrotron SED is on the level comparable to the IC $\gamma$-ray SED. On the other hand, for weak fields the IC $\gamma$-ray spectrum clearly dominates over the synchrotron spectrum. 
We also investigate the dependence of the spectra on the real distance of the MSP from the centre of the GC (we only know the projected but not the real distance). The spectra show rather weak dependence on the distance in the range between 0.12 pc to 8 pc from the centre of the GC (Fig.~6d). Note that the synchrotron spectra, expected in such a model, mostly peak in the optical energy range where the contribution from the stellar population in the GC dominates completely. Therefore, these models are poorly constrained by eventual observations of the GCs in the X-ray energy range. 

We have also performed calculations of the non-thermal emission from NGC~6624 in the case of leptons injected with a power law spectrum (see Fig.~7). Dependence of the spectra on similar parameters have been investigated as in the case of injection of the mono-energetic leptons. The general features of these spectra for different parameters are quite similar to the case of the mono-energetic injection of the leptons. The $\gamma$-ray spectra do not dependent very strongly within the considered range of parameters.
However, the photon spectra calculated in the case of the leptons injected with the power law spectrum extend to larger energies than for the mono-energetic leptons. Therefore, in the case of this model, in some cases the synchrotron spectra can extend up to the soft X-rays, allowing additional constraints of the model by sensitive X-ray observations. In the case of a strong magnetic field within NGC~6624, the synchrotron emission can even dominate over the IC emission. However, these large values of the magnetic fields seem unlikely. 
In the case of the power law model, we also consider the dependence of the TeV $\gamma$-ray flux on the spectral index of injected leptons. Even in the case of the steeper spectra (spectral index equal to 2.5), the TeV $\gamma$-ray spectrum does not differ significantly which is mainly due to a relatively small energy range of the leptons (between 0.1 TeV and 30 TeV, only 2-3 decades in energy).
However, the synchrotron fluxes change significantly since they are more sensitive to the power in the high energy end of 
the lepton spectrum.

The levels of $\gamma$-ray emission for the most of the considered parameters for both models are clearly above the sensitivity of the present Cherenkov telescopes and the planned Cherenkov Telescope Array (CTA, see Acharya et al.~2013), provided that the conversion efficiency from the pulsar to relativistic leptons is comparable to the efficiency of the GeV $\gamma$-ray production in the inner pulsar magnetosphere. 
From the comparison of the $\gamma$-ray spectra, calculated in terms of the mono-energetic model, with the 50 hr sensitivity of the MAGIC telescopes (see Fig.~6), we conclude that the Compton peak is up to an order of magnitude above the MAGIC sensitivity. Therefore, the injection rates of leptons with the power of the order of $\sim 1\%$ of the rotational energy loss rate of the MSP J1823-3021A can be potentially investigated. In the case of the power law injection of leptons, the TeV $\gamma$-ray spectra are in most cases a factor of $\sim 2$ above the MAGIC sensitivity for the mentioned above lower energy cut-off in the lepton spectrum. Therefore, in this model, the injection rate of leptons on the level of $5\%$ might be investigated. The constraints, which will be provided by the CTA, should be a factor of ten more restrictive.
We conclude that detailed constraints of the injection of leptons by the MSPs and on their radiation and propagation processes within GCs are  possible with extensive observations even by the present Cherenkov telescopes. Note that only radiation expected from a single pulsar (although exceptionally energetic) within GC NGC~6624 is considered.  Other MSPs could also contribute to the spectrum of leptons injected in the GC. Therefore, the TeV $\gamma$-ray fluxes, shown in Figs.~6 and~7, should be considered as the lower limits on the non-thermal emission from NGC~6624.

\begin{figure*}
\vskip 3.5truecm
\includegraphics{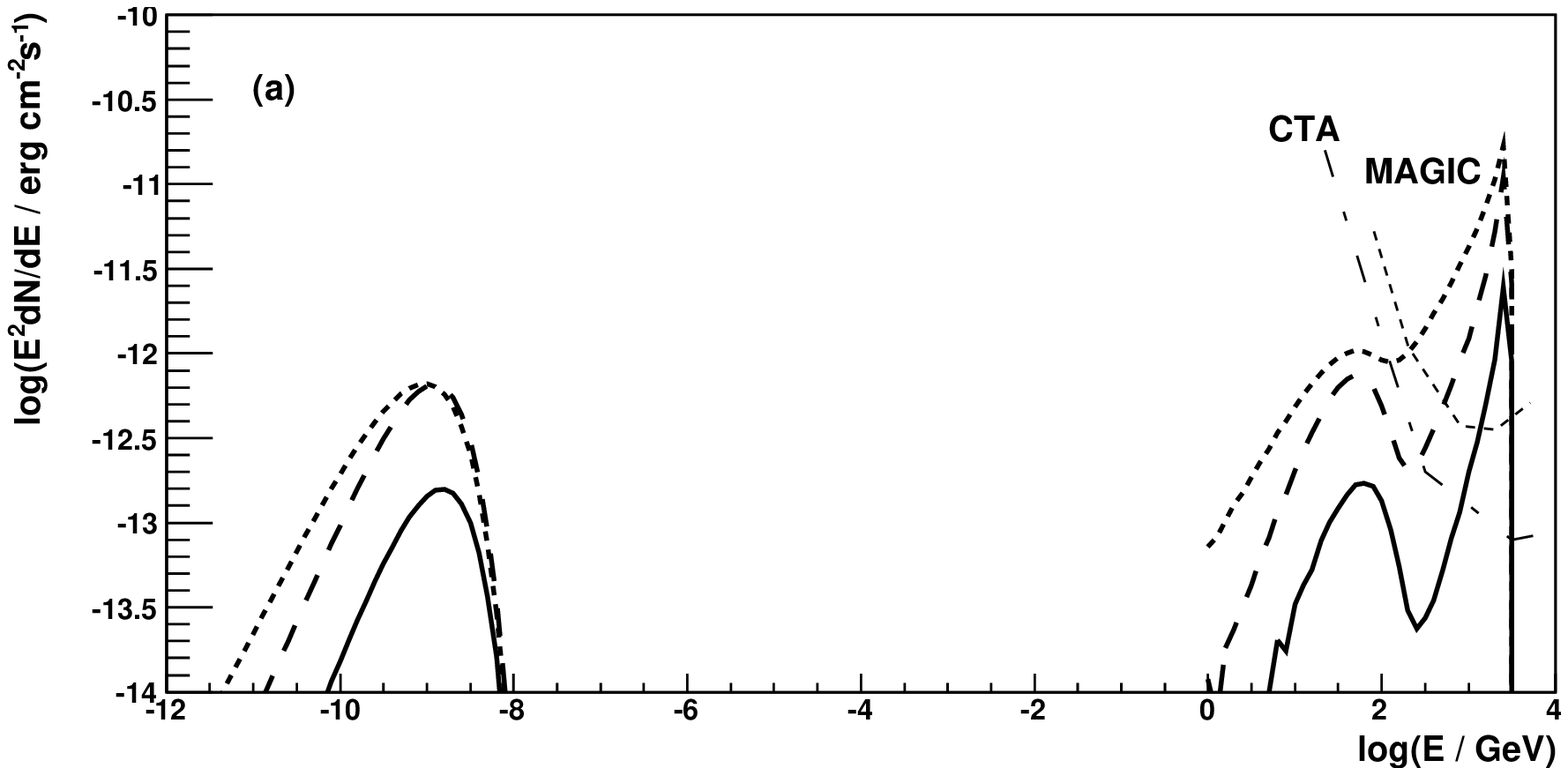}
\includegraphics{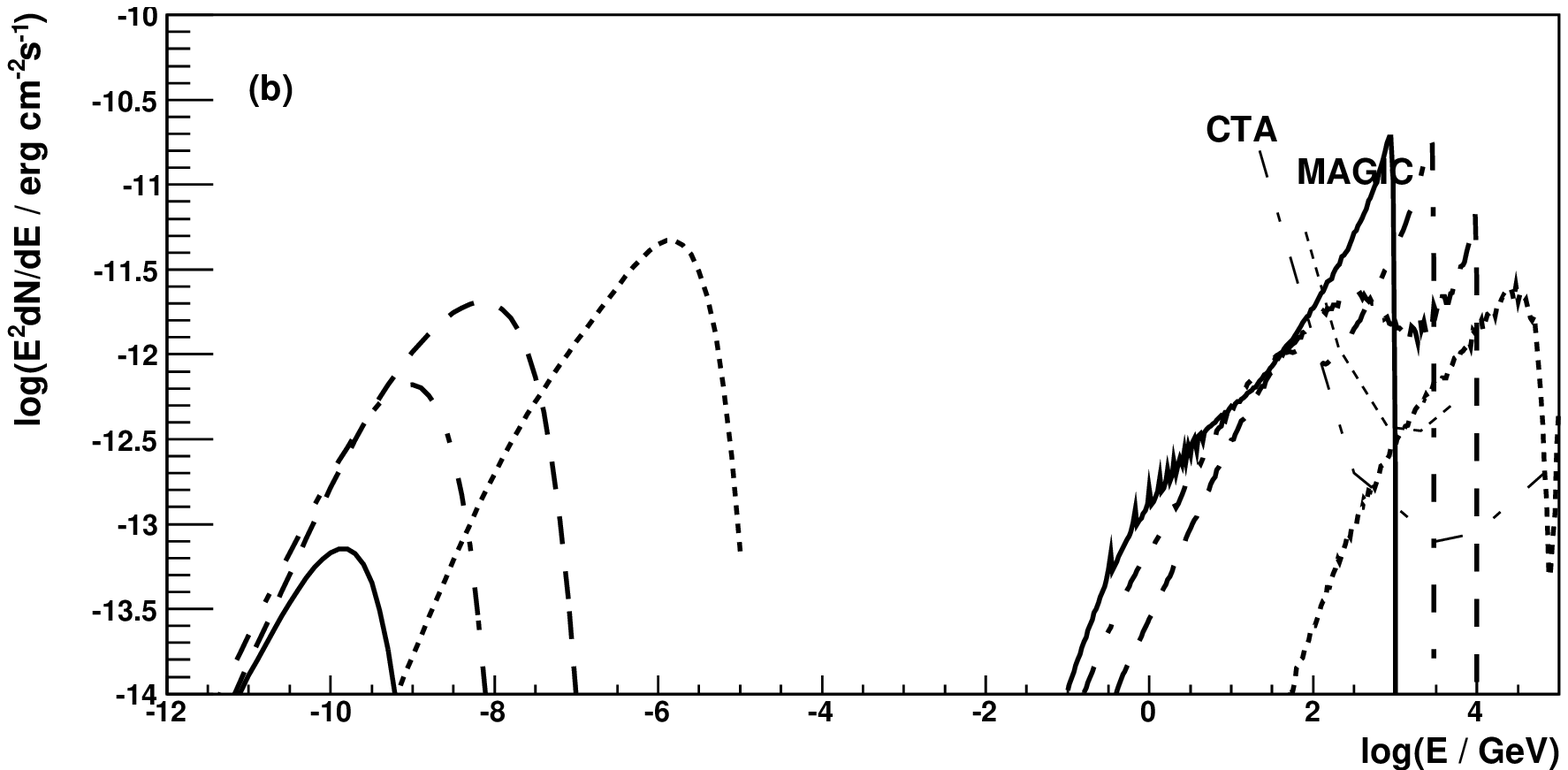}
\includegraphics{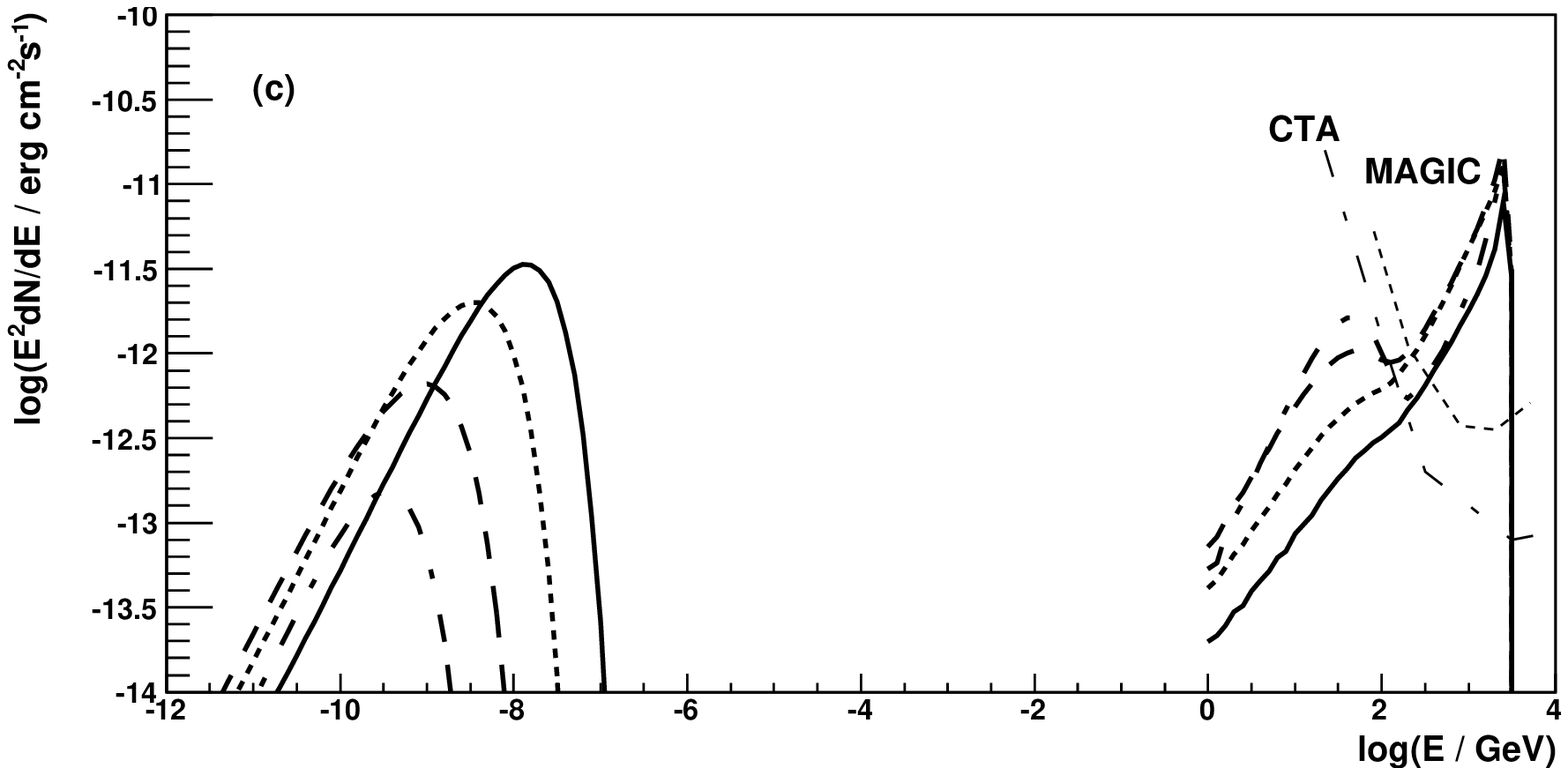}
\includegraphics{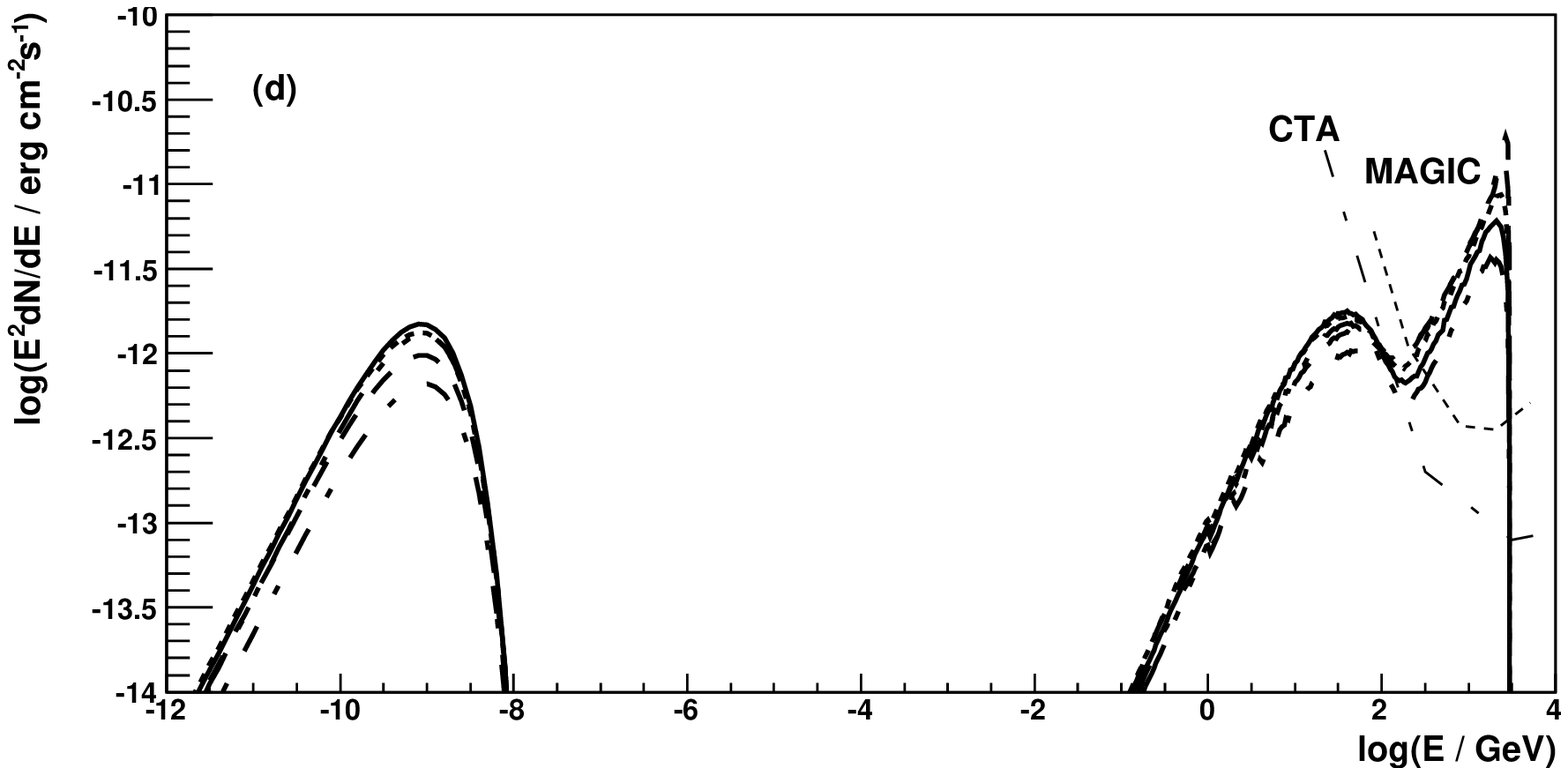}
\caption{SED from M28 produced by mono-energetic leptons injected from the millisecond pulsar B1821-24. The parameters are those same as considered for NGC~6624 in Fig.~6 unless specified otherwise.
The spectra are investigated as a function of the velocity of the GC wind for $v_{\rm adv} = 0$ (dotted), $10^7$ cm s$^{-1}$ (dashed), and $10^8$ cm s$^{-1}$ (solid) (figure (a)). Other parameters of the model are the following, the distance from the core $d = 0.33$ pc, the magnetic field strength $B = 3\mu$G, and the energy of the leptons 3 TeV. Dependence on energies of leptons are shown for $E_{\rm e} = 1$ TeV (dotted), 3 TeV (dot-dashed), 10 TeV (dashed), and 100 TeV (solid) (b). Dependence on the magnetic field strength is shown for $B_{\rm c} = 1 \mu$G (dot-dashed), 3$\mu$G (dashed), 10$\mu$G (dotted), and 30$\mu$G (solid) for other parameters as above and the distance of MSP from the core $0.33$ pc (c). Dependence on the real distance from the core of GC for $d = 0.33$ pc (dot-dashed), 2 pc (dashed), 4 pc (dotted), 6 pc (solid), and 8 pc (dot-dot-dashed) (d).}
\label{fig8}
\end{figure*}
\begin{figure*}
\vskip 6.5truecm
\includegraphics{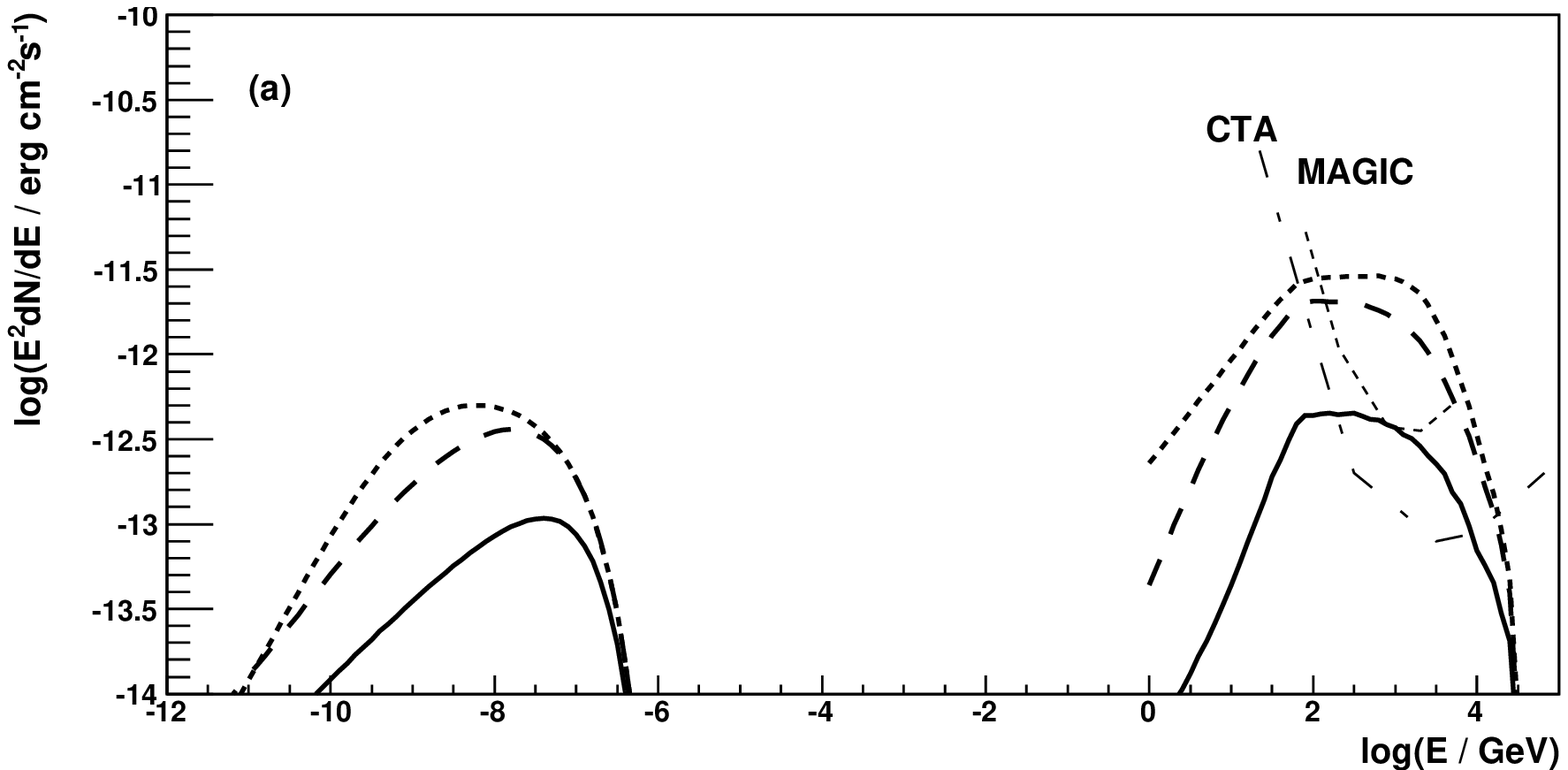}
\includegraphics{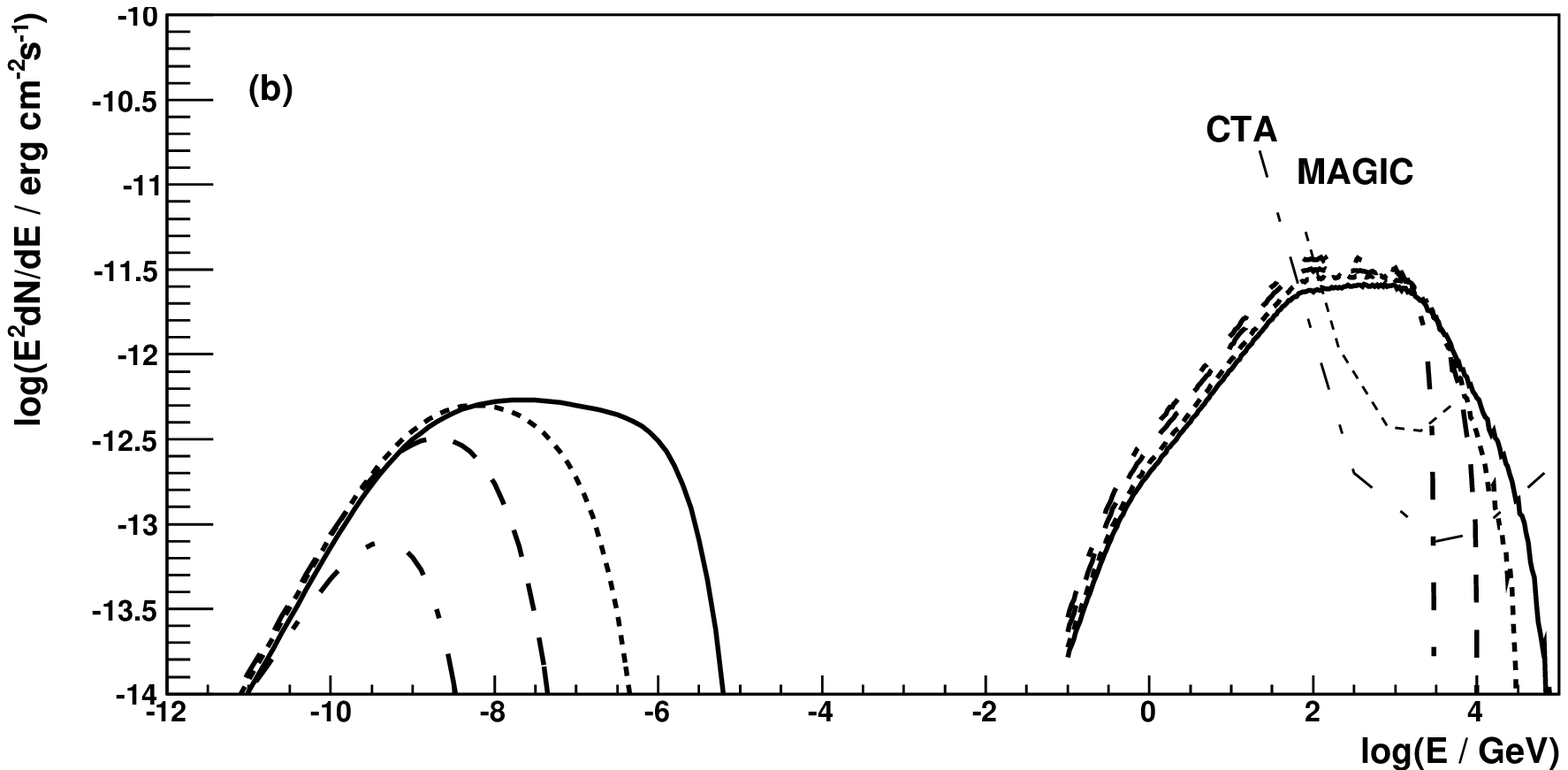}
\includegraphics{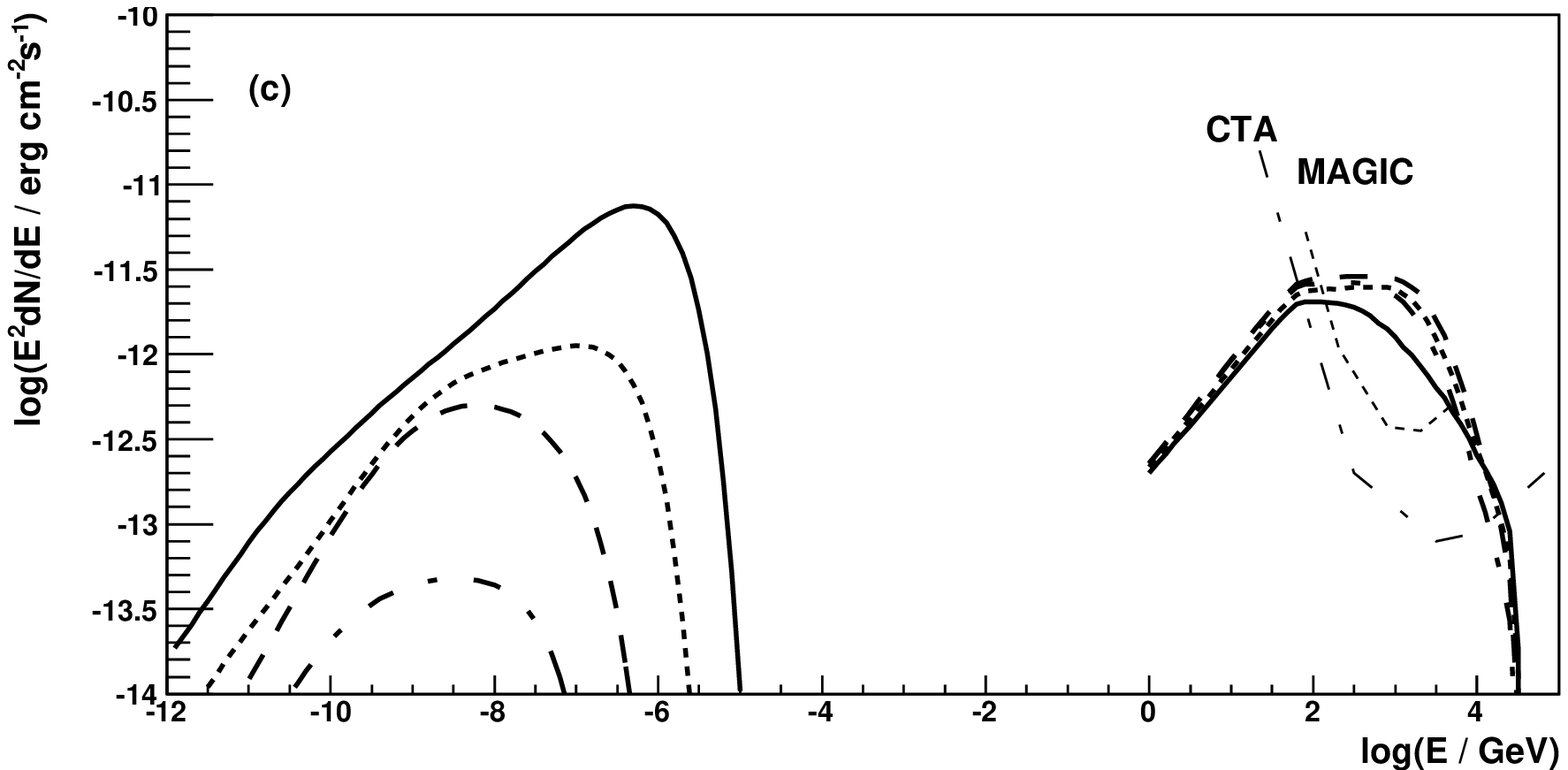}
\includegraphics{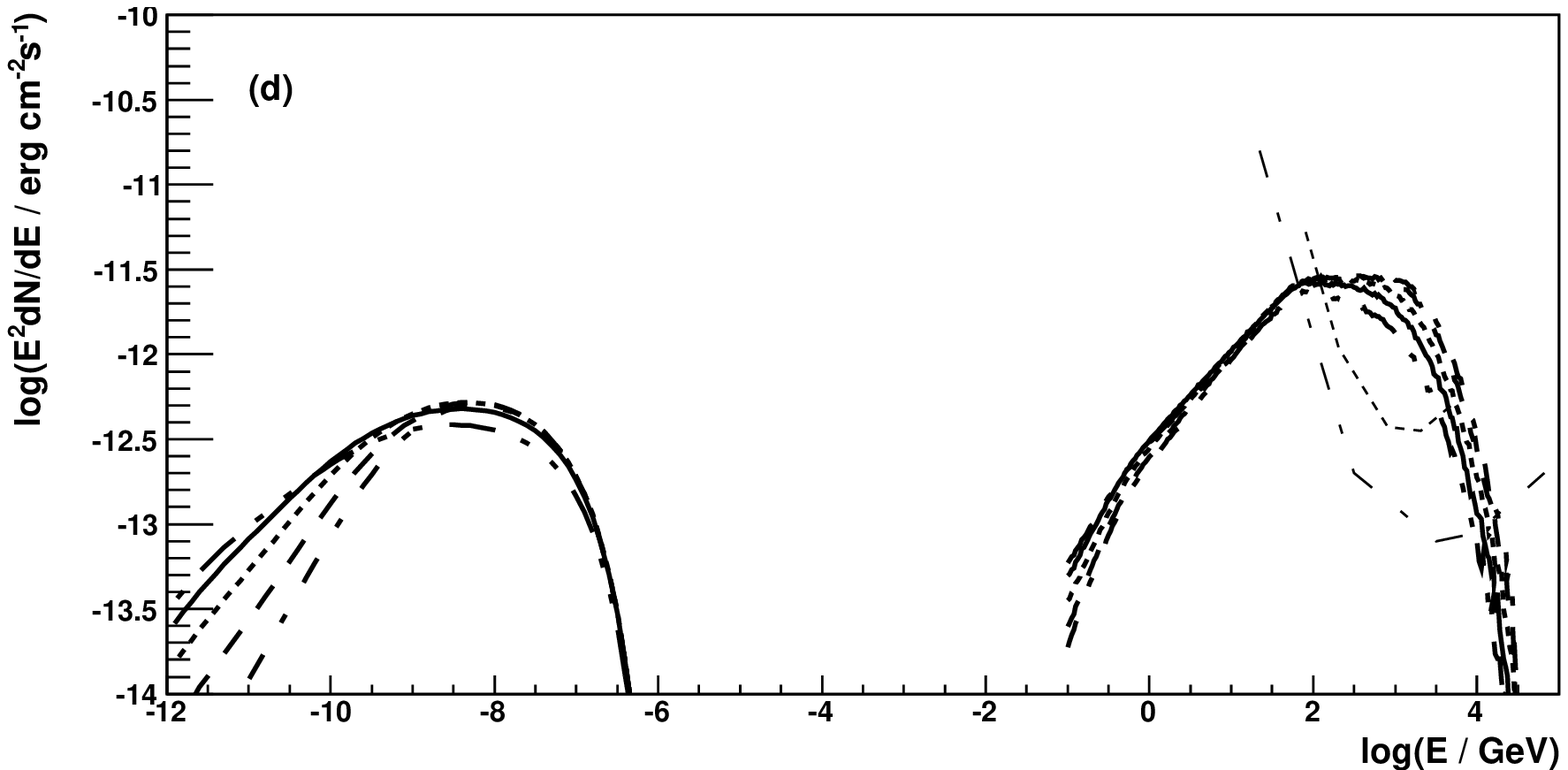}
\includegraphics{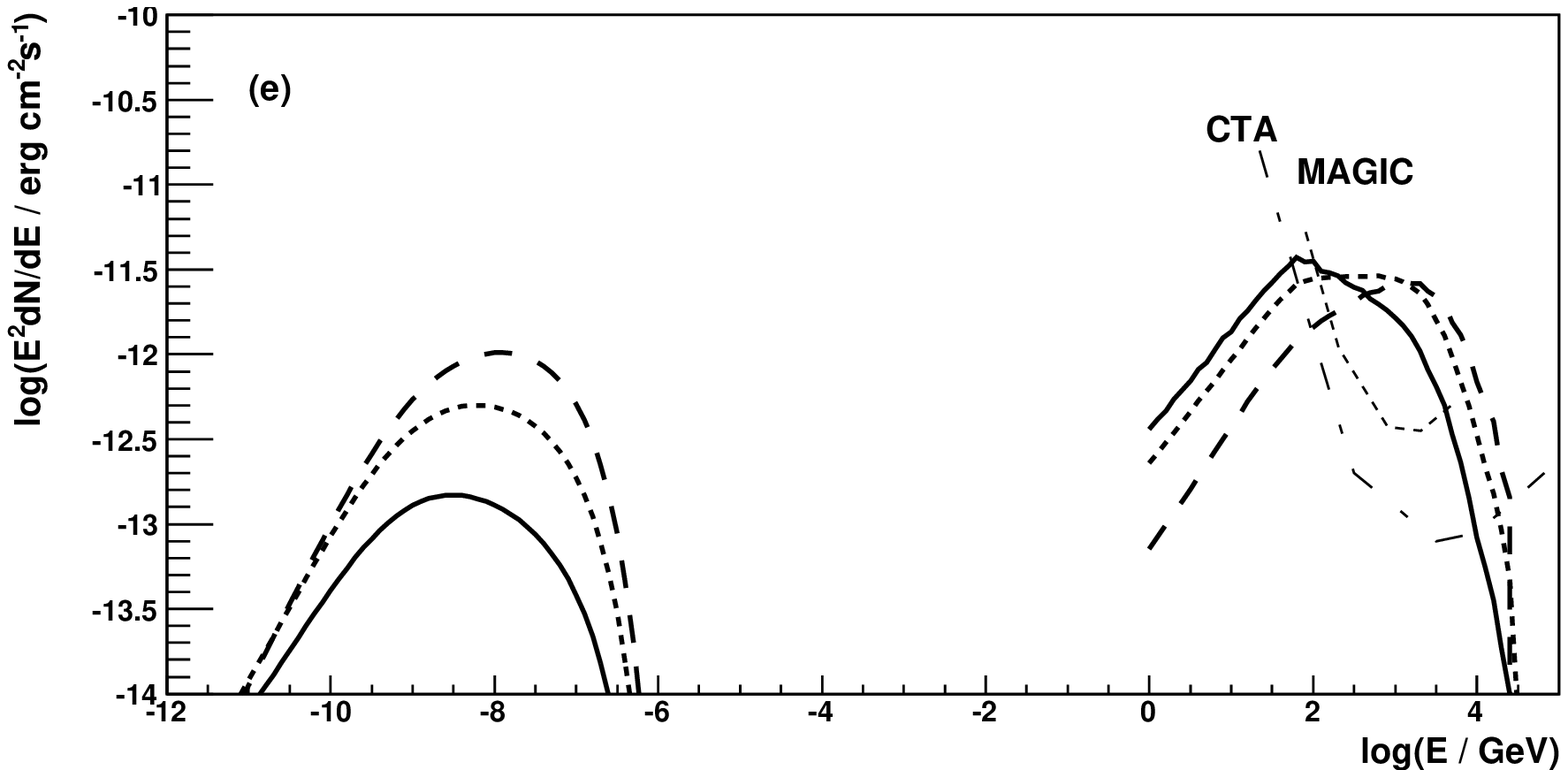}
\caption{As in Fig.~7 but for the GC M28. The dependence of the spectra on the advection velocity (a), on the maximum energy of injected leptons (b), on the magnetic field strength (c), and on the injection distance from the centre of the GC (d) for  $d = 0.33$ pc (dot-dashed). The dependence on the spectral index, equal to 1.5 (solid), 2.05 (dotted), and 2.5 (dashed), is shown in figure (e). The other parameters of the calculations are the same as in Fig.~7.}
\label{fig9}
\end{figure*}
\subsection{The case of PSR B1821-24 in M28}

The pulsed $\gamma$-ray emission from PSR B1821-24 in M28 has been discovered in the Fermi-LAT data with the power of $\sim$2$\%$ of
its spin down rate equal to $2.2\times 10^{36}$ erg s$^{-1}$ (Johnson et al.~2013, Wu et al.~2013). This pulsar has the period of 3.05 ms and the surface magnetic field of $\sim$2$\times 10^9$ G. It is offset from the centre of M28 by 0.2' which corresponds to the projected distance of $\sim$0.33 pc for the distance to M28 estimated on 5.6 kpc. M28 contains eleven other MSPs.

The observed pulsed GeV $\gamma$-ray power of PSR B1821-24 is approximately a factor of 2 lower than the power observed from PSR J1823-3021A. We normalize the power in relativistic leptons injected into the GC from PSR B1821-24 into M28 in a similar way as in the case of PSR J1823-3021A in NGC~6624 i.e., by normalizing this power to the observed pulsed $\gamma$-ray power,
$L_\pm = L_\gamma^{\rm GeV}$. With such assumption, we calculate the non-thermal emission from M28 for a similar range of 
parameters as discussed in the case of NGC~6624 (see Figs.~8 and 9). Interestingly, the expected synchrotron and IC $\gamma$-ray emission have very similar features as expected in the case of NGC~6624. Also, the fluxes of radiation in these energy ranges are very similar which is due to 
the closer location of M28 to the Earth. These two GCs differ significantly in the distance to the Galactic Centre, NGC~6624 at 1.2 kpc 
and M28 at 2.7 kpc. Therefore, the infra-red radiation field might be significantly stronger for leptons escaping from NGC~6624 than from M28 which may result in enhanced inverse Compton flux at sub-TeV energies.

\section{Possible constraints by current Cherenkov telescopes} 

Detection of the TeV $\gamma$-ray emission from specific GC (or an upper limit) will allow to constrain
the injection rate of leptons from the MSP inner magnetosphere or from their wind regions. 
The injection rate of leptons, $\eta_\pm$, and the production rate of GeV $\gamma$-rays, $\eta_\gamma$, produced directly in the inner pulsar magnetosphere, are the basic factors which should be consistent with the predictions of different pulsar models. Thus, TeV $\gamma$-ray observations of GCs could provide important tests of the radiation/acceleration processes in the MSP magnetosphere. At first, we discuss possible constraints obtained from the observations of the whole population of the MSPs within the GC M15. We consider close to the mono-energetic leptons injected from the inner pulsar magnetosphere assuming that they are not re-accelerated
in the pulsar surrounding. The power in injected the leptons can be expressed by, 
$L_\pm = \eta_\pm L_{\rm pul}N_{\rm MSP}$, 
where $\eta_\pm$ is the average part of the power of the MSP winds, $L_{\rm pul}$, converted to leptons in the inner pulsar magnetosphere, and $N_{\rm MSP}$ is the number of MSPs within M15. 
The MSP wind power is approximately equal to the MSP spin down power since the GeV $\gamma$-rays, emitted from the inner magnetosphere, takes only a small part of the pulsar spin down power of the MSP. In a similar way, we can express the pulsed 
$\gamma$-ray power emitted from the inner magnetosphere of the whole population of MSPs within M15, 
$L_\gamma^{\rm GeV} = \eta_\gamma L_{\rm pul}N_{\rm MSP}$, 
where $\eta_\gamma$ is the average part of the spin-down power of MSPs converted to $\gamma$-rays in their inner magnetosphere. The ratio of 
$\eta_\pm/\eta_\gamma = L_\pm/L_\gamma^{\rm GeV}$ is the parameter which is determined by the radiation/acceleration processes in the inner pulsar magnetosphere. This parameter is the basic output of the MSP modelling which can be in principle tested by Cherenkov observations. In the case of M15, we do not know the level of the $\gamma$-ray emission from the pulsar population $L_\gamma^{\rm GeV}$ yet. The available upper limit on the GeV $\gamma$-ray power from the {\it Fermi}-LAT observations is  $5.8\times 10^{34}$ erg s$^{-1}$ (Abdo et al.~2010). Therefore, based on our calculations of the diffusive TeV $\gamma$-ray flux from M15, we can only estimate the limit on $L_\pm$ in the case of observations with the present Cherenkov telescopes. We assume that the telescope arrays  with the sensitivity of the MAGIC observe M15 for 50 hrs producing a signal on the level of 5$\sigma$ (Aleksi\'c et al.~2016). The reachable limits on $L_\pm$, expected from such observations of M15, in the case of the mono-energetic injection of leptons for the range of the parameters describing the model, are shown in the upper part of Table~1. These values, confronted with the value of $L_\gamma^{\rm GeV}$ from M15 (which has to be below the upper limit reported by Abdo et al.~2010), could allow to constrain the parameter $\eta_\pm/\eta_\gamma$ clearly below unity. This means that the power injected from the inner pulsar magnetosphere in leptons should be clearly lower than the power escaping in GeV pulsed $\gamma$-rays.

\begin{table}
  \caption{Constraints on the power injected into M15 in leptons with the mono-energetic (upper part) and the power law (bottom part) spectra for selected model parameters obtained from the comparison of the TeV $\gamma$-ray spectra (shown in Fig.~4a,d,e) with the 50 hr sensitivity of the MAGIC array.}
  \begin{tabular}{llllllllllll}
\hline 
\hline 
$v_{\rm adv}$ (${{cm}\over{s}}$)   &  $10^{6}$  &  $10^{7}$ & $10^{8}$ & $3\times 10^8$ \\
\\
$L_\pm$ (${{erg}\over{s}}$) & 2.2$\times 10^{33}$ &  3.8$\times 10^{33}$  & 1.9$\times 10^{34}$ &  3.9$\times 10^{34}$ \\
\hline
$E_\pm$ (TeV)  & 0.3  &  1   &  3  & 10  \\
\\
$L_\pm$    & 5.7$\times 10^{33}$  &  2.2$\times 10^{33}$   & 2.2$\times 10^{33}$  &  1.3$\times 10^{34}$  \\
\hline 
\hline 
$v_{\rm adv}$ (${{cm}\over{s}}$)   & $10^{6}$  &  $10^{7}$ & $10^{8}$ & $3\times 10^8$  \\
\\
$L_\pm$ (${{erg}\over{s}}$) &  8$\times 10^{33}$  & 1.5$\times 10^{34}$  &  6.2$\times 10^{34}$  & 1.7$\times 10^{35}$ \\
\hline
$B$ (G)  &  $10^{-6}$  &  $3\times 10^{-6}$   &  $10^{-5}$  &  $3\times 10^{-5}$  \\
\\
$L_\pm$  (${{erg}\over{s}}$)   &  8$\times 10^{33}$ &  8$\times 10^{33}$   &  1.1$\times 10^{34}$  &  2.8$\times 10^{34}$ \\
\hline
$\alpha$   &  1.5  &  2   &  2.5  &   \\
\\
$L_\pm$         &  7.8$\times 10^{33}$ &  8$\times 10^{33}$   &  $1.2\times 10^{34}$  &   \\
\hline
\hline 

\end{tabular}
  \label{tab1}
\end{table}

Similar constraints on the parameter $\eta_\pm$ can be done in the case of injection of leptons with the power law spectrum as expected in the case of their re-acceleration process in the wind regions or termination shock regions of the MSPs. These limits are less restrictive than expected in the case of the mono-energetic injection of the leptons (see bottom part of Table~1). However, in this case the values of $\eta_\pm/\eta_\gamma$, clearly larger than unity, are quite possible as suggested by the observations of the nebulae around classical pulsars.  Note that $\eta_\pm/\eta_\gamma$, as defined above, in the case of the Crab Nebula is close to $\sim$10$^3$. In the case of MSPs, around which $\gamma$-ray nebulae have not been detected up to now, the acceleration and radiation processes are certainly not so efficient as around young classical pulsars. In opposite case, the TeV $\gamma$-ray sources in the direction of GCs should have been already detected by the available observations with Cherenkov telescopes. However, recent observations determined only the upper limits  (Aharonian et al.~2009, Anderhub et al.~2009, McCutcheon et al.~2009, Abramowski et al.~2011, Abramowski et al.~2013), except Ter 5 in which case the TeV $\gamma$-ray source is offset from the centre of the GC putting some doubts on their connection (Abramowski et al.~2011). 
We report the limits on the power in the injected leptons with the power law spectrum which could be potentially obtained by 50 hrs observations with the MAGIC array as a function of different parameters which determine this scenario such as the advection velocity of the GC wind, the magnetic field strength with the GC, and the spectral index of the injected leptons (see bottom part of Table.~1). In all considered cases the power in relativistic leptons is already lower than the average spin-down power of only 8 MSPs already observed in M15, assuming that the average spin down power of the MSP is $\sim$1.8$\times 10^{34}$ erg s$^{-1}$ (Abdo et al.~2009). Therefore, detection of the TeV $\gamma$-ray emission from M15 would provide important constraints on the acceleration/radiation processes in the medium around MSPs. 
The present upper limits on the TeV $\gamma$-ray emission from M15 reported by the VERITAS Collaboration (McCutcheon et al.~2009) based on only 6.5 hrs of observations (energy flux $<1.1\times 10^{-12}$ erg s$^{-1}$ cm$^{-2}$ above 600 GeV) allows to put constraints on the power in relativistic leptons on the level of approximately a factor of three larger than reported in Table~1.

More direct constraints on the parameters $\eta_\pm$ and $\eta_\gamma$ can be obtained when the spin-down luminosity and the GeV $\gamma$-ray luminosity of the pulsar injecting leptons is well determined.
As we mentioned above, two MSPs are known to dominate the GeV $\gamma$-ray emission from GCs NGC~6624 and M28, showing modulation of the $\gamma$-ray signals with their rotational periods. For these pulsars we know $L_\gamma^{\rm GeV}$ and $L_{\rm SD}$. Therefore, limits on the power injected in leptons can be directly transformed to the values of $\eta_\pm$. As an example, in Table~2, we show the potential constraints on $\eta_\pm$ which could be obtained from observations of NGC~6624 assuming that the leptons are injected mono-energetically and with the power law spectrum, respectively. These constraints are obtained from the comparison of the expected TeV $\gamma$-ray fluxes (shown in Figs.~6 and 7) with the 50 hr sensitivity of
the present Cherenkov telescopes such as MAGIC. Note that the future Cherenkov Telescope Array (CTA) is expected to have a few times better sensitivity allowing constraints of $\eta_\pm$ on the level of $\sim$$10^{-3}$. Very similar constraints are expected in the case of observations of the GC M28 since the MSP B1821-24 emits a factor of two less pulsed $\gamma$-rays than J1823-3021A (in NGC~6624) but M28 is a factor of $\sim$1.5 closer to the Sun than NGC~6624.   

\begin{table}
  \caption{Possible constraints on $\eta_\pm$ ($\sigma$) from observation of  NGC~6624 for selected model parameters assuming the mono-energetic (upper part) and the power law (lower part) injection spectra of leptons.}
  \begin{tabular}{llllllllllll}
\hline 
\hline 
$v_{\rm adv}$ (${{cm}\over{s}}$)   &  $10^{6}$  &  $10^{7}$ & $10^{8}$ & $3\times 10^{8}$ \\
\\
$\eta_\pm$ ($\sigma$)    &  0.004 (250)  &  0.005 (200) & 0.017 (58) &  0.042 (23) \\
\hline
$E_\pm$ (TeV)  & 1  &  3   &  10  & 30  \\
\\
$\eta_\pm$ ($\sigma$)   &  0.014 (70) &  0.004 (250)  &  0.023 (42) &  0.026 (37) \\
\hline 
\hline 
$v_{\rm adv}$ (${{cm}\over{s}}$)   &  $10^{6}$  &  $10^{7}$ & $10^{8}$ & $3\times 10^{8}$  \\
\\
$\eta_\pm$  ($\sigma$)       &  0.017 (58)   &  0.034 (28) & 0.1 (9)   &  0.26 (2.8) \\
\hline
$\alpha$   &  1.5  &  2   &  2.5    \\
\\
$\eta_\pm$ ($\sigma$)         & 0.003 (310)  &  0.004 (250)   &    0.006 (160)  \\
\hline
\hline 
\end{tabular}
  \label{tab2}
\end{table}

It is interesting to confront possible constraints on $\eta_\pm$, which could be reached by Cherenkov telescopes, with the expectations of some models for the acceleration and radiation processes within the MSP magnetosphere and their vicinity. 
For example, based on the fully 3D general-relativistic polar cap pulsar model, Venter \& de Jager~(2005) estimated the values of the parameter $\eta_\pm$ on $1\%\div 2.5\%$ and $\eta_\gamma$ on $2\%\div 9\%$ for the case of  PSR J0437-4715. These estimates have been generally confirmed in the analysis of the $\gamma$-ray emission from the population of MSPs in Tuc 47, $\eta_\pm\sim2\%$ and $\eta_\gamma\sim 7\%$ (Venter \& de Jager~2008). The estimates of $\eta_\gamma\sim 10\%$ are consistent with other modelling of processes in pulsar's inner magnetosphere based on a space charge-limited flow acceleration polar cap model (e.g. Harding et al.~2002), the outer gap model (Takata et al.~2010), 
and with the estimates based on the {\it Fermi} observations of the population of MSPs (Abdo et al.~2009a, 2009b, see also Fig.~9 in the pulsar catalogue by Abdo et al.~2013). As reported in Table~2 (see also Table~1), the value of $\eta_\pm$ can be clearly constrained below these predictions with the deep observations even with the present Cherenkov telescopes for most of the parameter space considered in our modelling.    

The parameter, $\eta_\pm$, can be also simply related to the well known parameter, $\sigma$, describing the magnetization of the plasma expelled by the pulsar. In fact, assuming that hadrons do not play energetically any role, 
\begin{eqnarray}
\sigma = {{(L_{\rm pul}-L_\pm)}\over{L_\pm}} = {{1}\over{\eta_\pm}}-1. 
\label{eq5}
\end{eqnarray}
\noindent
Note that with the Cherenkov observations of GCs, we can already obtain interesting limits on the value of $\sigma$ 
(see Table~2). According to our calculations for NGC~6624, $\sigma$ could be constrained to values over a few hundred to a few tens for the mono-energetic injection of leptons. The Fermi observations of MSPs indicate that $\sigma$ should be similar to values 
expected for classical pulsars, $\sim10^{3-4}$ (e.g. Timokhin \& Arons~2013). The constraints on a similar level could be obtained for the MSPs in the case of their observations 
with the next generation Cherenkov telescope array such as CTA. In the case of leptons injected with the power law spectra from the pulsar winds, expected lower limits on $\sigma$ with the present telescopes are on the level of a several to a few tens (see Table~2). The values of  $\sigma$ for the winds of classical pulsars are expected to be clearly below unity, e.g. for the Crab Nebula $\sigma_{\rm Crab}\sim 3\times 10^{-3}$ (Kennel \& Coroniti~1984) and for the Vela Nebula $\sigma_{\rm Vela}\sim 0.1$ (Sefako \& de Jager~2003).
Already observations with the present telescopes can constrain $\sigma$ in the MSP winds clearly above values expected for the nebulae around classical pulsars giving insight into the regions in the pulsar winds in which acceleration of leptons is efficient. 
In fact, such large values of $\sigma$ in the case of MSPs within GCs do not seem unreasonable. Note that many MSPs have low mass companions stars (Black Widow and Redback types binaries), which winds can already disrupt the relativistic winds relatively close to the MSPs. Moreover, in contrast to the winds around classical pulsars, even strongly magnetized MSP winds can still produce detectable IC $\gamma$-ray emission since the MSPs are immersed in the dense radiation field from the stars within the GC.

Most of the spin-down power of the MSPs is expected to be carried away in the form of the magnetized, relativistic pulsar wind as in the case of the classical pulsars. However, up to now, pulsar wind nebulae around the MSPs have not been discovered, neither unpulsed GeV $\gamma$-ray emission clearly detected (Abdo et al.~2013) nor the TeV $\gamma$-ray emission. However, similarities between GeV $\gamma$-ray light curves of these pulsars suggest that radiation processes in the inner MSP and classical pulsar magnetosphere should occur similarly.
Therefore, leptons with large multiplicities should be also produced in the MSP magnetosphere and their winds should have similar proprieties. This rises the question on the efficiency of energy transfer from the pulsar to the relativistic leptons already below the light cylinder radius. 
In fact, multiplicity of leptons in the MSP magnetosphere can increase in the case of a non-dipolar structure of the magnetic field,
e.g. introduced by the offset of the dipole axis in respect to the centre of the neutron star (e.g. Harding \& Muslinov~2011b) or multiple components close to the NS surface (e.g. Arons~1993, Takata et al.~2010).
MSPs within GCs are the best observational targets to answer to this basic question. In fact, the MSP winds are expected to find the target within the GC which should be able to dissipate efficiently the  wind energy. The MSP wind can be likely disturbed either in collisions with nearby winds from the stellar companions (many MSPs are within the binary systems of the black widow or redback type), or in collisions with a large number of nearby winds from stars within the GC (also red giants), or even in collisions between 
themselves (Bednarek \& Sitarek~2007) since a GC is expected to contain up to a hundred of MSPs. In these dissipation processes of the MSP wind, the leptons are expected to be accelerated with a power law spectrum. The power in the leptons is limited by the spin-down power of the MSP, i.e. $\eta_\pm$ should be less than unity. Our calculations of the TeV $\gamma$-ray emission from GCs, which contain identified MSPs (i.e. NGC 6624 and M28), show that with the present sensitivities of the Cherenkov telescopes the parameter $\eta_\pm$ can be already significantly constrained below this upper limit (see Table~2). Therefore, we encourage the Cherenkov Collaborations to include in their future plans deep observations of NGC~6624 and M28.

\section{Conclusions} 

We have developed a more complete model for the TeV $\gamma$-ray emission from globular clusters by introducing another mechanism for the transport of leptons across the GC, i.e. advection from the GC with the wind produced by the mixture of the winds from the population of the  millisecond pulsars and the winds from the red giant stars within the GC. We simulate the initial distribution of MSPs within the GC which are responsible for the injection of the leptons or their re-acceleration in the pulsar winds. Moreover, we consider not only the case of multiple injection centres of leptons (MSPs) but also the case of GC dominated by a single, very energetic MSP as recently discovered in the GCs NGC~6624 and M28. In the case of these two GCs, the parameters of the dominating MSP are well known which help to constrain a part of the free parameters defining considered scenario. We can determine the possible upper limits on the parameter $\eta_\pm$, which describes the ratio of the power in the leptons injected by the MSP to the spin down power of this MSP, based on the comparison of expected TeV $\gamma$-ray fluxes with the sensitivities of the modern Cherenkov telescopes such as H.E.S.S., MAGIC or VERITAS. The value of $\eta_\pm$ allows to test the models for radiation/acceleration processes within the inner pulsar magnetosphere and their wind regions or regions of their collisions with GC environment.

At first, we have calculated the TeV $\gamma$-ray production from the inverse Compton scattering of background radiation within the cluster and its surrounding (the optical radiation from GC stars, the infra-red radiation from the nearby galactic disk, and the Microwave Background Radiation) by leptons injected into GC by the MSPs. Also the synchrotron spectra, produced by these leptons in the GC magnetic field are calculated. The spectra are investigated as a function of a few parameters which determine this model. The whole population of MSPs within GC M15, a well known core collapsed northern sky GC, is considered as an example.
The injection of the leptons with the mono-energetic spectra, from the pulsar magnetopsheres, and the power law  spectra, obtained in collisions of pulsar winds with surrounding, are considered in detail. The TeV $\gamma$-ray spectra, expected for the likely range of the free parameters of the model, can be clearly tested with the present Cherenkov telescopes. Based on the comparison of calculations with the telescope sensitivities, we derive the achievable limits on the power in the mono-energetic leptons injected by the MSPs (Table~1). Unfortunately, these limits cannot be transferred directly to the limit on the parameter 
$\eta_\pm/\eta_\gamma$, since the level of the GeV $\gamma$-ray emission from M15 is not well known at present. If it is close to the available upper limit $L_\gamma^{\rm GeV}\approx 5.8\times 10^{34}$ erg s$^{-1}$ (Abdo et al.~2010), then $\eta_\pm/\eta_\gamma$ should be clearly below unity for the most of the considered parameter space (see Table~1). Therefore, the models for the radiation/acceleration processes within the inner MSP magnetosphere can be constrained already by deep observations with the present Cherenkov telescopes.
Note that the synchrotron spectra calculated for the mono-energetic injection of the leptons into M15, peaks at the optical/UV energy range. They are clearly on the level below optical emission from the stellar content of M15. Therefore, they are not expected to be very helpful for providing additional constraints on the parameters characterising M15 such as the energy of leptons or the magnetic field strength within M15. Similar limits on the power in leptons can be obtained in the case of the power law spectrum of the leptons (Table~1). However, the mechanism of acceleration of the leptons in the MSP wind could in principle transfer larger part of spin down power of the pulsar provided that the mechanism of acceleration is similar to that observed in the case of nebulae around young classical pulsars. The synchrotron spectra produced by these leptons for the extreme parameters (magnetic field strength, maximum energies of the leptons) can extend up to the soft X-rays which give a chance to test these extreme values by the observations with the X-ray satellites (see e.g. Eger et al.~2010, Clapson et al.~2011, Wu et al.~2014). 

We also calculated the TeV $\gamma$-ray and synchrotron spectra produced by leptons injected from the energetic pulsars which dominate the GeV $\gamma$-ray power observed from GCs NGC~6624 and M28. As for M15, the range of parameters defining the model has been investigated. The spectra obtained for these two GCs are very similar.  As an example, we show possible constraints on the parameter $\eta_\pm$, which might be obtained with the modern Cherenkov telescopes, based on the comparison of calculations performed for NGC~6624 with  50 hr sensitivities of these telescopes. 
In the case of the mono-energetic injection of leptons from the MSP inner magnetosphere, the value of $\eta_\pm$ can be potentially constrained with the modern telescopes below, $\eta_\pm\sim (4 - 42)\times 10^{-3}$ (see upper part in Table~2), depending on the model parameters. Therefore, such observations should already allow to test predictions of some pulsar models, such as the general relativistic polar cap model (Venter \& de Jager~2008). Less strict constraints on the values of $\eta_\pm$ are expected in the case of leptons injected in the acceleration process operating in the pulsar winds or the wind collisions (see bottom part of Table~2). However, in this case $\eta_\pm$, expected from the theoretical investigations and observations of nebulae around classical pulsars,
can have values close to unity, e.g. in the Crab Nebula. On the other hand, the non-thermal nebulae have not been discovered around MSPs yet. Therefore, the content of the MSP winds, and the acceleration processes occurring within them, might differ significantly. In fact, MSP winds in GCs are expected to dissipate energy much closer to the pulsar, then it is observed in the case of winds around classical pulsars, due to the presence of the nearby multiple stellar winds and the neighbouring pulsar winds. 
We stress however, that constraints on $\eta_\pm$, in the case of acceleration of the leptons in the surrounding of the MSPs, can be uniquely tested with the observations of GCs since their content is better determined (e.g. the 
parameters of MSPs responsible for the acceleration of leptons are well known) then the content of the nebulae around classical pulsars. 
Some of the free parameters of considered here model can be clearly limited which might in principle allow to constrain the parameters describing the propagation of the leptons through the GC medium. However, in reality the propagation of the leptons within GC can be very complicated depending on the location of the specific MSPs within the cluster with different intrinsic ejection rate and also due to the inhomogeneities in the advection process of the leptons due to a non-spherical distribution of the MSPs and the red giants. 
For example, in the present approach we only considered MSPs with this same average parameters and assumed that GC wind is axially symmetric.

\section*{Acknowledgements}
We would like to thank the Referee for many useful comments and suggestions.
This work is supported by the grant through the Polish Narodowe Centrum Nauki No. 2014/15/B/ST9/04043.
JS and TS thank for the partial support from the MNiSzW through the young scientist grant No. B1511500001045.02.


\label{lastpage}

\begin{thebibliography}{99}

\bibitem{ab09a} Abdo, A.A. et al. 2009a Science 325, 846
\bibitem{ab09b} Abdo, A.A. et al. 2009b Science 325, 848
\bibitem{ab10} Abdo, A.A. et al. 2010 A\&A 524, A75
\bibitem{ab13} Abdo, A.A. et al. 2013 ApJS 208, 17 
\bibitem{abr11} Abramowski, A. et al. (HESS Collab.) 2011, A\&A 531, L18
\bibitem{abr11b} Abramowski, A. et al. (HESS Collab.) 2011b, ApJ 735, 12
\bibitem{abr13} Abramowski, A. et al. (HESS Collab.) 2013, A\&A 551, A26
\bibitem{ach13} Acharya, B.S. et al. (CTA Collab.) 2013, APh 43, 3
\bibitem{ah09} Aharonian, F. et al. (HESS Collab.) 2009, A\&A 499, 273 
\bibitem{al16} Aleksi\'c, J. et al. (MAGIC Collab.)  2016 APh 72, 76 
\bibitem{an09} Anderhub, H. et al. (MAGIC Collab.) 2009,, ApJ 702, 266
\bibitem{ar93} Arons, J. 1993, ApJ 408, 160
\bibitem{bed12} Bednarek, W. 2012, JPhG 39, 065001
\bibitem{bs07} Bednarek, W., Sitarek, J. 2007, MNRAS 377, 920
\bibitem{bs14} Bednarek, W., Sobczak, T. 2014, MNRAS 445, 2842
\bibitem{bg70} Blumenthal, G.R., Gould, R.J. 1970, Rev.Mod.Phys. 42, 237
\bibitem{bo08} Boyer, M. L., McDonald, I., Loon, J. T.,Woodward, C. E., Gehrz, R. D., Evans,
A., Dupree, A. K. 2008 AJ 135, 1395
\bibitem{che10} Cheng, K.S. et al. 2010 ApJ 723, 1219
\bibitem{cl11} Clapson, A.C., Domainko, W., Jamrozy, M., Dyrda, M., Eger, P. 2011 A\&A 532, A47
\bibitem{dom11} Domainko, W. 2011, A\&A 533, L5
\bibitem{dy02} Dyks, J. 2002, PhD thesis, Centrum Astronomiczne im M. Kopernika PAN
\bibitem{ed12} Eger, P., Domainko, W. 2012 A\&a 540, A17
\bibitem{edc10} Eger, P., Domainko, W., Clapson, A.C. 2010 A\&A 513, A66
\bibitem{fre11} Freire, P.C.C. et al. (Fermi Collab.) 2011, Science 334, 1107
\bibitem{fre15} Freire, P.C.C. (2015) http://www.naic.edu/~pfreire/GCpsr.html
\bibitem{ha02} Harding, A.K., Muslinov, A.G., Zhang, B. 2002 ApJ 576, 366
\bibitem{ha11a} Harding, A.K., Muslinov, A.G. 2011a ApJ 743, 181
\bibitem{ha11b} Harding, A.K., Muslinov, A.G. 2011b ApJL 726, 10 
\bibitem{ha05} Harding, A.K., Usov, V.V., Muslimov, A.G. 2005, ApJ 622, 531
\bibitem{har82} Hartwick, F.D.A., Cowley, A.P., Grindlay, J.E. 1982 ApJ 254, L11 
\bibitem{hu11} Hui, C.Y., Cheng, K.S., Wang, Y., Tam, P.H.T., Kong, A.K.H., Chernyshov, D.O., Dogiel, V.A. 2011, ApJ 726, 100
\bibitem{joh13} Johnson, T.J. et al. 2013, ApJ 778, 106
\bibitem{ka07} Kabuki, S. et al. for the CANGAROO Collab. 2007, ApJ 668, 968
\bibitem{kc84} Kennel, C.F., Coroniti, F.V. 1984 ApJ 283, 694
\bibitem{kop13} Kopp, A., Venter, C., B\"usching, I., De Jager, O.C. 2013, ApJ 779, 126
\bibitem{kp06} Kuranov, A. G., Postnov, K. A. 2006, Astron. Lett. 32, 393
\bibitem{mc09} McCutcheon, M. et al. (VERITAS Collab.) Proc. 31st ICRC (Lodz, Poland) 2009 arXiv:0907.4974
\bibitem{me09} Meszaros Sz., Avrett E.H., Dupree A.K. 2009 AJ 138, 615
\bibitem{mi63} Michie, R. W. 1963, MNRAS, 125, 127
\bibitem{ok07} Okada, Y., Kokubun, M., Yuasa, T., Makishima, K. 2007 PASJ 59, 727
\bibitem{sd03} Sefako, R.R., de Jager, O.C. 2003 ApJ 593, 1013 
\bibitem{ta10} Takata, J., Wang, Y., Cheng, K. S. 2010 ApJ 715, 1318
\bibitem{ta13} Timokhin, A.N., Arons, A. 2013 MNRAS 429, 20
\bibitem{val07} Valenti, E., Ferraro, F. R., Origlia, L. 2007 AJ 133, 1287
\bibitem{ve05} Venter, C., de Jager, O.C.  2005, ApJL 619, L167
\bibitem{ve08} Venter, C., de Jager, O.C.  2008, ApJL 680, L125
\bibitem{ve09} Venter, C. et al. 2009, ApJ 696, L52
\bibitem{ve15} Venter, C., Kopp, A. 2015, arXiv:1504.04953
\bibitem{vk15} Venter, C., Kopp, A., Harding, A.K., Gonthier, P.L., B\"usching, I. 2015 ApJ 807, 130 
\bibitem{wu13} Wu, J.H.K., Hui, C.Y., Kong, A.K.H., Tam, P.H.T., Cheng, K.S., Dogiel, V.A. 2013 ApJ 765, 47
\bibitem{zai12} Zajczyk, A. 2012, PhD thesis, Centrum Astronomiczne im M. Kopernika PAN and Universit\^e Montpellier 2
\bibitem{zbr13} Zajczyk, A., Bednarek, W., Rudak, B. 2013, MNRAS 432, 3462

\end{thebibliography}
\end{document}